\documentclass[fleqn,usenatbib]{mnras}

\usepackage[T1]{fontenc}

\DeclareRobustCommand{\VAN}[3]{#2}
\let\VANthebibliography\thebibliography
\def\thebibliography{\DeclareRobustCommand{\VAN}[3]                  {##3}\VANthebibliography}

\usepackage{graphicx}	
\usepackage{amsmath}	

\newcommand{\HI}{{\ion{H}{1}}}

\newcommand{\kms}{$\,$km$\,$s$^{-1}$}
\newcommand{\WHz}{$\,$W$\,$Hz$^{-1}$}

\newcommand{\mJybeam}{mJy beam$^{-1}$}

\newcommand{\msun}{{M$_\odot$}}

\newcommand{\atlas}{{{ATLAS$^{\rm 3D}$}}}

\newcommand{\coOne}{{CO(1-0)}}

\newcommand{\degree}{\ensuremath{^\circ}}

\def\HI{\ion{H}{i}}

\newcommand{\ltsima} {$\; \buildrel < \over \sim \;$}
\newcommand{\gtsima} {$\; \buildrel > \over \sim \;$}
\newcommand{\lta} {\lower.5ex\hbox{\ltsima}}
\newcommand{\gta} {\lower.5ex\hbox{\gtsima}}

\title[ALMA CO(1-0) survey of local radio AGN]{An ALMA CO(1-0) survey of the 2Jy sample: large and massive molecular disks in radio AGN host galaxies}
\author[C. Tadhunter et al.]{C. Tadhunter,$^{1}$\thanks{E-mail: c.tadhunter@sheffield.ac.uk}
T. Oosterloo,$^{2,3}$ 
R. Morganti,$^{2,3}$
C. Ramos Almeida,$^{4,5}$
M. Villar Mart\'in,$^{6}$
\newauthor
B. Emonts,$^{7}$
D. Dicken$^{8}$
\\ \\
$^{1}$Department of Physics and Astronomy, University of Sheffield, Sheffield, S7 3RH, UK \\
$^{2}$ASTRON, the Netherlands Institute for Radio Astronomy, Oude Hoogeveensedijk 4, 7991 PD, Dwingeloo, The Netherlands \\
$^{3}$Kapteyn Astronomical Institute, University of Groningen, Postbus 800, 9700 AV Groningen, The Netherlands \\
$^{4}$Instituto de Astrof\'isica de Canarias, Calle V\'ia L\'actea,  s/n, E-38205 La Laguna, Tenerife, Spain\\
$^{5}$Departamento de Astrof\'isica, Universidad de La Laguna, E-38205 La Laguna, Tenerife, Spain \\
$^{6}$Centro de Astrobiolog\'ia (CSIC-INTA), Carretera de Ajalvir, km 4, 28850 Torrej\'on de Ardoz, Madrid, Spain \\
$^{7}$National Radio Astronomy Observatory, 520 Edgemont Road, Charlottesville, VA 22903, USA \\
$^{8}$UK Astronomy Technology Centre, Royal Observatory Edinburgh, Blackford Hill, Edinburgh EH9 3HJ, UK
}
\date{Accepted XXX. Received YYY; in original form ZZZ}

\pubyear{2024}

\begin{document}
\label{firstpage}
\pagerange{\pageref{firstpage}--\pageref{lastpage}}
\maketitle

\begin{abstract}
 The jets of radio  AGN provide one of the most important forms of AGN feedback, yet considerable
 uncertainties remain about how they are triggered. Since the molecular gas reservoirs of the
 host galaxies can supply key information about the dominant triggering mechanism(s), here
 we present Atacama Large Millimeter/sub-millimeter Array (ALMA) \coOne\  observations of a complete sample of 29 powerful radio AGN ($P_{1.4GHz} > 10^{25}$ W Hz$^{-1}$ and $0.05 < z < 0.3$) with an angular resolution of about 2 -- 3 arcsec (corresponding to 2 - 8\,kpc). We detect molecular gas with masses in the range  $10^{8.9} < M_{H_2} < 10^{10.2}$ \msun\ in the early-type host galaxies of 10 targets, while for the other 19 sources we derive upper limits. 
 The detection rate of objects with such large molecular masses  -- $34\pm9$\% -- is higher than in the general population of non-active early-type galaxies (ETG: $<$10\%). 
 The kinematics of the molecular gas are dominated in most cases by rotating disk-like structures, with diameters up to 25\,kpc. Compared with the results for samples of quiescent ETG in the literature, we find a larger fraction of more massive, more extended and less settled molecular gas structures.
 In most of the CO-detected sources, the results are consistent with triggering of the AGN 
 as the gas settles following a merger or close encounter with a
 gas-rich companion. However, in a minority of objects
 at the centres of rich clusters of galaxies,  the accretion of gas cooling from the hot X-ray halos is a plausible alternative to galaxy interactions as a triggering mechanism.
\end{abstract}
\begin{keywords}
galaxies: active -- galaxies: jets -- galaxies: interactions -- radio lines: galaxies
\end{keywords}


\section{Introduction}
\label{sec:Introduction}

The origin of the activity from a super-massive black hole (SMBH) and its impact on the host galaxy continue to be topics of extreme interest in extragalactic astronomy. This is due to the importance of understanding the phenomena involved and to the fact that growing SMBH, known as active galactic nuclei (AGN), are considered a key element in the evolution of galaxies of all types \citep[e.g.][]{Fabian99,Veilleux05,King15,Harrison18}. Not only are they likely to be triggered as a consequence of gas infall as galaxies evolve, but there is growing evidence that they also directly influence the evolution of their host galaxies via the feedback effects of their jets and winds. Clearly, if we are to properly incorporate AGN into galaxy evolution models, it is crucial to understand the symbiosis between AGN and their host galaxies and in particular the presence and properties of the fuelling gas \citep[e.g.][]{Burillo21,Ramos22}.

In this context, radio AGN ($P_{\rm 1.4~GHz} > 10^{24}$ W Hz$^{-1}$) are particularly important, because their relativistic jets are known to couple efficiently with the hot IGM/ICM on scales of 10~kpc – 1 Mpc and directly affect its cooling \citep{Best06,McNamara12}. This is arguably the most important form of AGN-induced feedback in the local universe in terms of overall impact on galaxy evolution; however, radio jets can also affect the interstellar medium (ISM) of their host galaxies on  smaller, (sub-)kpc scales   \citep[e.g. see][and many others]{Holt08,Mukherjee16,Morganti21,Murthy22,Girdhar22,Audibert23}. Moreover, radio AGN are invariably associated with massive ($M_* \gta 10^{11}$ \msun) early-type galaxies (ETG), making it easier to search for the signs of the triggering events such as tidal features, recent star formation activity, and accreted gas reservoirs.

Considerable recent progress has been made in understanding the triggering of radio AGN.  In particular, deep optical imaging studies have demonstrated that a majority (65 -- 95\%) of powerful 2Jy and 3CR radio sources with strong, quasar-like
emission lines -- here labelled strong-line radio galaxies (SLRG)\footnote{Formally SLRG are defined to have high [OIII]$\lambda$5007 equivalent widths ($EW_{[OIII]} > 10$\AA), in contrast to WLRG, which have low [OIII]$\lambda$5007 equivalent widths ($EW_{[OIII]} < 10$\AA) \citep{Tadhunter98}. The SLRG and WLRG classes are similar to, but not exactly the same as high-excitation radio galaxies (HERGs) and low-excitation radio galaxies (LERGs) respectively (see \citealt{Tadhunter16}).} -- show high-surface-brightness tidal features or double nuclei, consistent with the triggering of their AGN in galaxy interactions \citep{Ramos11,Pierce22}. In contrast, weak-line radio galaxies (WLRG) show a
lower incidence of tidal features, and
may be fuelled instead by accretion of gas cooling from the hot halo \citep{Ruffa19a,Maccagni21,Oosterloo24}, or
by the direct accretion of hot gas from the halo \citep{Hardcastle07}.

Further information about the triggering mechanisms is provided by deep {\it Herschel} far-IR observations of the 2Jy
sample
which allow the 
cool ISM contents (using dust masses as a proxy) and star formation rates (SFR) of the host galaxies to be quantified. Consistent with the trends seen in the optical morphology results, it is found that SLRG have
both higher dust masses and SFR than WLRG \citep{Bernhard22}, reinforcing the idea that
the SLRG are triggered by cool gas accreted in galaxy interactions, but that the fuelling gas in many WLRG may have a 
different origin. 
However, for most of the 2Jy objects the {\it Herschel} detections only extend to 160$\,\mu$m, and may miss
the emission from the coolest dust ($T_{dust} <20$\,K), especially in the presence of synchrotron emission from the radio 
cores, which can extend to mm and far-IR wavelengths and be strong in some cases \citep{Dicken23}. Moreover, 
the low spatial resolution  {\it Herschel} observations ($FWHM \sim$12\, arcsec at 160$\,\mu$m) give no information on the distribution and kinematics of the cool
ISM, which are key for understanding the origins of the gas and the triggering of the activity. Therefore, 
for a full picture of the triggering of radio AGN, it is important to obtain CO observations of the molecular gas, which directly probe
the masses, distributions and kinematics of the cool ISM.

Pre-ALMA  CO surveys of samples of powerful radio AGN in the local universe detected substantial molecular gas reservoirs in some objects \citep{Evans05,Ocana10,Smolcic11}, but lacked the spatial resolution necessary to map the
distribution and kinematics of the gas, which are key for understanding its origins and interplay with the radio plasma. Therefore, the higher spatial resolution and sensitivity offered by mm interferometers like ALMA have the potential to lead to a major step forward in this field. 
Spatially resolved observations of molecular gas in powerful radio AGN are still limited to a small number of local objects in modest samples (e.g. \citealt{Ruffa19a, Audibert22}) or studies of single objects (e.g. some of the more recent additions are \citealt{Morganti15,Espada17,Oosterloo19,Rose19,Maccagni21,Morganti21,Carilli22,Oosterloo24}),  in some cases observed with high (sub-arcsec) angular resolution.
In addition,  particular attention has been given to tracing the molecular gas in and around radio AGN in (cool core) clusters (see e.g. \citealt{Tremblay16,Russell19,Rose20} and refs therein). In such systems, the mass and radial extent of the molecular gas can be particularly large and only in a minority of cases displays a clear and regular rotation (see e.g. \citealt{Rose19}).

These observations have already shown the presence of massive molecular gas disks (see e.g. \citealt{Rose19,Ruffa19a,Ruffa19b,Ruffa20,Audibert22}) in a significant fraction of nearby radio AGN, but also gas with disturbed kinematics suggestive of the infall of clouds
in some cases (e.g. \citealt{Tremblay16,Maccagni18,Maccagni21}), and signs of molecular outflows  \citep[e.g.][]{Morganti15,Oosterloo17,Oosterloo19,Morganti21,Murthy22} in others. However, CO studies of radio AGN have sometimes been done in a piecemeal way, with most concentrating on nearby ($z<0.05$) radio AGN  which have relatively low radio powers ($P_{\rm 1.4GHz} < 10^{25.5}$ \WHz). It is now important to extend such studies to larger samples that encompass more powerful radio sources, to identify whether differences in the occurrence and properties of the molecular gas are present, thereby allowing a systematic investigation of how the molecular gas contents relate to host galaxy and AGN properties.

To reach this goal, here we present ALMA Cycle 7 \coOne\ observations of a complete sample of 29 nearby ($0.05 < z < 0.3$), powerful ($10^{25} < P_{1.4GHz} <10^{27.5}$ W Hz$^{-1}$) 2Jy radio AGN that are a factor of $\sim$3 -- 100 deeper in rms sensitivity than previous CO(1-0) observations of samples of powerful radio AGN, and have an angular resolution ($\sim$2.1 -- 2.9 arcsec) that is high enough to trace the gas distribution and kinematics. We target the CO(1-0) transition, because it is less affected by excitation conditions than the higher CO transitions, and is thus the most reliable tracer of the overall molecular gas mass. The sample and observations are described in Sections 2 and 3 respectively, and the main results in Section 4. This is followed by a comparison with the results from other samples of radio AGN, quasars and early-type galaxies (ETG) in Section 5, and a discussion that puts the results in the context of the triggering of radio AGN in Section 6. Finally, the conclusions are presented in Section 7. All distances, spatial scales and luminosities were calculated assuming a flat Universe with $\Omega_{\rm M} = 0.286$, $\Omega_{\rm vac} = 0.714$, and $H_\circ = 69.6$ \kms\ Mpc$^{-1}$.

\begin{table*} 
\caption{Basic target information and redshifts. The redshifts of most of the observed galaxies are taken from \citet{Dicken08}, but those marked with $a$, $b$, $c$, $d$ and $e$ are taken from the 6dF redshift survey, \citet{Santoro20}, \citet{Hamer16}, \citet{Inskip07} and \citet{Koss22} respectively. The 3th and 4th columns 
give the optical and radio classifications of the sources: WLRG (weak-line radio galaxy: $EW[OIII] < 10$\AA); SLRG (strong-line radio galaxy: $EW[OIII] > 10$\AA); FRI (Fanaroff-Riley class I); FRII (Fanaroff-Riley class II); FRI/FRII (hybrid radio morphology between FRI and FRII); CSS (compact steep spectrum radio source); and GPS (gigahertz-peaked radio source).  The 5th column gives an indication of the large-scale environments of the sources: objects with  extended
ICM X-ray luminosities $L_{2-10\,\rm keV} > 5\times10^{43}$\,erg s$^{-1}$ (as measured by \citealt{Ineson15}) and/or an angular clustering amplitude $B_{gq} > 600$ (as measured by \citealt{Ramos13}) are taken to be in relatively rich, cluster-like environments (indicated by "C"); whereas objects not fulfilling these requirements are assumed to be in lower-density group, or isolated, environments (indicated by "G/I"). Note that the [OIII] emission-line luminosities (6th column) are taken from \citet{Dicken23}, and the CO rms values (10th column) were calculated for a channel width of 30\,\kms. The final column gives the major and minor axes of the restoring beam.}
\begin{center}
\begin{tabular}{lclllclcccccc} 
\hline\hline 
   &Alt. & Optical & Radio &Env. & L$_{\rm [OIII]}$ & $z$ &T$_{\rm obs}$   & rms$_{\rm cont}$ & rms$_{\rm CO}$ & Beam \\
   &Name &Class  &Class & & log(W) &  &(sec)   & (mJy/b) &  (mJy/b)  & (arcsec)    \\
\hline
PKS0034-01 & 3C~15 &WLRG &FRI/FRII &G/I  & 33.5 & 0.073 & 1663 & 0.10 & 0.25 & 2.26$\times$1.82  \\
PKS0035-02 & 3C~17 &SLRG &FRII  &G/I &35.1 & 0.22  & 695 & 0.49 & 0.46 &  2.60$\times$2.04 \\
PKS0038+09 & 3C~18 &SLRG &FRII  &G/I &35.2 & 0.188  & 1482 & 0.11 & 0.28 & 2.23$\times$1.75 \\
PKS0043-42 & &WLRG &FRII &G/I &33.7 & 0.116  & 1512 & 0.03 & 0.33 & 2.42$\times$1.80 \\
PKS0213-13 & &SLRG &FRII &G/I  &35.1 & 0.147  & 1421 & 0.04 & 0.32 & 2.40$\times$1.96  \\
PKS0349-27 & &SLRG &FRII &G/I &33.7 & 0.066 & 847 & 0.03 & 0.44 & 2.51$\times$1.70 \\
PKS0404+03 & 3C~105&SLRG &FRII &G/I &34.5 & 0.089 & 847 & 0.02& 0.41 & 2.27$\times$1.97 \\
PKS0442-28 & &SLRG &FRII &G/I &34.8 & 0.147 & 1391 & 0.05 & 0.28 & 2.31$\times$1.98 \\
PKS0620-52 & &WLRG &FRI &C &$<$32.4 & 0.051 & 1935 & 0.16 & 0.33 & 2.18$\times$1.96 \\
PKS0625-35 & &WLRG &FRI &C &33.5 & 0.055 & 1724 & 0.15 & 0.24 & 2.17$\times$1.89 \\
PKS0625-53 & &WLRG &FRI &C &$<$33.1 & 0.054 & 1905 & 0.05 & 0.33 & 2.12$\times$2.00  \\
PKS0806-10 & &SLRG &FRII &G/I &35.8 & 0.10899$^a$ & 393 &0.03 & 0.59 & 2.93$\times$2.38 \\
PKS0945+07 & 3C~227 &SLRG &FRII &G/I &34.9 & 0.086 & 1663 & 0.02 & 0.31 &  2.22$\times$1.87 \\
PKS1151-34 & &SLRG &CSS &G/I &35.5 & 0.2579$^b$ & 696 & 0.35 & 0.44 &  2.84$\times$1.82 \\
PKS1559+02 & 3C~327 &SLRG &CSS &G/I &35.3 & 0.104 & 393 & 0.05 & 0.60 & 2.44$\times$2.03 \\
PKS1648+05 & 3C~348 &WLRG &FRI/FRII &C &33.7 & 0.1547$^c$ & 756 & 0.08 & 0.39 & 2.78$\times$1.73 \\
PKS1733-56 &  &SLRG &FRII &G/I &34.8 & 0.098 & 424 & 0.17 & 0.56 & 2.22$\times$1.88 \\
PKS1814-63 & &SLRG &CSS &G/I &33.6 & 0.06374$^b$ & 514 & 0.07 & 0.64 & 2.27$\times$1.93 \\
PKS1839-48 & &WLRG &FRI &C &$<$32.4 & 0.112 & 1572 & 0.14 & 0.32 & 2.35$\times$1.65 \\
PKS1932-46 & &SLRG &FRII &G/I &35.4 & 0.2307$^d$ & 726 & 0.06 & 0.39 & 2.35$\times$1.87 \\
PKS1934-63 & &SLRG &GPS &G/I &35.1 & 0.1826$^b$ & 1572 & 0.10 & 0.28 & 2.20$\times$1.97 \\
PKS1949+02 & 3C~403 &SLRG &FRII &G/I &34.9 &0.0584$^e$ & 454 & 0.05 & 0.64 &  2.37$\times$1.77\\
PKS1954-55 & &WLRG &FRI &C &$<$32.0 & 0.06 & 1875 & 0.10& 0.31 & 2.17$\times$2.04 \\
PKS2135-14 & &SLRG &FRII &G/I &36.1 & 0.2 & 1361 & 0.07 & 0.33 & 2.84$\times$1.60 \\
PKS2211-17 & 3C~444 &WLRG &FRII &C &33.4 & 0.153 & 726 & 0.03& 0.43 & 2.82$\times$1.56 \\
PKS2221-02 & 3C~445 &SLRG &FRII &G/I &35.2 & 0.057 & 1753 & 0.02& 0.34 & 2.45$\times$1.73 \\
PKS2314+03 & 3C~459 &SLRG &FRII &G/I &35.2 & 0.2199$^b$ & 363 & 0.05 & 0.58 & 2.33$\times$1.92 \\
PKS2356-61 & &SLRG &FRII &G/I &35.0 & 0.096 & 877 & 0.06& 0.40 & 2.09$\times$1.88 \\
\hline
PKS0915-11 & 3C~218 &WLRG &FRI &C &33.5 & 0.054 & 2700 & -- & 0.21 & Rose et al.    \\
\hline\hline 
\end{tabular}
\end{center}
\label{tab:info}
\end{table*}

\section{The sample and its multi-wavelength properties}
\label{sec:Sample}

The sample used in this work is a complete sub-set of the full 2Jy sample of 46 southern radio galaxies with intermediate redshifts $0.05 < z < 0.7$, high 2.7\,GHz radio fluxes $S_{\rm 2.7~GHz}>2$ Jy, southerly declinations $\delta < +10^\circ$, and steep radio spectral indices $\alpha > 0.5$ (for $F_{\nu} \propto \nu^{-\alpha}$), as presented in
\citet{Tadhunter98} and \citet{Dicken09}. 
The radio source, AGN and host galaxy properties of this sample are well characterised by a rich set of multi-wavelength data:
all objects have deep ESO2.2m, ESO3.6m, NTT or VLT optical spectra \citep{Tadhunter93,Holt07}, Gemini r$^\prime$-band images \citep{Ramos11,Ramos13}, {\it Spitzer}/MIPS and {\it Herschel}/PACS mid- to far-IR photometry \citep{Dicken08,Dicken09,Tadhunter14,Dicken23}, and high and low frequency VLA/ATCA radio maps \citep{Morganti93,Dicken08}; 98\% have deep Chandra or XMM X-ray data \citep{Mingo14}; 90\% have deep {\it Spitzer}/IRS mid-IR spectra \citep{Dicken12}. 

For this project to observe  CO(1-0), we have selected a complete sub-set of all 29 2Jy-sources with redshifts in the range $0.05 < z < 0.3$. 
For one target, Hydra~A (PKS0915--11), sufficiently deep \coOne\ ALMA observations already existed and have been presented in \citep[][ref: ADS/JAO.ALMA\#2016.1.01214.S,  ADS/JAO.ALMA\#2017.1.00629.S]{Rose19}.
We observed the remaining 28 objects in our Cycle 7 ALMA project (ref: ADS/JAO.ALMA\#2019.1.01022.S). Basic information about the targets and their ALMA  observations is given in Table \ref{tab:info}.

The upper redshift limit is chosen to ensure that the objects are at sufficiently low redshift to be confident of detecting CO(1-0) in reasonable integration times in ALMA Band 3, while the lower redshift limit ensures that the objects host genuinely powerful radio sources. On the basis of their optical [OIII] emission-line and 24\,$\mu$m mid-IR continuum luminosities, the sample covers 2 orders of magnitude in bolometric luminosity for a given radio power, with 19 of the targets classified as strong line radio galaxies (SLRG) and 10 objects classified as weak-line radio galaxies (WLRG)  based on their optical spectra. In terms of radio classifications, 21\% are FRI, while the remaining 79\% are FRII, FRI/FRII, CSS (compact steep spectrum radio source) or GPS (gigahertz-peaked radio source), see \cite{ODea21} for a review of peaked-spectrum sources.

The angular resolution obtained by the observations presented here corresponds to $\sim$2\,kpc for the lowest redshift objects in our sample, and increases to $\sim$8\,kpc for the highest redshift in the sample. This resolution is sufficient to separate the CO emission of the radio AGN host galaxies from that of the companion galaxies that are present in many cases, thus providing a major advantage compared with previous single-dish CO observations.
For one object, PKS2314+03 (3C~459), we also make use of higher spatial resolution (0.238 arcsec) data  available in the archive\footnote{ADS/JAO.ALMA\#2018.1.00739.S, PI Balmaverde}. The latter data are used only for the modelling of the kinematics of the gas (see Section \ref{sec:Kinematics}), whereas all other molecular gas properties for this
source (e.g. molecular gas mass and extent) were derived from the lower resolution data.

\section{Observations}
\label{sec:Observations}

The observations of \coOne\ and the 3-mm continuum were obtained in Cycle 7 using ALMA with 46 antennas in its most compact configurations (C43-1, C43-2), with a shortest baseline of 15\,m and a maximum baseline of about 500\,m. The observations were done during the period 2019-10-30 to 2019-11-14.   
The angular resolution of the observations ($\sim$2.1 -- 2.9 arcsec)
was chosen  to allow the observations to be sensitive to the detection of extended and low surface-brightness molecular gas in a relative short integration time. The on-source exposure times were set based on pre-existing information on the cool ISM contents in the targets provided by {\it Herschel}-derived dust mass estimates \citep{Tadhunter14,Bernhard22},  and assuming a  factor 100 conversion factor from dust mass to molecular gas mass\footnote{This is consistent with the total gas-to-dust ratios of nearby galaxies of solar metallicity \citep[e.g.][]{Ruyer14,DeVis19,Casasola20} and a molecular-to-neutral hydrogen mass ratio of $M_{H_2}/M_{HI} \sim 1$.}
The \coOne\ observations of Hydra~A, taken from the ALMA archive, were obtained with a comparable resolution (1.8 arcsec) and reach a noise level of 0.7 mJy beam$^{-1}$ for a  2.7\,km s$^{-1}$ \citep{Rose19} velocity channel, equivalent to 0.22 mJy beam$^{-1}$ for a 30\,km s$^{-1}$ velocity channel -- slightly more sensitive than typically obtained for the rest of the sample.

The field of view (FoV) of $\sim$60$^{\prime\prime}$ and relatively low spatial resolution allow the observations to be sensitive to extended emission (both line and continuum; most of the objects are extended radio galaxies). However, in some cases the FoV was not large enough to trace the entire continuum emission. This does not affect the interpretation of the \coOne\ results. The observations were done in Band 3, making use of the correlator in Frequency Division Mode using four spectral windows. The central frequency of the line spectral window was set to the redshifted frequency of the \coOne\ observations (rest frequency 115.3\,GHz),  with a total bandwidth of 1.875 GHz.  With the 3840 channels used we had a native velocity resolution of 1.4\,\kms, but in the subsequent data reduction channels were combined to make image cubes with 15\,km s$^{-1}$ velocity channels, and a velocity resolution better matching the observed line widths (see below). The remaining spectral windows, each 2 GHz wide covered by 128 channels,  were centred on frequencies surrounding the line spectral window and were used for continuum imaging.
The calibration was done in CASA (v5.1.1; \citealt{CASA22}), which includes the ALMA
pipeline (v. r40896; \citealt{Hunter23}), using the reduction scripts provided by the ALMA observatory.  

The products (continuum and line cube) provided by the ALMA pipeline were found to be of sufficient quality to perform the analysis. In Table \ref{tab:info} we summarise, for each target, the observing times, restoring beams and rms noises of the final continuum images and line data cubes. 

The cubes provided by the ALMA pipeline were made with a robust weighting of 0.5, resulting in an average size of the restoring beam of about 2 arcsec, and a rms noise ranging from 0.24 to 0.64  \mJybeam\ for a channel width of 30\,\kms. The numbers are summarised in Table \ref{tab:info}.  The velocity resolution of $\sim$30\,\kms (with the exact value depending on the redshift of the source) is the result of binning 10 channels and then Hanning smoothing the original cubes. 
For the modelling of the kinematics of the gas for the stronger detections, we have produced cubes with a lower velocity resolution (60\,\kms) in order to increase the signal-to-noise. 

The  cubes were visually inspected for the detection of \coOne. However, to make the analysis systematic, they were also run through the Source Finding Application (SoFiA)\footnote{https://github.com/SoFiA-Admin/SoFiA}, see \cite{Serra15} for details. This software performs spatial and velocity smoothing in order to optimise the search for detections. The set of smoothing kernels used comprises all combinations of smoothing with 0, 3, and 7 pixels spatially and in velocity, and we built the detection mask by using those voxels above the 3.5-$\sigma$ level in any of the smoothed cubes. All mask regions smaller than 3 pixels in  any direction were discarded as being noise peaks. Following this, all masks were dilated with 2 spatial pixels and 1 velocity channel in order to include faint emission at the edges of the masks.  
The moment images shown in Figure \ref{fig:VeloField} are those obtained by SoFiA after visually checking that the  procedure gave reliable results.



The continuum images were produced by the ALMA pipeline by combining the three continuum spectral windows, 
The images were also produced using a robust weigthing of 0.5, resulting in a restoring beam of $\sim$2  arcsec and with rms noise ranging from $\sim$0.03 to $\sim$0.5 \mJybeam\ (see Table \ref{tab:info}). The objects with high noise often have emission extending over the entire field of view (and in many cases beyond). 

\begin{figure*}
   \centering
   
\includegraphics[height=7.5cm]{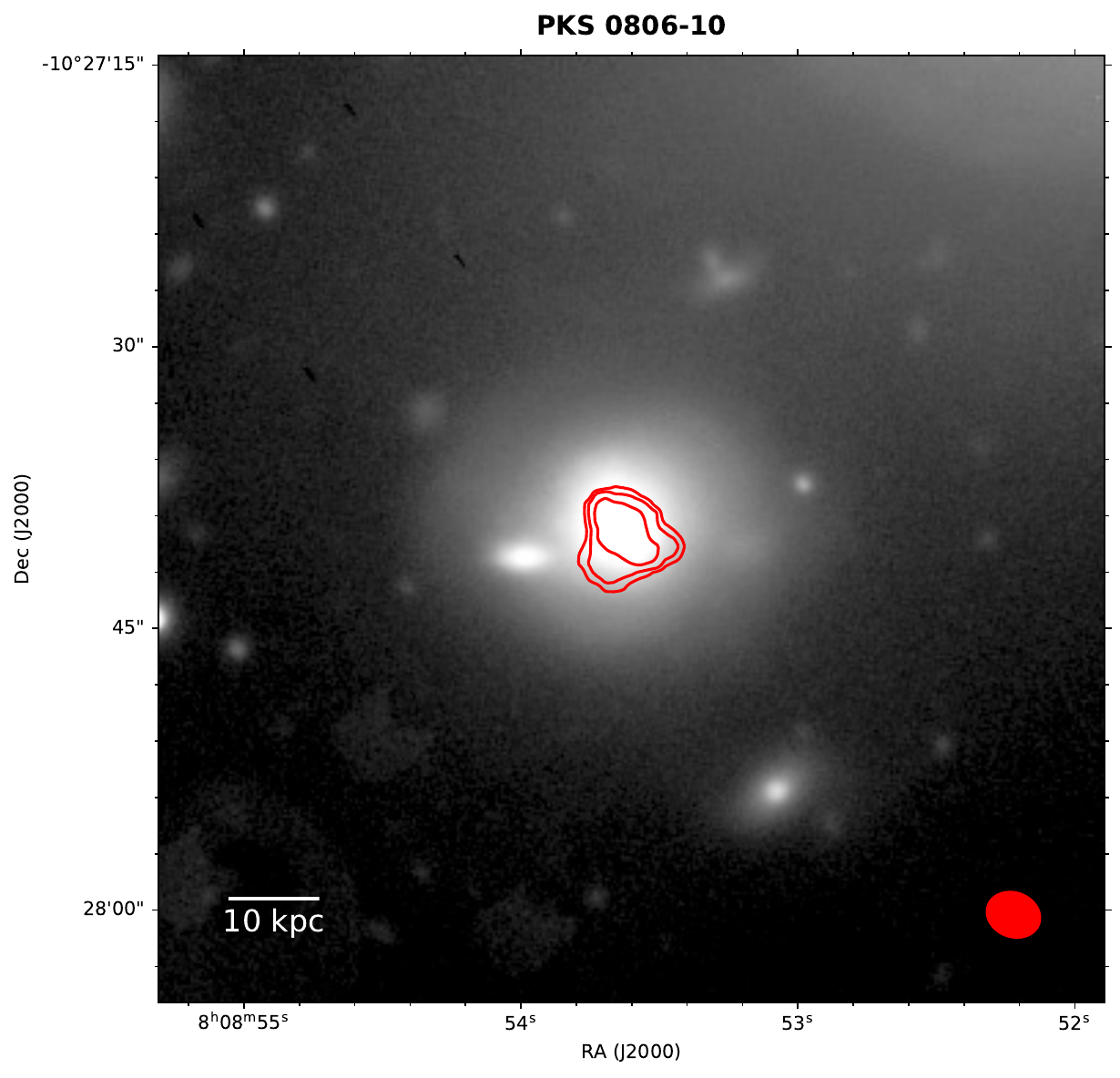}
\includegraphics[height=7.5cm]{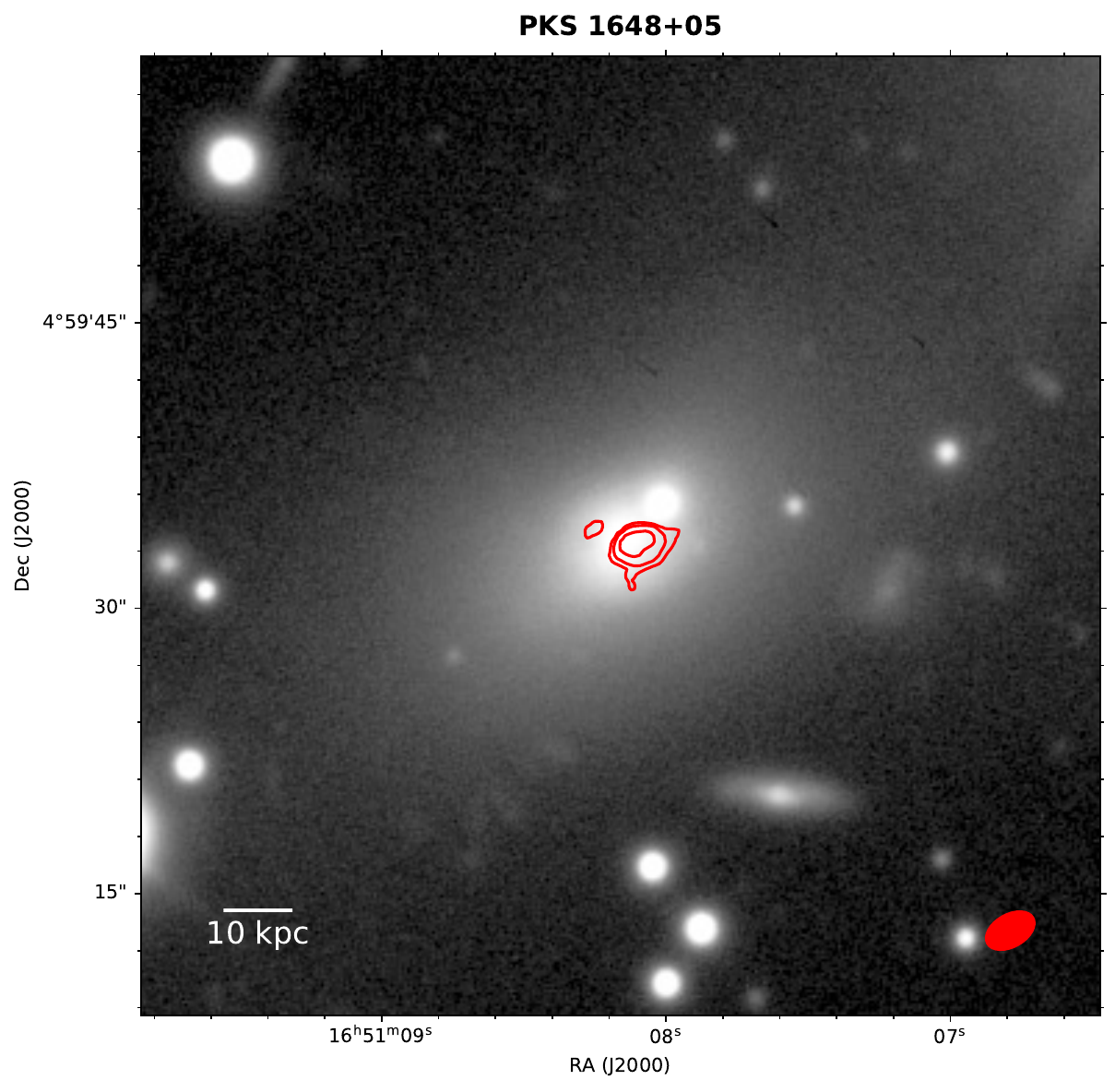} 

\includegraphics[height=7.5cm]{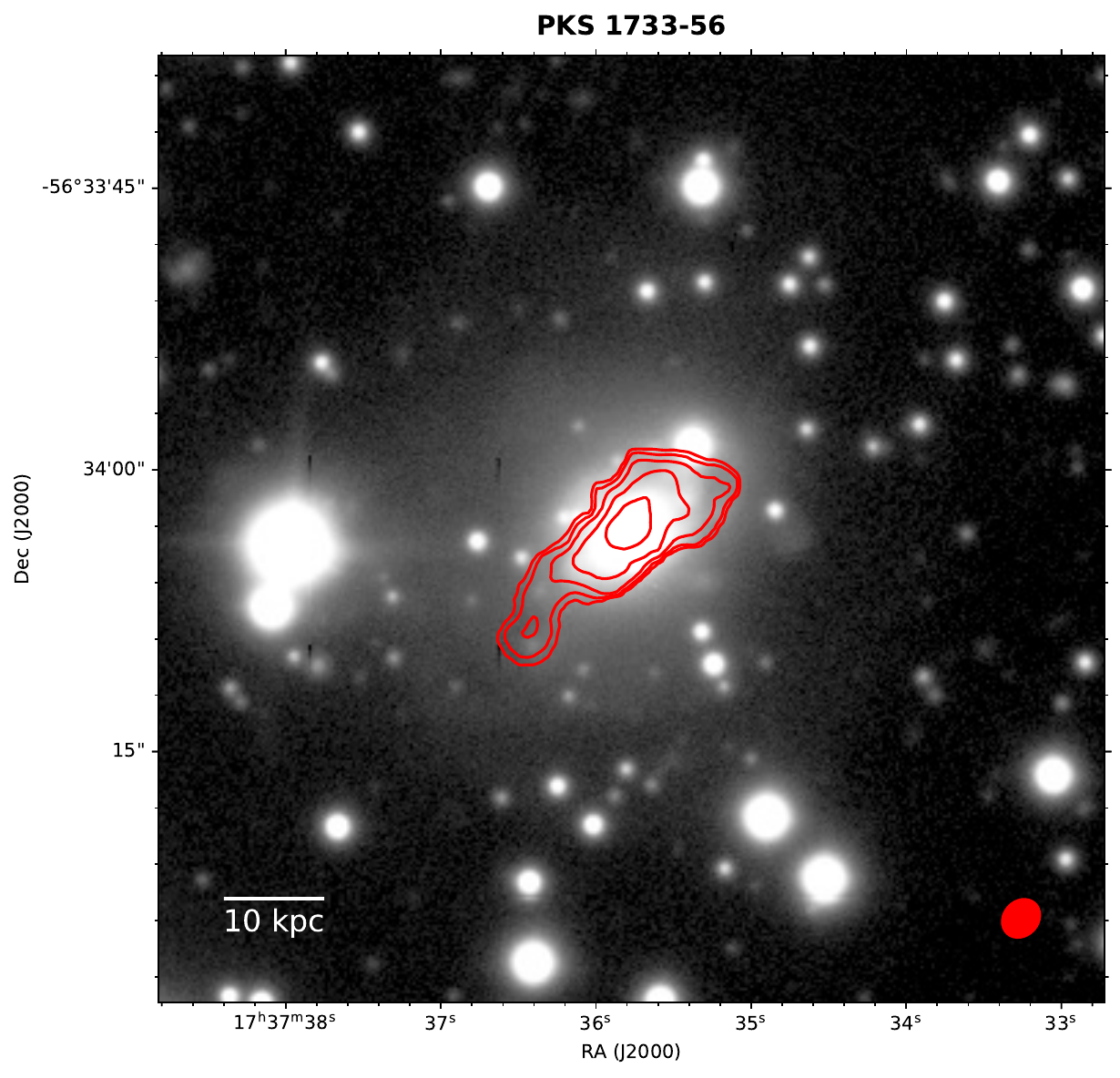}
\includegraphics[height=7.5cm]{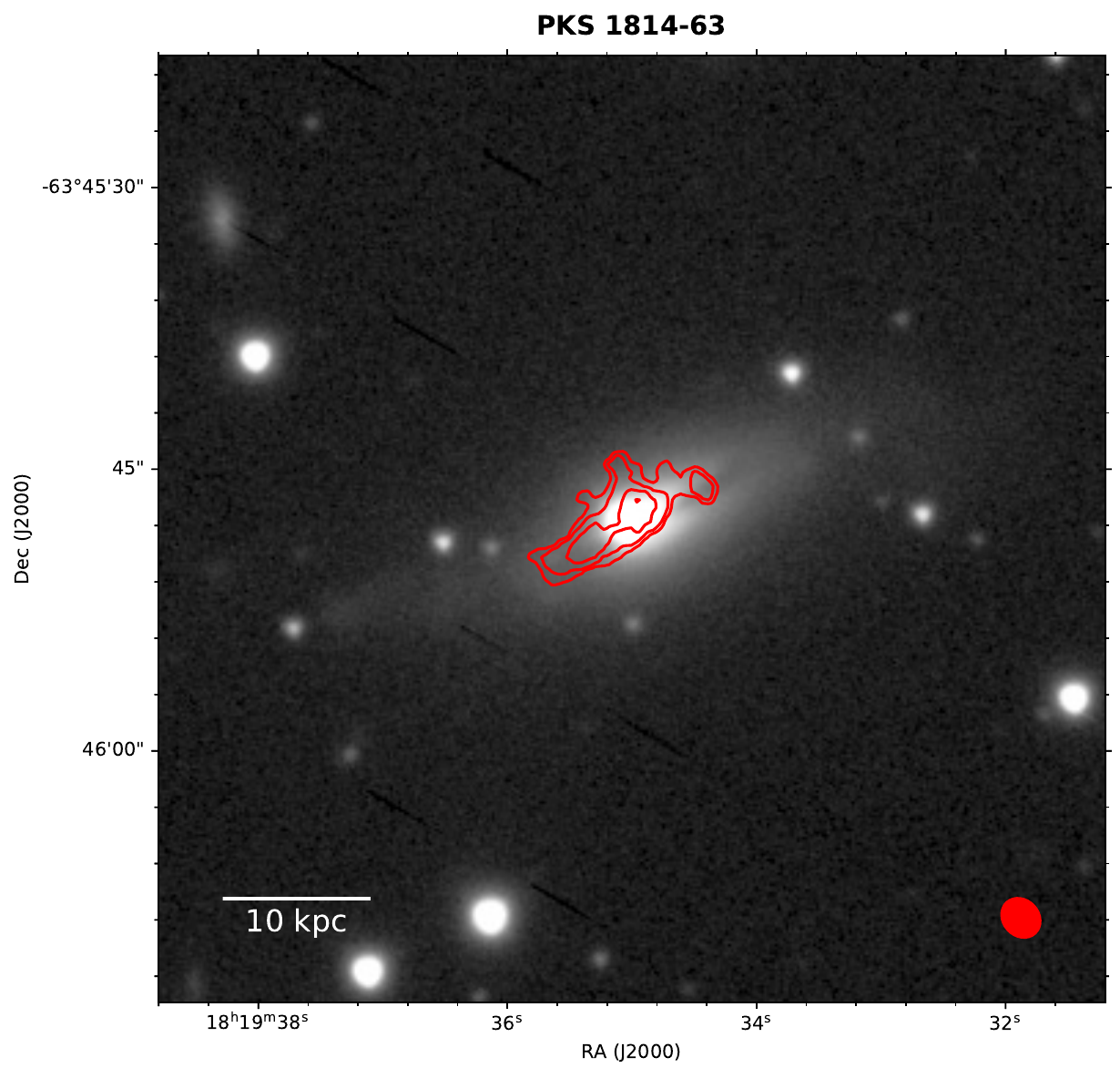}

\includegraphics[height=7.5cm]{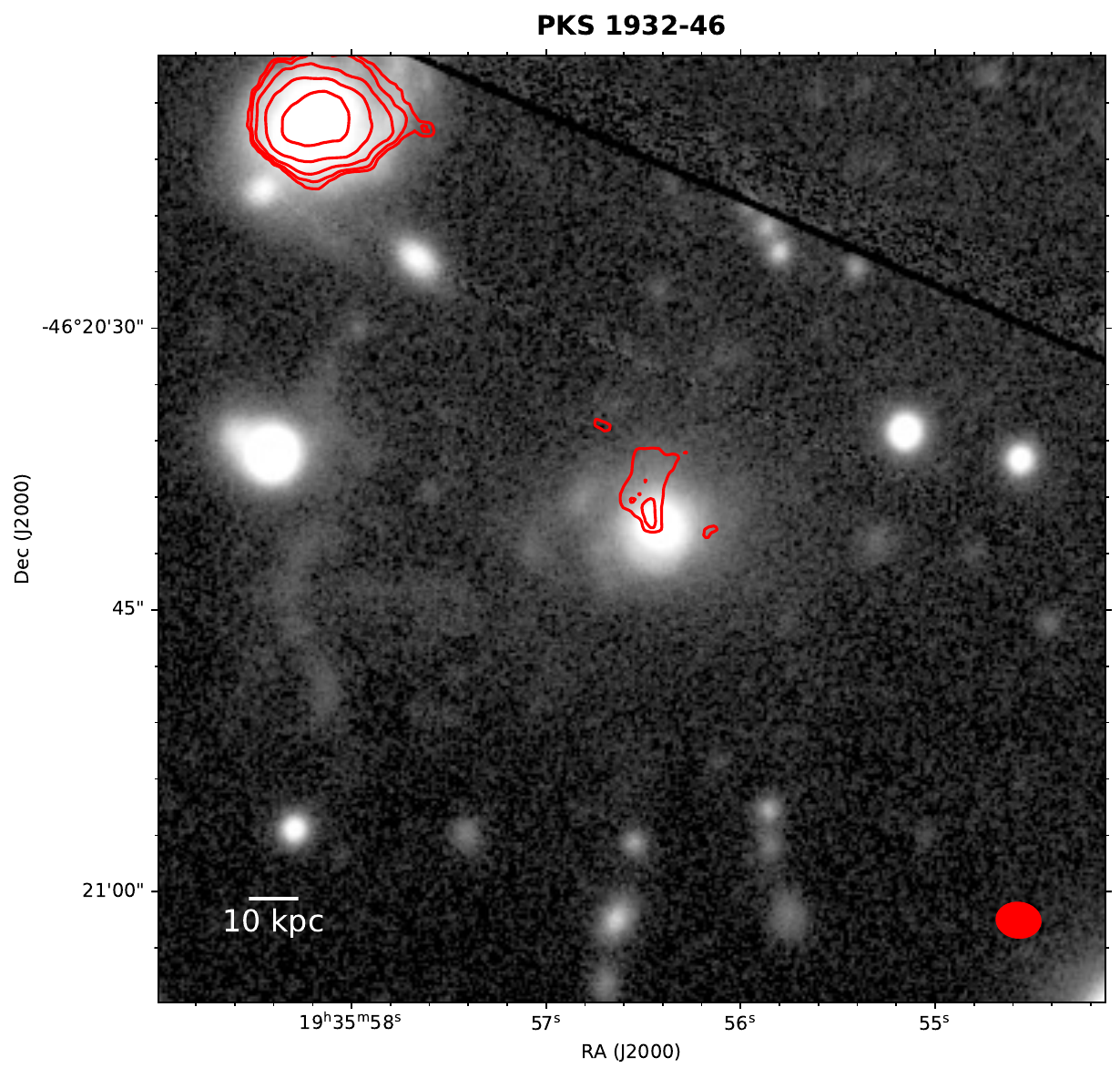}
\includegraphics[height=7.5cm]{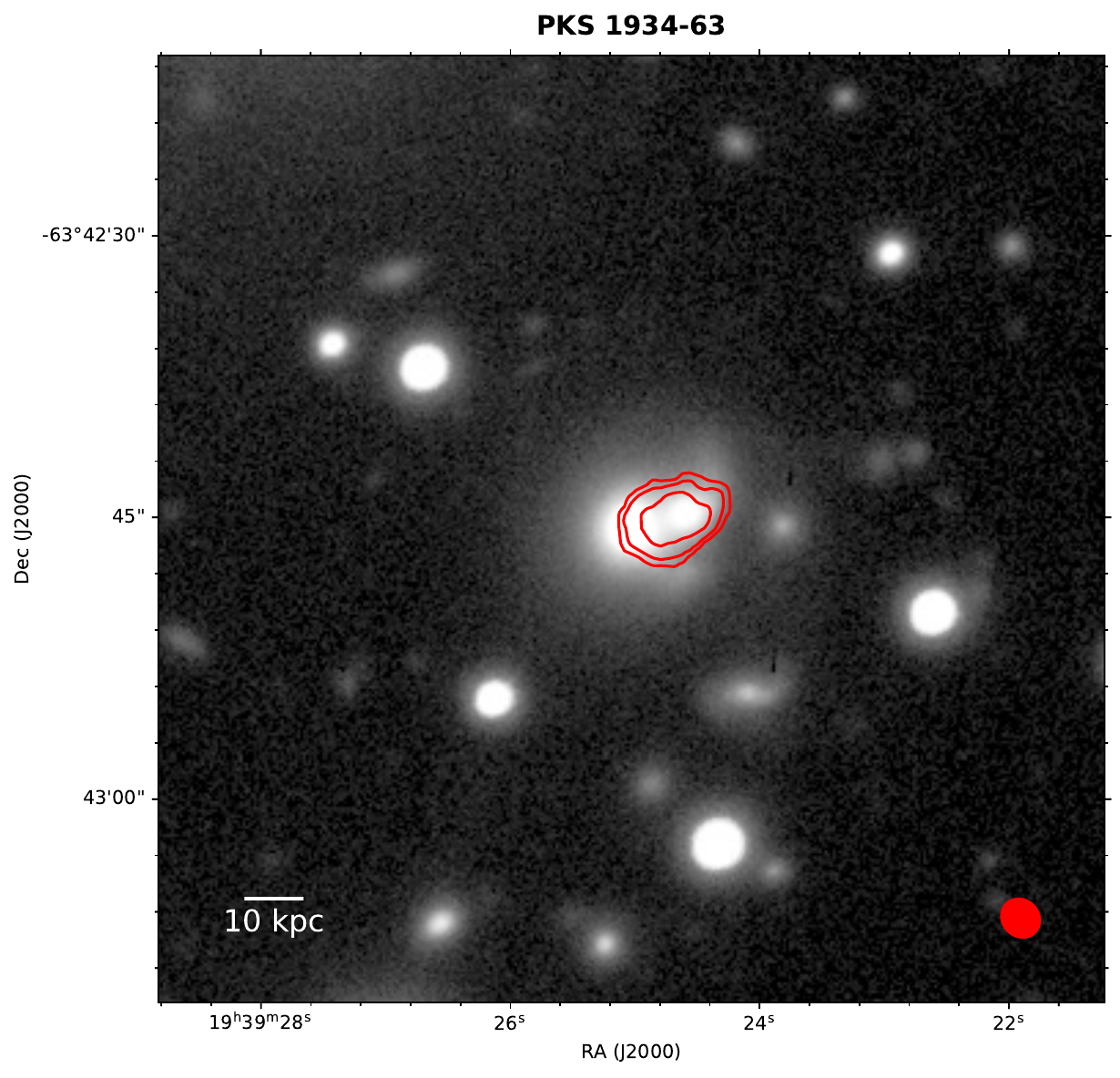}

\caption{Total H$_2$ column density contours superimposed on Gemini GMOS r$^{\prime}$ images from \citet{Ramos11} for the CO-detected sources. H$_2$ contour levels are 1.5, 3, 6,  \ldots $\times 10^{21}$ cm$^{-2}$. The red ellipses indicate the ALMA beam.} 
\label{fig:TotIntensity}
\end{figure*}

\begin{figure*}
   \centering
 \includegraphics[height=7.5cm]{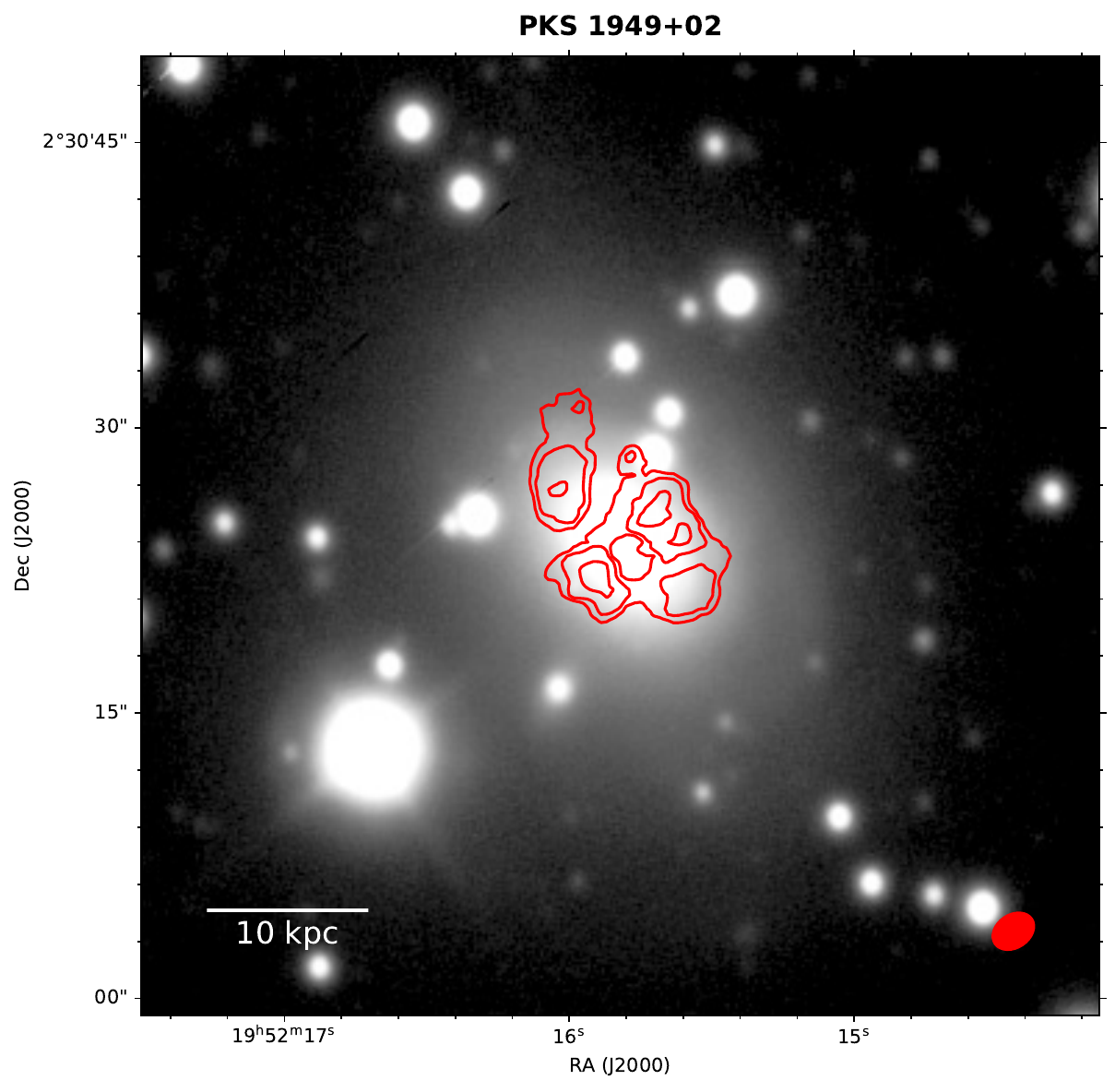}
\includegraphics[height=7.5cm]{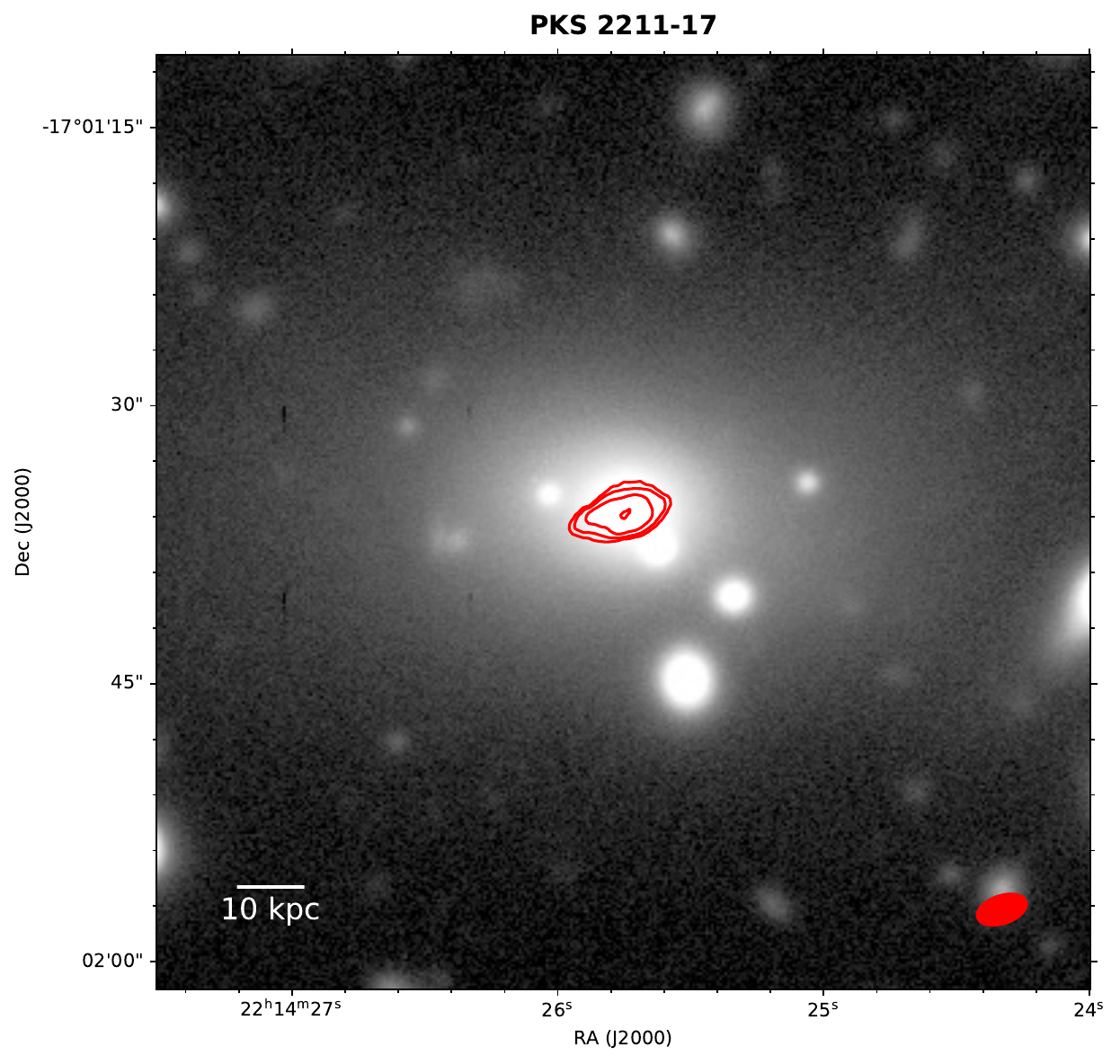}

\includegraphics[height=7.5cm]{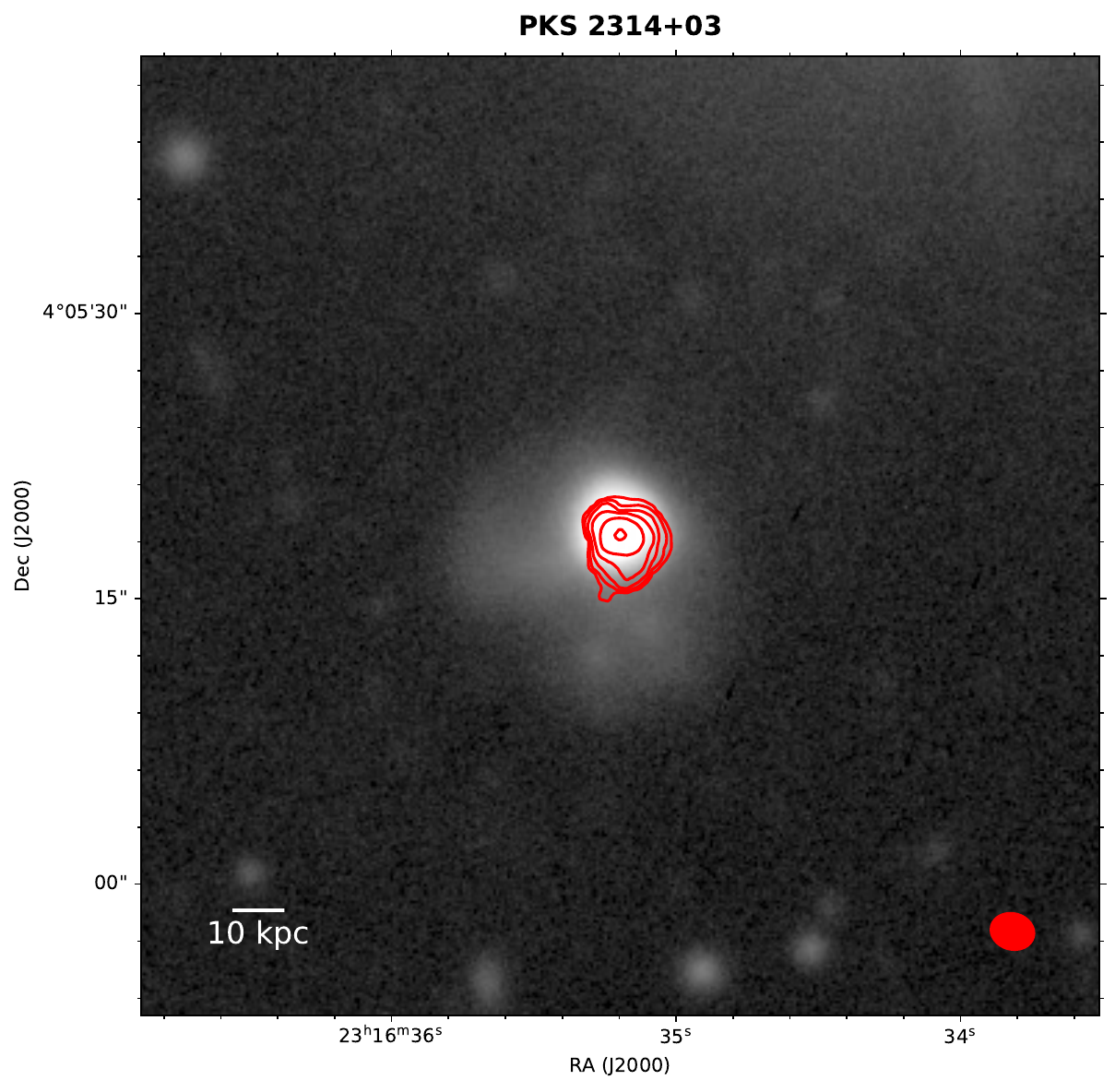}  

\contcaption{Total H$_2$ column density contours superimposed on Gemini GMOS r$^{\prime}$ images from \citet{Ramos11} for the CO-detected sources. H$_2$ contour levels are 1.5, 3, 6,  \ldots $\times 10^{21}$ cm$^{-2}$. The red ellipses indicate the ALMA beam.} 
\end{figure*}

\begin{figure}
\vglue -1cm
\hspace*{-6mm}
\includegraphics[width=11cm]{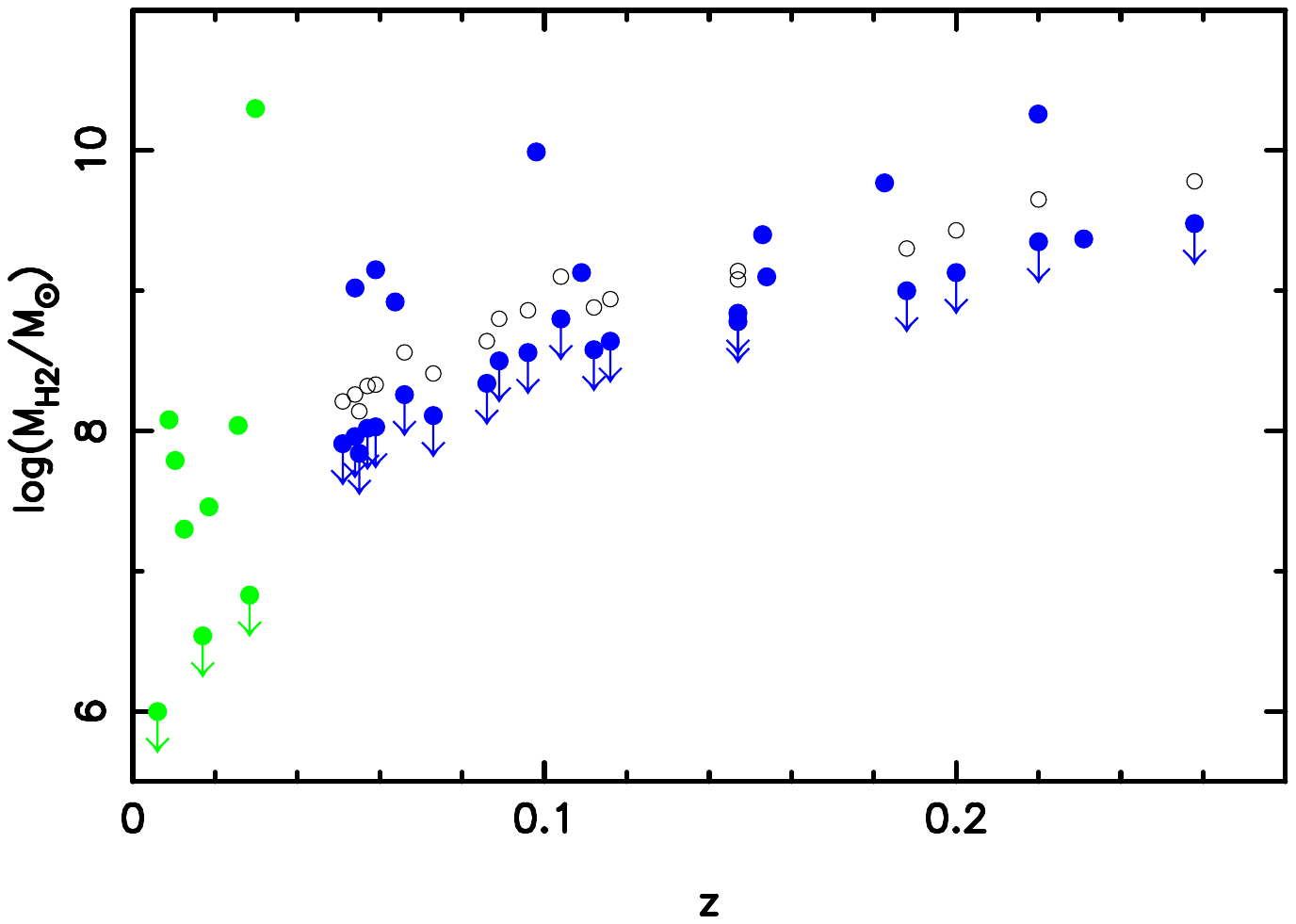}
 \caption{Distribution of the H$_2$ masses and limits as function of the redshift. The filled blue symbols represent the 2Jy
 detections, filled blue symbols with downward arrows  represent 2Jy upper limits calculated assuming
 the CO emission covers 1 beam, and open grey symbols represent 2Jy upper limits calculated assuming 4 beams. Note that PKS1934--63 has been marked as detected in this and following plots, see Section \ref{sec:Results} for details.  The plot also includes the objects from \citet{Ruffa19a}
(filled green symbols) -- a comparison between the two samples is presented in Section \ref{sec:ComparisonLiteratureSamples}.}
\label{fig:plot1}
\end{figure}

\section{Results}
\label{sec:Results}

Of the twenty-eight targets of the new ALMA observations, we have 8 detections of \coOne\ emission (and, in one case, also absorption) at the location of the host galaxy. The targets for which we have detected \coOne\ are: PKS0806--10, PKS1648+05, PKS1733--56, PKS1814--63 (where both emission and absorption are detected), PKS1942--46, PKS1949+02,  PKS2211--17 and PKS2314+03. The case of PKS1934--63 is more uncertain (see below).
For the CO detected sources, the total (velocity-integrated) H$_2$ column densities derived from the CO maps are shown superimposed on the Gemini GMOS r$^{\prime}$-band images of \citet{Ramos11} in Figure  \ref{fig:TotIntensity}, and the velocity field and position-velocity (PV) images are presented in Figure \ref{fig:VeloField} in the Appendix.

The mm continuum images of all targets (excluding the known CSS/GPS sources which are unresolved at the angular resolution of our observations) are shown in Figure \ref{fig:imagesCont}, while the fluxes of the unresolved cores measured from these
images are presented in \citet{Dicken23}.
A large fraction of the objects show extended continuum structures, consistent with those found at GHz frequencies  \citep[e.g.][]{Morganti93,Morganti99,Dicken08}. In some cases the emission clearly extends beyond the field of view of ALMA, confirming the large structures already seen at lower frequencies.

We also detect \coOne\ in one system (PKS1934--63, see Figure \ref{fig:TotIntensity}) in which an on-going interaction is taking
place between the host galaxy of the radio AGN and a close companion (separated by only about 3\,arcsec [9\,kpc] --  see \citealt{Roche16,Santoro18}). At the resolution of our observations it is difficult to disentangle how much of the CO is actually on the target, and most of the CO is located between the two objects; however, a tail reaching PKS1934--63 (including its central regions) appears to be present. Indeed, the end of the tail  closest to the centre of PKS1934--63 has a CO
velocity within $\sim$50 \kms\ of the systemic velocity of
the radio galaxy host from \cite{Santoro18} (see Fig. \ref{fig:VeloField}).

Based on these detections, and adding the literature results for Hydra~A, which is part of the sample, we derive  a detection rate for the confirmed cases of 10/29 = 34$\pm$9\%.\footnote{We have used binomial statistics to calculate the uncertainties on the proportions throughout this paper.} It is important to note that this detection rate is for objects with relatively high molecular masses, i.e. above 10$^{8.9}$\msun, as covered by the present observations --- see discussion in Section \ref{sec:H2masses} ---  and this should be considered when making the comparison with other samples of radio AGN and ETG, see below and Section \ref{sec:ComparisonLiteratureSamples}.


As mentioned above, we detect a deep absorption against the continuum of PKS1814--63. This is a CSS source with two lobe-like structures separated by about 400\,pc  (\citealt{Tzioumis02,Morganti11}) and unresolved in our ALMA observations. A discussion of the properties of the absorption and what we can learn from it is presented in Section \ref{sec:Abs}.  

Finally, in 10 fields we detect \coOne\ from associated companion galaxies ($\vert \Delta V \vert < 2500$\,\kms). 
Particularly interesting is the case of PKS0620--52, where a companion galaxy is detected in \coOne\ --- with velocity blueshifted by about 1200 \kms\ compared to the systemic velocity of the target --- at the location where the jet bends,  $\sim$10\,kpc ($\sim$10\,arcsec) from the nucleus of the target (see Figure \ref{fig:CompanionsTentative}). 
Determining whether an ongoing jet-cloud interaction is present, or the alignment with the bending of the jet is only a projection effect, will require deeper and higher resolution observations. The CO profile does not show extreme velocities, which might otherwise suggest the presence of an interaction. However, the limited S/N of the detection does not allow to completely rule out the presence of kinematically disturbed gas. 

In 11 fields we do not clearly detect any \coOne\ emission either in the targets or in companion galaxies, although
in one of these fields (PKS2221--02) there are two tentative detections of companions.

\begin{table*} 
\caption{Molecular gas and cool dust mass estimates. Where CO(1-0) emission is detected, the uncertainties on the molecular gas masses
include contributions from noise (estimated as $\delta(S_{\rm CO} \Delta V)_{\rm noise} = \sigma_{\rm chan} \Delta V_{\rm chan} \sqrt{\Delta V_{\rm FWHM}/\Delta V{\rm chan}}$, with
$\Delta V_{\rm FWHM}/\Delta V{\rm chan} = 10$) and an assumed 5\% systematic flux calibration uncertainty \citep{Vankempen14}, with these uncertainties
added in quadrature. Note that uncertainties in the assumed value of $\alpha_{\rm CO}$ have not been
taken into account. For the Gas/Dust mass ratios in column 7, where both the dust mass and the gas mass are upper limits, the number is shown in brackets.}
\begin{center}
\begin{tabular}{lccccccl} 
\hline\hline 
   &   F (CO)   & M$_{\rm  H_2}$ & Limits H2$^{\dagger}$ & Diameter & Dust mass$^*$ & Gas/Dust & Detections in field \\
   &                 Jy km/s & \msun/$10^9$ & \msun\ & kpc(")  & log(\msun)       &         & \\
\hline
PKS0034-01    &   -- & -- & 1.3E+08 & -- & $<$5.9 & (160) & \small{One companion}\\
PKS0035-02    &   -- & -- & 2.3E+09 & --  &$<$7.3 & (130)  &  \small{No detection}\\
PKS0038+09    &  -- & -- & 1.0E+09 & --  &$<$7.0 & (100)  & \small{One companion} \\
PKS0043-42    &  -- & -- & 4.4E+08 & -- &6.5 & $<$140  & \small{No detection} \\
PKS0213-13    &  -- & -- & 6.9E+08 & -- &6.5 & $<$220 & \small{No detection} \\ 
PKS0349-27    &  -- & -- & 1.8E+08 &  -- &6.8 & $<$32 & \small{No detection}\\
PKS0404+03 &  -- & -- &  3.2E+08 & -- &6.5 & $<$100  &\small{No detection}\\
PKS0442-28 &  -- & -- &  6.2E+08 & -- &$<$6.5 & (190) & \small{One companion} \\
PKS0620-52 &  -- & -- & 8.2E+07 & -- &$<$5.6 & (210) & \small{CO in companion along jet}  \\
PKS0625-35 &  -- & -- & 7.0E+07 & -- &$<$6.3 & (34) & \small{One companion }\\
PKS0625-53 &  -- & -- & 9.2E+07 & -- &$<$5.4 & (370) & \small{One companion (+two tentative)}\\
PKS0806-10 &  0.55$\pm$0.06 & 1.35$\pm$0.15 & -- & 10.5(5.2) &7.5 & 45 &  \small{Detection}\\
PKS0945+07 &   -- & -- &  2.2E+08 & -- &6.8 & $<$40 & \small{No detection}\\
PKS1151-34 &   -- & -- & 3.0E+09 & -- &7.6 & $<$76 & \small{No detection}\\
PKS1559+02 &   -- & -- & 6.4E+08 & -- &7.4 & $<$24 & \small{One companion}\\
PKS1648+05 &  0.25$\pm$0.04 & 1.25$\pm$0.19 & -- & 11.2(4.5) & 7.3 & 63 & \small{Detection } \\
PKS1733-56 &  4.91$\pm$0.25 & 9.72$\pm$0.49 & -- & 25.2(13.8) &7.9 & 130 & \small{Detection } \\
PKS1814-63 &  1.02$\pm$0.09 & 0.84$\pm$0.08 & -- & 13.3(10.8) &7.2 & 64 & \small{Detection with absorption} \\
PKS1839-48 &  -- & -- &  3.8E+08 & -- &$<$6.2 & (240) & \small{One companion}\\
PKS1932-46 &  0.20$\pm$0.04 & 2.33$\pm$0.44 & -- & $16.4$(4.4) &$<$7.2 & $>$150 & \small{{ Detection}, one companion} \\
PKS1934-63 &  0.82$\pm$0.05& 5.83$\pm$0.35 & -- & 20.6(6.6) &7.6 & 150 & \small{CO in tail with companion} \\
PKS1949+02 & 2.02$\pm$0.12 & 1.42$\pm$0.08 & -- & 13.5(11.8) &7.3 & 73 & \small{Detection }\\
PKS1954-55 &  -- & -- &  1.1E+08 & -- &$<$5.5 & (340) & \small{No detection} \\
PKS2135-14 &  -- & -- & 1.3E+09 & -- &7.3 & $<$69 & \small{No detection}\\
PKS2211-17 &  0.51$\pm$0.06 & 2.53$\pm$0.25 & -- & 14.2(5.3) &7.7 & 56  & \small{Detection }\\
PKS2221-02 &  -- & -- & 1.1E+08 & --  &6.5 & $<$35 & \small{Two tentative companions} \\
PKS2314+03 &  1.73$\pm$0.10 & 18.1$\pm$1.1 & -- & 11.7(3.2)$^{\dagger\dagger}$ &8.5 & 67 & \small{Detection } \\
PKS2356-61 &  -- & -- & 3.6E+08 & -- &6.5 & $<$110 & \small{No detection}\\
PKS0915-11 & 1.81$\pm$0.09     & 1.06$\pm$0.05   &  --   &4.5    & 6.9    & 140 &  From \cite{Rose19} \\
\hline
\end{tabular}
\end{center}
$^*$ Dust masses from \citet{Bernhard22}.\\

$^\dagger$ Mass upper limits (5$\sigma$, 1 beam, $\Delta V_{\rm FWHM} = 300$ \kms), see text for details. \\
$^{\dagger\dagger}$ The disk observed in the high spatial resolution cube is smaller, with a diameter of about 1 arcsec, i.e. 3.6 kpc. 
\label{tab:H2masses}
\end{table*}

\subsection{Masses of the molecular gas}
\label{sec:H2masses}

The masses of the molecular gas were calculated using the standard formulae:
\begin{equation}
M_{\rm mol} = \alpha_{\rm CO} L'_{\rm CO}
\end{equation}
and 
\begin{equation}
L'_{\rm CO} = 2453 \  S_{\rm CO} \  \Delta V\  D^2_{\rm L}/(1+z)
\end{equation}
with $S_{\rm CO}$ in Jy, $D_{\rm L}$ in Mpc, and $\Delta V$ in \kms\ in the emitters frame (i.e.\ de-redshifted velocities). The values of the H$_2$ masses presented in Table \ref{tab:H2masses} have been obtained using $\alpha_{\rm CO} = 4.3$ \msun (K \kms\ pc$^2$)$^{-1}$ (corresponding to X$_{CO} = 2 \times 10^{20}$ cm$^{-2}$ K \kms)$^{-1}$. The choice of this value, which is typically assumed in the case of quiescent gas \citep{Bolatto13}, has been motivated by the fact that, at least to first order, the structures observed in our sample appear to be mostly large disks with relatively regular kinematics (see Section \ref{sec:Kinematics}); the same or a similar value for  $\alpha_{\rm CO}$ has been assumed in many other CO studies of radio AGN \citep[e.g.]{Ocana10,Ruffa19a} and quasars \citep{Jarvis20,Shangguan20,Ramos22}. When making comparisons with other samples, all H$_2$ masses for objects from the other samples have been converted to this value.

The upper limits were derived using the relation:
\begin{equation}
S_{\rm CO} \Delta V\ < 5 \sigma_{\rm chan} \Delta V_{\rm chan} \sqrt{\frac{\Delta V_{\rm FWHM}N_{\rm beam}}{\Delta V_{\rm chan}}},
\end{equation}
where we assumed  5-$\sigma$ upper limits, a width of the expected profile of $\Delta V_{\rm FWHM} = 300$ \kms\ for a channel width of $\Delta V_{\rm chan} = 30$ \kms, and $N_{\rm beam}$ is the number of beams,
initially assumed to have a value of 1.  The 5-$\sigma$ could be considered quite conservative compared to other studies (e.g. both \citealt{Smolcic11} and \citealt{Ruffa19a} use 3-$\sigma$ upper limits).

Figure \ref{fig:plot1} shows how the molecular masses of the detections and upper limits are distributed as function of redshift. This plot demonstrates that the detections are not limited to the targets at low redshift.  
The figure also includes targets from \cite{Ruffa19a} which fill the low redshift range. The comparison with this and other samples will be discussed in detail in Section \ref{sec:ComparisonLiteratureSamples}.

Considering the lower end of the redshift range ($z < 0.12$), it is clear that the upper limits estimated by assuming the emission is spatially
confined to 1 beam fall significantly below the detections of objects at similar redshifts. One possible explanation for
this is that 
the assumption of 1 beam 
may be too optimistic. To illustrate the effect of a more extended CO distribution, we also show as open black circles upper limits calculated 
by assuming that the molecular gas covers 4 beams --- more consistent by the typical size of the observed \coOne\ structures
in the detected sources. With this even more conservative assumption, the upper limits fall
closer to the detections, especially at the higher redshifts; however, there remains a gap
between the upper limits and detections at the low redshift end. This may suggest a bimodal shape for
the molecular gas mass distribution of radio AGN at low redshifts, but deeper observations of a larger sample would
be required to confirm this. In the following, we assume that the emission is extended over 1 beam when calculating the
upper limits for molecular gas masses. 

\begin{figure}
   \centering
\vglue -1cm
\hspace*{-6mm}
\includegraphics[width=11cm]{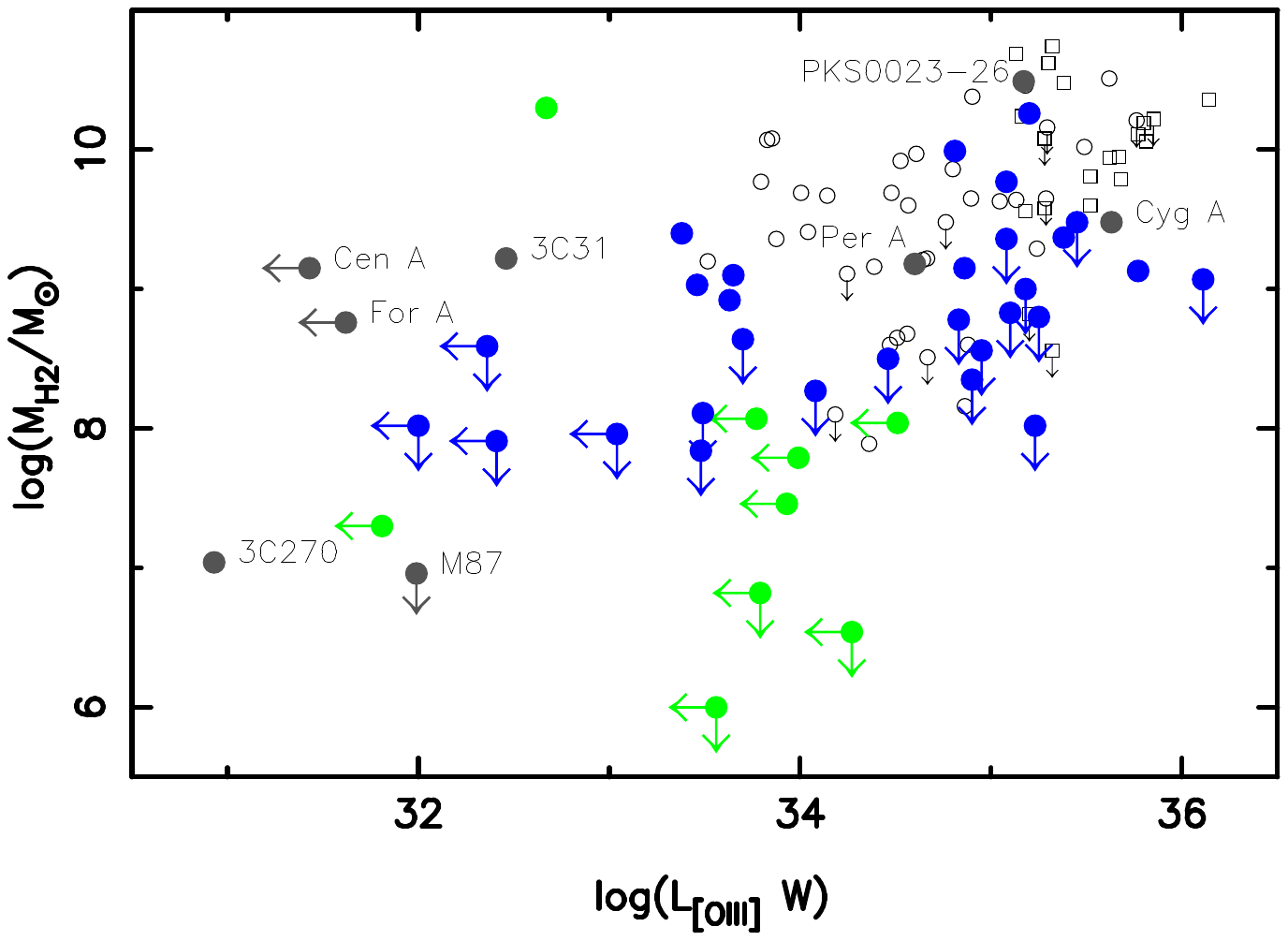}
\vglue -1.3cm
\hspace*{-6mm}
\includegraphics[width=11cm]{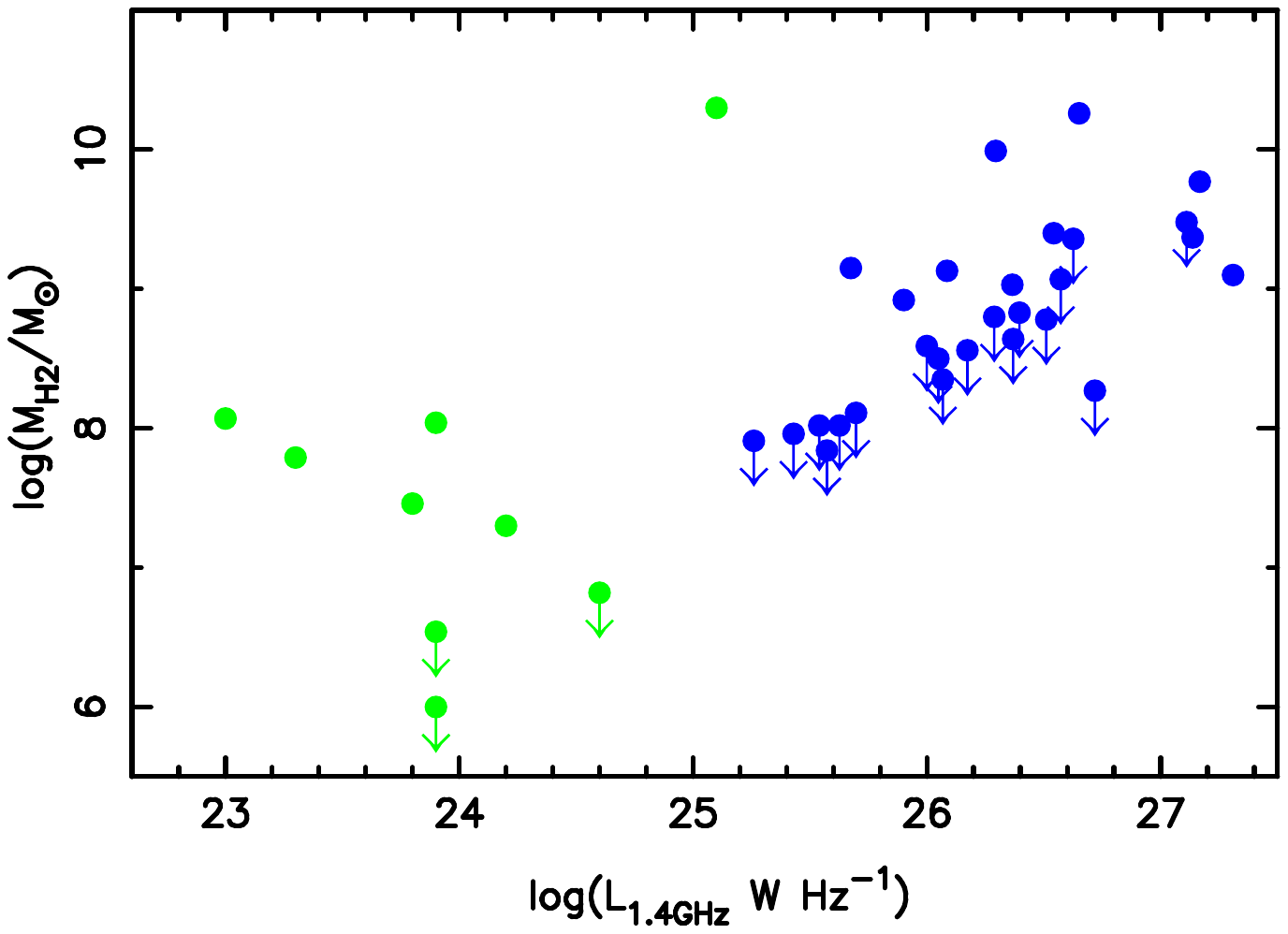}
 \caption{ 
 {\sl Top:} Molecular gas mass vs [OIII]$\lambda$5007 luminosity  for the 2Jy sample. In addition to the 2Jy (solid blue circles)
 and \citet{Ruffa19a} (solid green circles) objects, for comparison we also show results for nearby type 2 quasars
 from \citet{Ramos22} and \citet{Molyneux24} (open black squares), type 1 quasars from \citet{Shangguan20}  (open black circles),
 and some well-known radio AGN from the literature (labelled, filled dark-grey cicles). The references
 for the individually-labelled radio AGN are as follows:  Cygnus A \citep{Carilli22},  Fornax A \citep{Maccagni21}, 
 Centaurus A \citep{Parkin12},  3C270 \citep{Boizelle21}, Perseus A \citep{Bridges98},  3C31 \citep{North19}, M87 \citep{Salome08}, and
PKS0023-26 \citep{Morganti21}. {\sl Bottom:} Molecular gas mass vs radio luminosity
for the 2Jy and \citet{Ruffa19a} samples (same symbols as in the top plot). For the 2Jy sample, the [OIII]$\lambda$5007 luminosities have been taken from \citet{Dicken09}, and the 1.4\,GHz radio luminosities have been converted from the 5.0\,GHz radio luminosities presented in that paper assuming a spectral index of $\alpha=0.7$ (for $F_{\nu}\propto\nu^{-\alpha}$).}
\label{fig:plots2}
\end{figure}

Figure \ref{fig:plots2} presents the distribution of the masses of the molecular gas (or limits) as function of [OIII] luminosity (a proxy for the AGN bolometric luminosity: \citealt{Heckman04}) and radio luminosity. 
There is no obvious
trend with either [OIII] or radio luminosity, and clearly radio AGN of given luminosity encompass a
wide range of molecular gas mass ($\sim$1 -- 2 orders of magnitude), echoing previous results based on {\it Herschel}-derived dust masses
\citep{Bernhard22}. This suggests that the mass of the molecular gas reservoir is not the sole
determinant of the level of AGN and jet activity; other factors (e.g. the distribution of gas,
and degree to which it is dynamically settled) must also be important.

For comparison, Figure \ref{fig:plots2} top also includes the targets studied by \cite{Ruffa19a}, 
some well-known individual radio AGN, and results for samples of nearby radio-quiet type 1 
and type 2 quasars \citep{Ramos22,Shangguan20,Molyneux24}.
At high [OIII] luminosities, the radio AGN overlap with radio-quiet quasars: clearly the hosts
of some radio AGN are as gas-rich as radio-quiet quasars of similar [OIII], perhaps reflecting
a common origin for the gas in galaxy mergers \citep{Husemann17,Ramos22,Pierce23}. However, it is also notable that the 
the 2Jy radio AGN sample shows a much a higher proportion of non-detections than the radio-quiet
quasar samples, despite that fact that the rms sensitivity was similar to, or better than, than
that of those samples (see Section \ref{sec:ComparisonLiteratureSamples}).

\subsection{Molecular gas detected in absorption}
\label{sec:Abs}

\begin{figure}
\hspace*{-6mm}
\includegraphics[width=11cm]{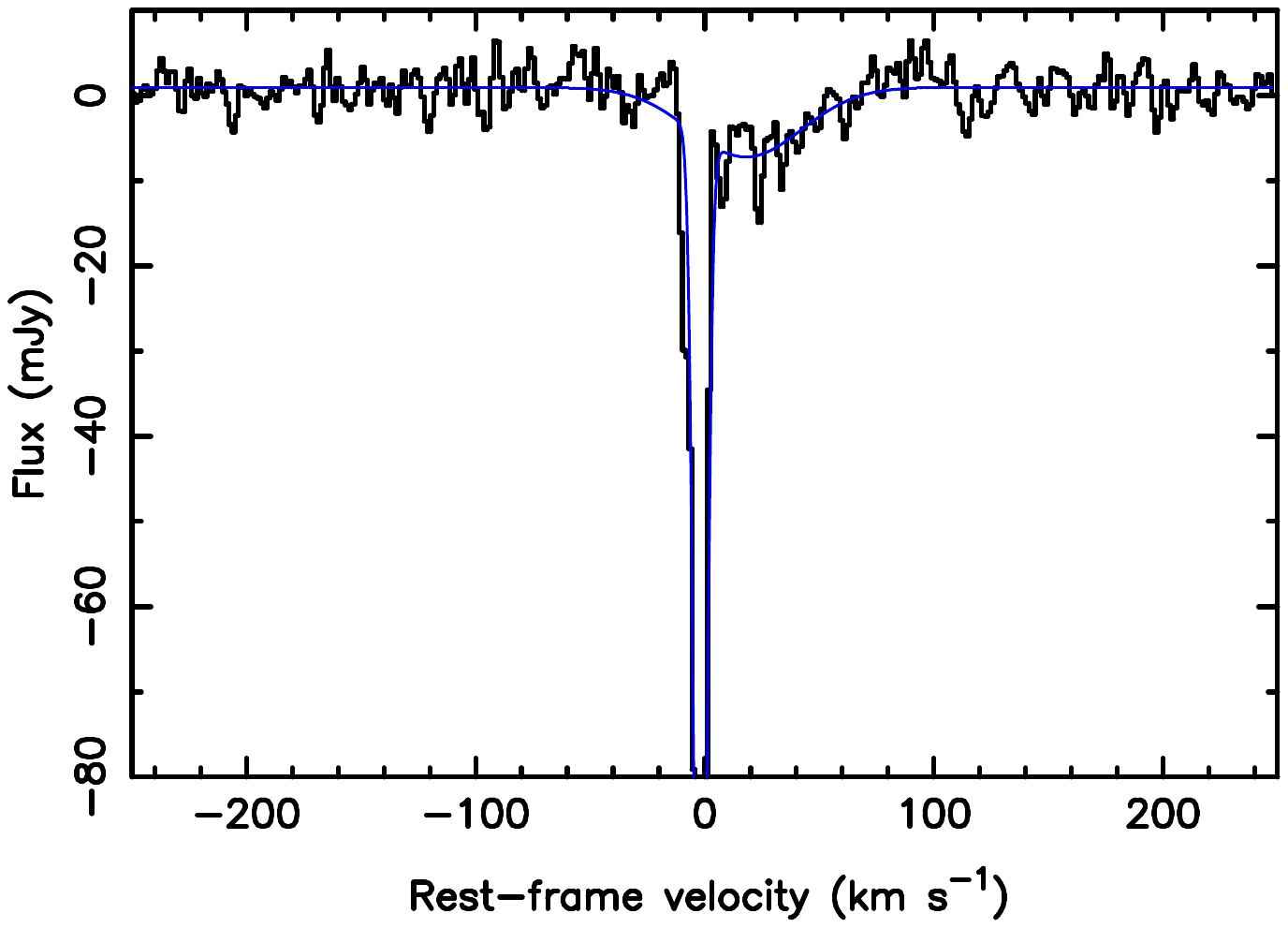} \\
\caption{CO absorption profile extracted at the location of the peak continuum of PKS1814--63 (black line), compared with a double-Gaussian fit to the profile (blue line). The profile has been obtained from the cube made at full velocity resolution (1.4\,\kms) in order to minimise the presence of emission, and the continuum emission has been subtracted. Note that the full depth of the narrow
absorption is not shown in order to highlight the shallow, broad feature.}
\label{fig:pks1814Abs}
\end{figure}

We detect  deep and narrow \coOne\ in absorption at the location of the peak of the continuum emission in PKS1814--63. Figure \ref{fig:pks1814Abs} shows the absorption profile from the cube at full velocity resolution.   
Based on a double-Gaussian fit to the profile, the full width at half maximum of the deep absorption is  $FWHM =6.1\pm0.1$\,\kms, and it is centred on the systemic velocity. 
The fitted depth of the absorption is 140\,mJy against the continuum peak flux of 340\,mJy. This corresponds to a peak optical depth of $\tau_{\rm peak} = 0.53$. Interestingly, the profile also shows a broad component, with $FWHM = 57.3\pm5.3$\,\kms\, offset compared to the systemic velocity by $+18.0\pm3.7$\,\kms. The peak depth of this feature is 8\,mJy, corresponding to an optical depth of $\tau_{\rm peak} = 0.024$. 

The column density of the absorption was estimated following the standard formulae in \citet{Bolatto13} and the approach described in \cite{Maccagni18} and \cite{Morganti21}. The values were derived by assuming two extreme values for the  excitation temperature ($T_{\rm ex}$) of the gas.  On the low side, the typical temperature of the CO gas in conditions of thermal equilibrium in the Milky Way is $\sim$16 K (e.g., \citealt{Heyer09}). However, this temperature may be higher if the gas is affected by the AGN, as expected for gas located in the circum-nuclear region. Temperatures up to $T_{\rm ex}$ $\sim$60 K have been reported in some cases (see \citealt{Dasyra14,Oosterloo17}). For example, a value of $T_{\rm ex}$ $\sim$42\,K has been estimated by \cite{Rose19} for Hydra~A, a radio galaxy that belongs to the 2Jy sample. The H$_2$ column density derived for the deep absorption ranges from $2 \times 10^{20}$ to $2 \times 10^{21}$ cm$^{-2}$ for excitation temperatures between 16 and 60 K, and a conversion factor CO-to-H$_2 = 10^{-4}$.\footnote{Note that the lower end of this range is consistent with the column density derived from the \HI\ data for the deep, narrow component of $\sim  3 \times 10^{20}$ cm$^{-2}$, assuming a spin temperature  $T_{\rm spin}$ = 100 K.} Similarly, we derive column densities in the range $9 \times  10^{19}$ to $7 \times 10^{20}$ cm$^{-2}$ for the broad component.

Intriguingly, the shape of the \coOne\ absorption profile is similar (but not identical) to that of the \HI\ absorption profile (\citealt{Morganti11}), which shows  both  deep (narrow) and shallow (broad) absorption features.
However, the broad component seen in CO is less broad than  observed in \HI\ ($FHWM_{HI} > 100$\,km s$^{-1}$) and, unlike the \HI, is only red-shifted.  This could result from differences in the structure of the extended continuum emission at the two (very different) frequencies: at 1.4~GHz the source consists of two lobes with a total extent of about 400 mas, ($\sim$480\,pc \citealt{Tzioumis02}), but the diffuse southern lobe (associated with blue-shifted HI broad absorption) is fainter and has a steeper spectral index than the northern lobe (associated with red-shifted HI broad absorption), so is probably too faint to contribute significant flux at the frequency of our ALMA observations. Therefore, the broad, red-shifted \coOne absorption is most likely observed against 
with the northern lobe, which is relatively compact ($<120$ pc, see \citealt{Morganti11}).

Absorption by molecular gas has also been seen in other radio galaxies (see e.g.\ \citealt{Maccagni18,Rose19,Ruffa19a,Morganti21,Rose24} for some recent examples). In particular, the \coOne\ absorption detected in  Hydra~A by  \citet{Rose19,Rose20}, has a profile that is strikingly similar to that of PKS1814-63, with both broad and narrow components detected, and similar velocity widths and velocity shift amplitudes.  

From their study of the \coOne\, absorption detected against the radio cores of a sample of nearby massive galaxies, \citet{Rose24} suggest that the absorption profiles indicate the presence of two distinct populations of molecular clouds: on the one hand, clouds with low velocity dispersion, likely belonging to a galaxy-scale, edge-on disc of molecular gas; on the other, clouds with high velocity dispersion located in circumnuclear discs and/or tracing bulk movement of molecular gas towards the galaxy centre. 

A similar situation may be present in PKS1814-63: \citet{Morganti11} suggest that the broader HI absorption (which, as described above, shows both a red and blue wings) is due to gas located in a circum-nuclear disk moving under gravity, with perhaps some of the line width accounted for by
projection effects. 
However, given the relatively large velocity width ($FWHM \sim 60$\,\kms) of the broad CO absorption and the fact that it is seen against such a compact lobe, an alternative possibility is that the large width results from interaction
between the gas and the expanding radio jet.

We further note that the similarities between the CO and  \HI\  absorption profiles,  also highlighted by \cite{Rose19} for Hydra A, suggest that the \HI\ and the molecular gas may be spatially coincident. This is consistent with the hypothesis that, when heated, cold molecular gas regions produce an encompassing skin of warm atomic gas (e.g.\ \citealt{Jaffe05}).



\subsection{Structure and kinematics of the molecular gas}
\label{sec:Kinematics}

In the detected targets we find that the molecular gas is distributed over large, extended structures. We measure the diameters of the structures from the lower contour of the total intensity images, and check them using PV diagrams.  The estimated full sizes (diameters) are  between 4.5 and 25\,kpc (see Table \ref{tab:H2masses}). These structures are  larger than those typically found in nearby low-power radio galaxies and ETG (see Section \ref{sec:ComparisonLiteratureSamples}), but similar to the extents of the molecular gas found in radio AGN in rich cluster environments. 

In at least five of the detected targets (Hydra~A [PKS0915--11], PKS1733--56, PKS1814--63,  PKS1949+02, PKS2314+03) most of the molecular gas appears distributed in disks/rings with relatively regular kinematics (see Figure \ref{fig:VeloField}). It is worth noting that Hydra~A represents a rare case of a molecular gas disk found in a cool-core cluster (\citealt{Rose19,Russell19}). In the case of PKS2314+03 (3C~459) the disk is more clearly seen in the high resolution data available from the archive (see Section \ref{sec:Observations}). However, tails or clouds of unsettled gas are observed in the outer parts of the distribution of molecular gas in all these objects.  

We have investigated the molecular gas kinematics in more detail by modelling the CO using the 3D-Based Analysis of Rotating Object via Line Observations (3D-Barolo), a tool for fitting 3D tilted-ring models to emission-line data-cubes, described in \cite{DiTeodoro15}. The main purpose of this modelling was to determine whether rotational motions in a disk or ring provide a good description of the measured gas kinematics.

For this analysis, we concentrated on the four cases where the distribution of the gas  is seen to be clearly extended over several beams. 
The plots resulting from the modelling are shown in Figure  \ref{fig:Barolo}, and Table \ref{tab:BaroloParameters} presents the parameters. 
Given the quality of the data, we fixed as many parameters as possible in the fitting. In particular, the position angle and inclination used in the fitting were taken from available information like the presence of dust-lanes, see e.g. \cite{Ramos11}. The velocity dispersion was set to 10\,\kms, but the rotation velocity was always left free in the fitting.
The plots in Figure  \ref{fig:Barolo} show that the models describe to first order the data in all the targets, confirming that most of the CO is observed in a regularly rotating structure. However, the plots also show that the limited sensitivity of the observations does not allow a detailed comparison between data and models. 


The estimated PAs of the major axes of the disks are compared with the PAs of the (projected) radio axes in Table \ref{tab:BaroloParameters}. Since we do not know the orientation of the jet relative to the line of sight, we cannot derive precise values for the de-projected angle of the radio axis relative to the
plane of the disk. However,
comparing the projected jet and disk PAs, it clear that in a significant number of cases the jet is not perpendicular to the gas disk (see also  Fig.\ref{fig:VeloField} where the directions of the radio axes are compared with those of the major axes of
the gas disks).  We note that this is similar to the situation
found in a number of objects studied in literature, e.g. \cite{Morganti15,Ruffa19a} and references therein. 

\begin{table*} 
\caption{Parameters from the modelling of the CO cube with 3D-Barolo, see text for details. The second column gives the de-projected rotation velocity, the third column the position angle (PA) of the major axis of the disk, the fourth column the inclination of the disk rotation axis relative to the line of sight,  the fifth column the difference between the systemic velocity of the disk and the galaxy rest frame defined
by the redshift, the sixth column the radio axis PA, and the seventh column the difference between the disk and radio axis PAs. The gas velocity dispersion assumed was 10\,\kms. For Hydra A we list only the PAs of the molecular disk
(from \citealt{Rose19}) and of the radio jet.
}
\begin{center}
\begin{tabular}{ccrcrrcl} 
\hline\hline 
             & $V_{\rm rot}^a$ & PA$^b$   & Incl  & $V_{\rm sys}$ & PA radio & $\Delta$PA$^c$ (projected)  & Notes  \\
             &  \kms\          &  deg     & deg   &  \kms\        & deg  & deg     &        \\
\hline
PKS1733-56    &  343            & --54     & +70   & 80            & +38      & 91 & CO aligned with dust-lane \\ 
PKS1814-63    &  245            & --51     & +74   & --3            & +7      & 58 & CO aligned with dust-lane \\ 
PKS1949+02    &  327            & --135     & +53   & --165            & +80      & 35  & Aligned with dust-lane?  \\ 
PKS2314+03    &  247            & --75     & +75   & --60            & +90      & 15 &
Disturbed galaxy, optical halo/fans \\ 
\hline
$0915-11$   &                  &     --80     &       &             &     +10         &   90     & from \cite{Rose19}  \\
\hline
\end{tabular}
\end{center}
$^a$ From the outer ring \\
$^b$ Angle of the approaching side \\
$^c$ $\Delta$PA = $ \vert$ PA$_{\rm CO}$ -- PA$_{\rm jet}\vert$
\label{tab:BaroloParameters}
\end{table*}

In the case of PKS0806--10, the molecular gas structure is only marginally resolved, so we cannot perform a meaningful modelling. 
However, the velocity map and PV diagram for this source (see Figure \ref{fig:VeloField}) suggest that at least part of the molecular gas is distributed in a rotating structure and, like some
of the objects modelled with 3D-Barolo, in projection
the radio axis appears to be aligned close to the major axis of the disk, rather than perpendicular to it.

In two cases -- PKS2211--17 (only marginally resolved) and PKS1648+05 -- we do not detect velocity gradients across the distribution of \coOne\ that are suggestive of orderly rotation. Interestingly, both of these radio AGN are located at the centres of  galaxy clusters \citep{Croston11,Ramos13,Ineson15}, and the apparent lack of evidence for regular rotation in a disk is consistent with what is typically observed for the molecular gas associated with
central cluster galaxies \citep[e.g.][]{McNamara14,Olivares19} -- see Section \ref{sec:discussionOccurrence} for further 
discussion.

It is also notable that the CO emission in PKS1932-46 has a highly asymmetric distribution, with all the detected
molecular gas located to the north of the nucleus, where it is redshifted by up to $\sim$200\,\kms relative to the host galaxy rest frame (see Figure \ref{fig:VeloField}). Interestingly, this emission is extended roughly in the direction of a morphologically disturbed, gas-rich companion galaxy at a similar redshift to PKS1932-46 and situated $\sim$100\,kpc to the NE in projection (see Fig. \ref{fig:TotIntensity}). It is possible in this case that the molecular gas has been accreted in a close encounter between the radio AGN host galaxy and its companion in the past (see also discussion
in \citealt{Inskip07}).

Another example of a ongoing interaction with a companion galaxy is provided by 
the case of PKS1934--63, where the host is apparently linked to its companion by a tail of molecular gas, and distribution of the CO emission peaks between the two galaxies (see Figure \ref{fig:TotIntensity}). The PV plot  (Figure \ref{fig:VeloField})  shows that the kinematics of the molecular gas has a gradient, with velocities ranging from the systemic velocity of PKS1934--63 to that -- about 250 \kms\ redshifted -- of the companion (see Figs. 14 and 17 in \citealt{Roche16}). Interestingly, the PKS1934--63 (\citealt{Santoro18}) is a young radio source that consists of two lobes separated by about 130\,pc, and with an estimated kinematic age of $\sim$10$^3$\,yr (derived by monitoring the lobe separation, \cite{Ojha04}). This close interaction and the presence of molecular gas is particularly interesting in terms of understanding  the triggering of the young radio AGN, but much higher spatial resolution observations will be required to investigate this further. 

Although we detect some molecular gas with disturbed kinematics, we do not detect fast molecular outflows in \coOne\ emission in any of our targets. This is likely due to a combination of the relatively shallow depth and limited spatial resolution of the present observations. Molecular outflows in radio AGN have so far been found mainly in young radio sources \citep{Oosterloo17,Oosterloo19,Murthy22,Holden24}, and their detection has required deep observations combined with high spatial resolution in order to disentangle the outflow from the rest of the molecular gas (which may not be regularly rotating either, e.g. \citealt{Murthy22}). Interestingly, we do not find signatures of extreme outflows even in the galaxies that host luminous, quasar-like AGN in addition to the powerful radio jets. 



\begin{table*} 
\caption{Comparison of CO studies of molecular gas in samples of radio AGN, nearby early-type galaxies (ETG), and radio-quiet quasars. Note that, for the sake of comparison, the rms sensitivities in column 4 have been calculated for a 100\,km s$^{-1}$ spectral channel in all cases.}
\begin{center}
\begin{tabular}{llllll} 
\hline\hline 
Study        &N &Redshift  &rms       &Detect. &Selection \\
             &  &range (z)     &mJy b$^{-1}$   &rate (\%)      & \\
\hline
{\bf Radio AGN pre-ALMA} & & & & & \\
\citet{Evans05}  &33  &0.02 -- 0.15     &7 -- 17   &27$\pm8$\% &Detected by {\it IRAS} (3C, 4C, PKS, B2) \\ 
\citet{Ocana10}  &52 &$<0.077$  &1.0 -- 1.7   &52$\pm7$\%  &$P_{\rm 1.4~GHz} > 10^{22.5}$\,W Hz$^{-1}$ (3C, B2, NGC) \\ 
\citet{Smolcic11} &21 &$<0.1$     &2.1 -- 5.1   &62$\pm11$\% &Type 2 radio AGN observed in X-ray (mainly 3C)  \\ 
\hline
{\bf Radio AGN ALMA} & & & & & \\
\citet{Ruffa19a}  &9 &$<0.03$     &0.13 -- 0.58    &66$\pm16$\% &Volume and flux-limited ($S_{\rm 2.7~GHz} > 0.25$\,Jy), ETG \\ 

Current study &29 &0.05 -- 0.3 &0.13 -- 0.35 &34$\pm9$\% &Complete 2Jy sample, $S_{\rm 2.7~GHz} > 2$\,Jy, $\delta <10$ deg. \\
\hline
{\bf Nearby ETG} & & & & & \\
\citet{Young11} &259 &$<0.01$ &6.5 -- 20 &22$\pm3$\% &ETG from ATLAS$^{\rm 3D}$ sample \\
\citet{Davis19} &67 &$<0.025$ &3.4 -- 10  &25$\pm5$\% &ETG with $M_* >10^{11.5}$\,M$_{\odot}$ from MASSIVE sample \\
\hline
{\bf Quasar-like AGN ALMA} & & & & & \\
\citet{Shangguan20} &23 &$<0.1$ &1.1 -- 1.7 &91$\pm6$\% &PG quasars, radio-quiet \\
\citet{Ramos22} &7 &$<0.14$ &0.12 -- 0.27 &71$\pm17$\% &Type 2 quasars, radio-quiet \\
\citet{Molyneux24} &17 &$<0.2$ &1.7 -- 4.4 &82$\pm9$\% &Type 2 quasars, radio-quiet\\
\hline
\end{tabular}
\end{center}
\label{tab:sample_comparison}
\end{table*}

\section{Comparison with other samples}
\label{sec:ComparisonLiteratureSamples}

In order to put our results in a broader context, we now compare them with those obtained for other samples of radio galaxies, nearby quasar-like AGN, and quiescent ETG observed in CO for which statistics have been derived for the occurrence of molecular gas.

\subsection{Samples of radio AGN}

Details of the main molecular gas studies of samples of radio AGN are summarised in Table \ref{tab:sample_comparison}. The comparison is, of course, limited by differences in the depth and spatial resolution of the various observations, as well as the redshift ranges of the samples studied. 

At first sight it may seem surprising that all the pre-ALMA studies of samples of radio AGN \citep{Evans05,Ocana10,Smolcic11} achieve CO detection rates that are comparable with, or significantly higher than, that achieved in the current study, despite that fact that their rms sensitivities are a factor $\sim$3 -- 100 lower. However, the samples for these studies have median redshifts  that are significantly lower ($z <$0.05) than that of the 2Jy sample used here. This point is
further emphasised by the \citet{Ruffa19a} ALMA interferometric study that has a similar rms sensitivity to that of our 2Jy survey\footnote{Note, however, that the \citet{Ruffa19a} observations were
made of CO(2-1) rather than CO(1-0).}, but targets much
lower redshift objects ($z<0.03$, as illustrated in Figure  \ref{fig:plot1}), and achieves a detection rate that is more than a factor of two higher (66\% vs 34\%), albeit with a small sample.

In addition, the samples used for the pre-ALMA studies are often heterogeneous, incomplete or selected using criteria that may favour CO detection. For example, \citet{Evans05} studied radio AGN that were detected by the {\it IRAS} far-IR satellite. Since the overall detection rate of radio AGN in the {\it IRAS} survey was low, this selection automatically favours objects with larger than average masses of cool, CO-emitting ISM.

One further factor to consider is the dependence on the host galaxy morphology. In a recent study of a small sample of seven type 2 quasars, \cite{Ramos22} found an high detection rate of molecular gas (71$\pm$17\%). However, all the detections are associated with disk and strongly merging systems, while the two ETG in the sample are undetected in CO. Similar results were found by \citet{Husemann17} for
a sample of nearby type 1 quasar-like objects. This suggests that care should be taken in the comparison of samples of objects in which the galaxies have different morphological types.  This has been already seen in the studies of the other cold phase of the gas, i.e. \HI\, see \cite{Serra12}.



\subsection{Quiescent ETG samples}

We can also expand the comparison to samples of non-active ETG. This is particularly useful for determining whether the molecular
properties of the ETG hosts of radio AGN are different from their quiescent counterparts, which might provide clues on how they are triggered.

One of the first systematic molecular CO studies of ETG was performed by \cite{Wiklind95} using IRAM and SEST. Given the selection of far-IR detected ETG galaxies, the reported (high) CO detection rate ($\sim$55\%) cannot be considered representative of this group of galaxies. 
Indeed, later observations of a representative sample 43 galaxies from the SAURON survey of nearby ETG that were not selected on the basis of of their far-IR emission had a lower CO detection rate (28\%: \citealt{Combes07}).

More recent observations of molecular gas in larger samples of ETG were reported for the \atlas\ sample by \citealt{Young11} and \citealt{Alatalo13}, and for the MASSIVE sample by \citet{Davis19}. These samples can be use as reference point for the properties of molecular gas in massive galaxies without strong AGN. However, they focus on objects at much lower redshifts than our 2Jy sample -- up to a  distance of 107 Mpc ($z \sim 0.025$). In that sense, they are similar to the sample of radio AGN presented by \cite{Ruffa19a}. While the stellar masses of the \atlas\ sample objects are lower on average than those of the 2Jy sample, with median of $\sim$3$\times 10^{10}$ \msun, the MASSIVE sample covers higher stellar masses (M$_* > 3\times10^{11}$\,M$_{\odot}$). 

\cite{Young11} found  a  CO detection rate of 56/259 or 22$\pm3$\% for the \atlas\ sample (observed with IRAM-30m) over a range of molecular mass $10^{7} < M_{H_2} < 10^{9.2}$\,\msun. For the MASSIVE sample a similar detection rate of CO was found (25$\pm5$\%), with the molecular masses
of the detected sources in the range  $10^{8.5} < M_{H_2} < 10^{9.5}$\,\msun. 
No dependence on the stellar masses of the host galaxy was found, 
confirming that stellar mass loss is not the dominant source of 
the H$_2$ gas in ETG,   but rather that (at least a large fraction of) the molecular gas  is likely to have an external origin (see \citealt{Davis19} for an overview).

The total CO detection fractions for the ETG samples are similar to the 34$\pm$9\% we find for the 2Jy sample.
However, if we consider
the proportion of ETG that have molecular gas masses comparable to those we find for the 2Jy sample (i.e. $\ge 10^{8.9}$\,M$_{\odot}$), large differences are seen: only 1.2$\pm$0.7\% of the \atlas\ and 9$\pm3$\% of the MASSIVE ETG respectively have molecular
gas masses comparable with those of the detected 2Jy objects\footnote{Note that, for the sake of this comparison, we have re-calculated the molecular
gas masses from the \atlas\ and MASSIVE ETG samples using X$_{C0} = 2 \times 10^{20}$ cm$^{-2}$ (K km s$^{-1}$)$^{-1}$
to be consistent with our calculations for the 2Jy sources.}; clearly, the hosts of a significant proportion of the 2Jy radio
AGN sample are unusually gas-rich compared with the majority of quiescent ETG.

Follow-up observations of \atlas\ CO detected sources with CARMA interferometer (with a few arcsec resolution, \citealt{Alatalo13}), and of small samples of massive ETGs from the WISDOM and MASSIVE samples observed using ALMA \citep{Williams23,Dominiak24},
show that in the majority of the cases the molecular gas is distributed in kinematically settled structures (disks, rings, etc.). These structures have diameters in the range 0.25 -- 10\,kpc\footnote{Similar to the technique we have used for the 2Jy objects, here we have estimated the diameters of the molecular disks in the ETG from the lowest contours visible in
the maps presented in the relevant papers.}; however, 24/30 or 80$\pm$7\% of the \atlas\ sample in \citet{Alatalo13}  have molecular
disk diameters less than
5\,kpc, and the median diameter for the whole sample is 3.6\,kpc. Similarly, all 9 of the ETGs from the \citet{Williams23} and \citet{Dominiak24} studies that were not already included in the study of \citet{Alatalo13}, have molecular disk
diameters smaller than 5\,kpc, and the median disk diameter is 1.6\,kpc. For these samples, the dynamically unsettled cases are only mildly disrupted. Since the detected CO structures are rather compact spatially, with a average  radial extent of $\sim$0.8 --1.8 kpc, they have correspondingly small dynamical time-scales ($\sim$10$^7$ -- $2\times10^8$\,yr). The molecular gas structures are thus expected to relax and reach equilibrium rapidly.

In comparison, the molecular structures that we detect in the 2Jy radio AGN are notably more extended, with diameters covering the range 4.5 -- 25\,kpc (median 13.5\,kpc). On the other hand, the properties of the molecular disks detected in lower radio- and optical-luminosity radio AGN by \citet{Ruffa19a} are closer to those of the ETG samples: disk diameters in the range 0.4 -- 15\,kpc (median $\sim$2\,kpc), and only one object --NGC~612\footnote{Interestingly, this is also the only FRII source in the \citet{Ruffa19a} sample to be detected in CO.} -- with a diameter greater than 5\,kpc.




\subsection{Local radio-quiet quasar samples}

In order to determine whether there is any difference between the molecular gas reservoirs of 
nearby radio-loud and radio-quiet AGN, we also compare our results with those for samples of the PG type 1 quasars
from \citet{Shangguan20} and the type 2 quasars from \citet{Ramos22} and \citet{Molyneux24}, which are at similar redshifts to the 2Jy sample. Whereas the type 1 quasars in the \citet{Shangguan20} cover a similar range of [OIII] emission line luminosity to the SLRG in the 2Jy sample (see Figure \ref{fig:plots2}), the type 2 quasars
have [OIII] luminosities at the upper end of the range for the 2Jy SLRG. Both the type 1 and the type 2
quasar samples are dominated by radio-quiet AGN with $P_{\rm 1.4~GHz} < 10^{24.5}$\,W Hz$^{-1}$.

It is striking that all three of the radio-quiet quasar samples have higher CO detection fractions (71 -- 91\%) than the 2Jy sample; however, it is important to be cautious when making this comparison because 
the quasar observations have different rms sensitivities to that of the current study, and the
\citet{Shangguan20} and \citet{Ramos22} observations targeted the CO(2-1) line rather than the CO(1-0) line.
Therefore, we concentrate on comparing the molecular gas masses. We find that 65$\pm10$\% of the 23 type 1
quasars in \citet{Shangguan20}, and 90$\pm7$\% of the 20 type 2 quasars in the sample obtained by combining the 
those of \citet{Ramos22} and \citet{Molyneux24}, have molecular gas masses comparable to or greater than those of the detected 2Jy sources ($M_{\rm H_2} \ge 10^{8.9}$\,\msun).
For comparison, the detection rate of CO for the 19 SLRG in the 2Jy sample -- the radio AGN most comparable with the
quasar samples in terms of [OIII] luminosity -- is only 37$\pm$12\%. 
This provides evidence that the host galaxies of local radio-quiet quasars, and in particular local type 2 quasars, are more gas-rich on average than the 2Jy SLRG. However, some of this apparent contrast may be
due to differences in host galaxy morphologies (see discussion in Section 5.1). Certainly, the type 1 quasar sample of \citet{Shangguan20} has a much higher proportion of disk galaxies (>62\%) than the 2Jy radio AGN sample considered in this paper (3\%). On the other hand,
the proportion of disk galaxies in the full QSOFEED sample of nearby ( $z < 0.14$)
type 2 quasars -- which has higher [OIII] luminosities on average  than the \citet{Shangguan20} type 1 quasar sample -- is relatively low
\citep[<21\%:][]{Pierce23}.

\section{Discussion}
\label{sec:discussion}

Here we connect the detection rate, distribution and kinematics of the molecular gas with the properties of the AGN, radio sources and their host galaxies. 

\subsection{Occurrence, morphology and origin of the molecular gas}
\label{sec:discussionOccurrence}

We have detected large amounts of molecular gas in 10 of our targets. For these objects, the mass of the detected H$_2$  ranges between  $8 \times 10^8$ and $2 \times 10^{10}$\,\msun.
The upper limits to the H$_2$ masses from our observations are in the range $\sim7 \times 10^7$ to a few $10^9$\,\msun, and strongly redshift-dependent, as shown in Figure  \ref{fig:plot1}. This results in a detection rate -- for these high molecular gas masses -- of 34$\pm$9\%. 



One point of interest is whether the detection rate of molecular gas depends on the optical spectroscopic
and radio morphological classifications of the AGN, perhaps related to different triggering/fuelling
mechanisms. Considering first the radio morphology, we find that
1/6 (17$\pm$15\%) of the FRI and 9/23 (39$\pm$10\%) of the FRII sources in our sample have CO(1-0) detections. While this
difference is consistent with the idea that the two groups are triggered in different ways, for example FRII sources by mergers
and FRI sources by hot gas accretion, the result has a low statistical significance due to the small
sample sizes. Moreover, when we consider the optical spectroscopic classifications, we find no
evidence for significant differences in the detection rates that might be related to different accretion modes
(e.g. \citealt{Smolcic11}): 3/10 (30$\pm$14\%) of the WLRG and 7/19 (37$\pm$11\%) SLRG have molecular detections.



As already noted, in terms of the spatial distribution of molecular gas, where we detect CO emission in the 2Jy objects, it tends to be more extended and show a more irregular structure than found in the spatially resolved studies of lower-luminosity (mainly FRI) radio AGN by \citet{Ruffa19a} and nearby ETG by \citet{Alatalo13}, \citet{Williams23} and \citet{Dominiak24}.

At first sight, the relatively large extents and irregular distributions of the molecular gas appear similar to those found in
CO observations of cool-core clusters \citep[e.g.][and references therein]{Tamhane22,Russell19}. However, the majority of
the detected 2Jy sources show a higher degree of rotation in the extended molecular gas than observed in most 
cool core clusters, and number-count and X-ray studies suggest that most of these objects are also in relatively
low density, group-like environments, rather than clusters \citep[see Table\,\ref{tab:info};][]{Ramos13,Ineson15}. Therefore, it seems unlikely
that the molecular gas has condensed from the hot X-ray halos in such cases, and a galaxy merger or interaction 
origin -- which could also explain the extent and irregular distribution of the molecular gas -- is more plausible.

Although the majority of the detected 2Jy sources are in relatively low-density group-like environments, a minority -- comprising PKS0915-11 (Hydra A), PKS1648+05 (Hercules A) and PKS2211-17 -- are at the centres of galaxy clusters based on number count analysis and X-ray observations \citep{Ramos13,Ineson15}. Two of them (PKS1648+05, PKS2211-17) show a lack of clear evidence for rotation in the molecular gas, while the third (PKS0915-11) shows a 
regularly rotating disk with a diameter of 4.5\,kpc \citep{Rose19}. Certainly, for the former two cases cooling from the X-ray haloes
is a plausible alternative to galaxy mergers/interactions as an origin for the molecular gas. Indeed, although they
both show evidence for patchy dust absorption in optical images \citep{Ramos11}, neither object
shows tidal features suggestive of galaxy interactions. On the other hand, of the 8 objects in our sample that
are in relatively rich, cluster-like environments (see Table 1) -- all WLRG and many with tight upper limits on their molecular gas masses -- only 38$\pm$17\% are detected in \coOne. This is similar to the \coOne\ detection rate for the sample as a whole.


We also caution that using the
kinematics and distribution of the molecular gas is not a foolproof method for distinguishing a cooling-flow origin. This is illustrated by the case of higher-redshift 2Jy object PKS0023-26 ($z = 0.32$), which shows a massive reservoir of molecular gas
($M_{\rm H_2} = 3.1\times10^{10}$\,M$_{\odot}$) that is extended over a region with diameter 24 kpc and has an irregular distribution, with no clear sign of regular rotation \citep{Morganti21}. Despite the similarities with
the molecular gas properties of cool core clusters, X-ray observations reveal a relatively low X-ray luminosity
($L_{2-10\rm keV} < 3\times10^{43}$\,erg s$^{-1}$) for this source that is inconsistent with the object being in a rich cluster; rather, the observations are more consistent with interactions in a galaxy
group as the most plausible origin for the molecular gas.

Overall, our results suggest that accretion via galaxy interactions is the origin of the molecular gas for the SLRG  
objects in our sample. 
Indeed, all the CO-detected SLRG show tidal features in deep optical images that are suggestive of merger events in the recent past (\citealt{Ramos11}). However, while two of the sources have large molecular gas masses ($\sim$10$^{10}$\,M$_{\odot}$) that are comparable with those of ULIRG-like major gas-rich mergers \citep[e.g.][]{Solomon97}, most have much lower molecular masses, suggesting that more minor or less-gas-rich mergers play an important role in the triggering of this type of AGN, consistent with the dust mass results \citep{Tadhunter14,Bernhard22}.



 We further note that the one CO-detected FRI source in our sample -- PKS0915-11 (Hydra~A) -- has the most compact and regular CO structure ($D\sim 4.5$\,kpc, \citealt{Rose19}) of all the objects in our sample, despite being in a cluster, and its molecular gas disk is close to perpendicular in projection to its radio axis.  All the other CO-detected sources have an FRII or similar radio morphology, and more extended and irregular gas distributions. This is consistent with the HST optical imaging study of 3C radio galaxies by \citet{Dekoff00}, who found that the detected dust absorption features (dust lanes, patches) are less regular, more extended on average, and show less evidence for alignment perpendicular to the radio axes in FRII than FRI radio galaxies in their sample. Potentially, it is also consistent with the recent finding by \citet{Balmaverde22} that FRII galaxies show ``superdisk'' structures in ionized gas that are much larger than the compact emission-line structures seen in the FRI sources in their sample of 3CR radio galaxies.



\subsection{Molecular gas and the evolution of the radio sources} 

The radio properties provide information on the stage of evolution of the radio sources, allowing us to explore the 
relationship between this and the molecular gas. In particular, young radio sources can be identified as Peaked Sources (see \cite{ODea21} -- sometimes labelled as Compact Steep Spectrum (CSS) or Giga-Hertz Peaked Sources (GPS)) -- and they are  often found embedded in a particularly rich medium compared to the large (and more evolved) sources. 
In the sample studied here there are  3 targets classified as CSS/GPS: PKS1151--34, PKS1814--63, PKS1934--63 and a further object -- PKS2314+03 -- represents an intermediate case of a radio source on galactic scales (diameter $\sim$50\,kpc) that also
has a steep-spectrum radio core. 
Of these objects, only PKS1151--34  is not detected in \coOne, but it is the highest redshift
object in the full sample, and consequently has a relatively high molecular gas mass upper limit ($M_{H2} < 3\times10^9$\,M$_{\odot}$). All the others are detected, although the fraction of molecular gas associated with PKS1934--63 is unclear at the resolution of the present data.

These findings can be combined with those available in the literature. 
For example, two more 2Jy radio galaxies classified as CSS/GPS sources have recently been observed and studied in detail with ALMA: PKS1549--79 and PKS0023--26 \citep[see][]{Oosterloo19,Morganti21}). Both show massive molecular gas reservoirs ($M_{H_2} > 10^{10}$\,M$_{\odot}$), that would have been readily detected if observed as part of the
current survey. Overall, all 6 of the CSS/GPS and related objects from the wider 2Jy sample of \citet{Tadhunter98}
at redshifts $z < 0.4$ have now been observed with ALMA, and 5 (83$\pm$17\%) of them are detected in CO. This high
rate of detection reinforces the idea that the host galaxies of such sources have a richer cool ISM  on average
than the more common extended radio sources in the same sample, consistent with the fact that they
also show higher incidence of recent star formation activity  \citep{Tadhunter11,Morganti11,Dicken12}.


\subsection{Gas-to-dust ratio}
\label{sec:Dust}

The detection of CO emission in our sample objects shows a strong correlation with the
detection of dust features (lanes, patches) in existing ground-based optical images. \citet{Ramos11} found that 7 of the 29 objects in the 2Jy sample
with redshifts in the range$0.05 < z < 0.3$ show certain or possible dust features in Gemini r$^{\prime}$-band optical images: PKS0915-11 (Hydra A), PKS1559+02 (3C327), PKS1648+05 (Her A), PKS1733-56, PKS1814-63, PKS1949+02 (3C403),  PKS2211-17 (3C444). Of these, all but one -- PKS1559+02 -- are detected in CO(1-0). In contrast, PKS0806-10 and
PKS1942-46 were detected in CO(1-0) but show no clear dust features in the optical images.

Given that the detection or not of dust features depends on a number of factors (e.g. dust mass, geometry, orientation, spatial resolution),  the optical images provide only a lower limit on the true rate of occurrence of dust {\it at some level} in the host galaxies. Indeed, {\it Spitzer} and {\it Herschel} observations of our sample reveal the presence of dust in the form of thermal bumps at far-IR wavelengths in several objects for which no dust features were detected in optical images \citep{Dicken08,Dicken23}. Based on the far-IR observations, robust dust masses have been estimated by \citet{Bernhard22} using grey-body fits to the SEDs of the thermal dust emission. The derived dust masses depend to some extent on the assumptions made in making the fits (e.g. the $\beta$ index\footnote{The $\beta$ index is the spectral index of the power law that modifies
the Planck curve black body spectrum for optically-thin, grey-body dust emission.}), but \citet{Bernhard22} show that they are likely to be accurate to within
a factor of 2. The dust masses presented in Table 2 are those calculated by \citep{Bernhard22} for the cool dust assuming  $\beta=2$.

\begin{figure}
\vglue -1cm
\hspace*{-6mm}
\includegraphics[width=11.0cm]{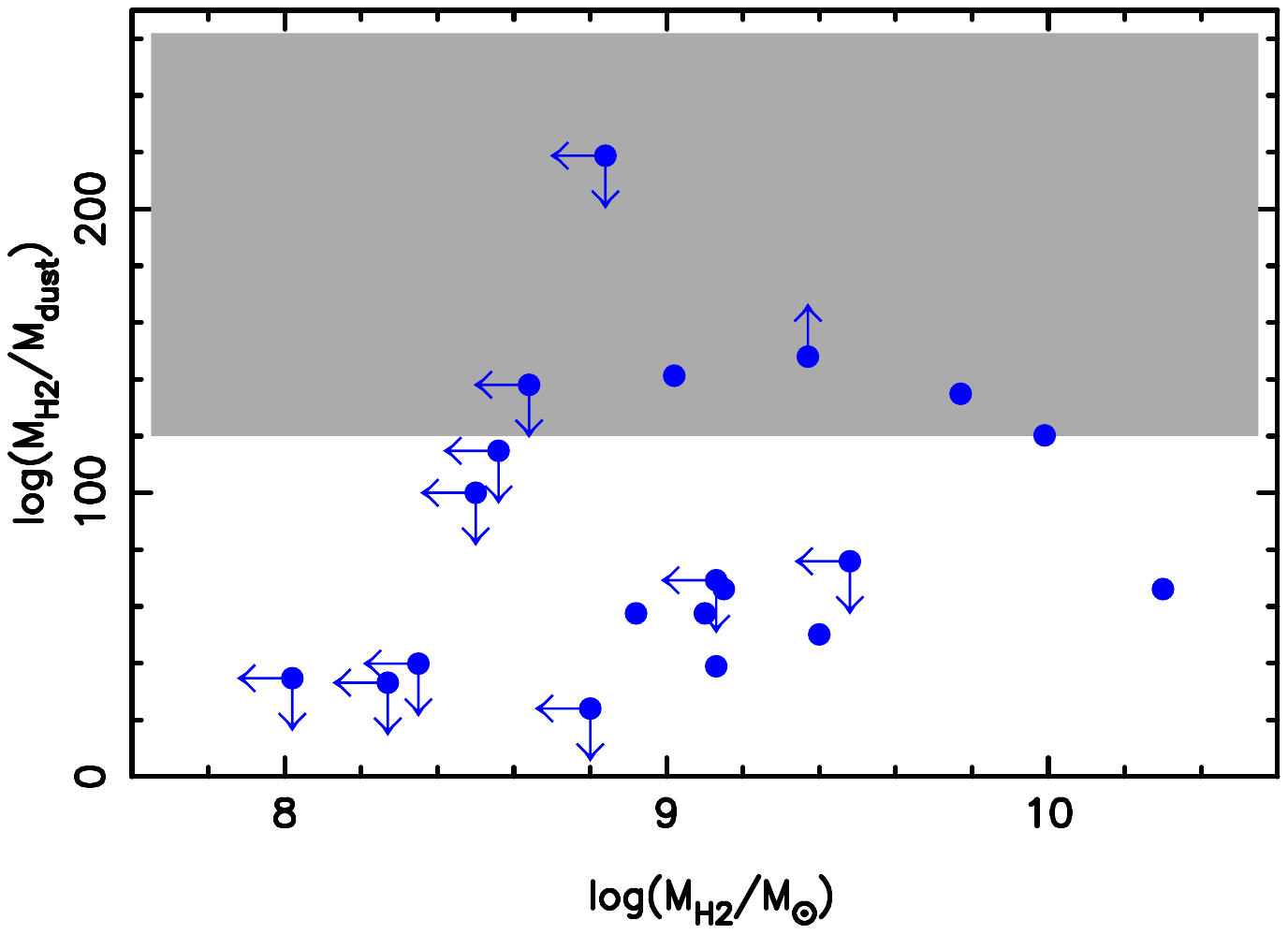}
 \caption{Distribution molecular-gas-to-dust ratio as a function of $H_2$ mass
 for the 2Jy sample. Blue filled circles represent objects detected 
 in dust and $H_2$, while blue filled circles with arrows represent
 objects with upper or lower limits on the gas-to-dust ratio. Objects that have upper limits in both dust and gas masses are not included. The shaded region represents the lower end of the range of typical total gas--to-dust ratios for disk galaxies.}
\label{fig:plot3}
\end{figure}

In Figure  \ref{fig:plot3} we show the distribution of gas-to-dust mass ratios as function of the molecular gas mass. These were calculated by simply dividing the molecular gas mass by the
dust mass.
The gas-to-dust ratios of the detections -- with an mean value of $M_{gas}$/$M_{dust} \sim 90$ -- are often below typical Galactic values and those found for disk galaxies and ETG in general ($120 < M_{gas}$/$M_{dust} < 450$; \citealt{Li01,Zubko04,Draine07,Ruyer14,DeVis19,Casasola20}), but this likely to be due in part to the fact that we have not accounted for the contribution of neutral HI gas, which could increase the total gas mass by a factor of two or more. 
Furthermore, as expected, most of the CO-detected objects also have dust detections at far-IR wavelengths. The one exception is PKS1932-46, which is detected in \coOne\ but has an uncertain detection of dust emission at far-IR wavelengths, due to
the potential contamination by non-thermal radiation from the core of the radio source, which is relatively
strong in our $\sim$100\,GHz continuum image, and is also detected by APEX/LABOCA at 870$\,\mu$m (344\,GHz) \citep{Dicken23}.
However, apart from this one object, there are no objects in our sample with high-mass, cool ISM components (detected in CO) that were missed in the {\it Herschel} observations because of non-thermal contamination, or because of a lack of observations at the longer far-IR wavelengths ($>160$\,$\mu$m) combined with low dust temperatures.

On the other hand, we note that five objects  -- PKS0806-10, PKS0349-27, PKS0945+07, PKS1559+02 and PKS2221-02 -- show notably low values or upper limits for the gas-to-dust ratio: $M_{gas}$/$M_{dust}< 50$. Of these objects, the case of PKS1559+02 (3C327) is particularly interesting. In many ways, it shows very similar properties to PKS0806-10: redshift $z\sim0.1$, quasar-like [OIII] emission-line and 24\,$\mu$m continuum luminosities, and relatively high dust mass ($M_{dust} \sim 10^{7.5}$\,M$_{\odot}$). However, unlike PKS0806-10, it is not detected in CO(1-0), and its gas/dust ratio 
is a factor of $\sim$2$\times$ or more lower ($M_{gas}$/$M_{dust} < 24$).
Such low gas-to-dust ratios  are rare but not unique. A comparable low value (gas-to-dust ratio = 28) has been reported by \cite{Morokuma19} in the case of Fornax~A. One possible explanation in the case of PKS1559+02 is that
a massive gas reservoir is present, but is dominated by non-molecular phases of the ISM, such as the cool HI or warm ionized phases, perhaps as a consequence of the heating and ionizing effect of the quasar-like AGN present in this object.


Aside from physical explanations, an alternative possibility is that the CO emission is significantly more extended than the single ALMA beam we have used to estimate the upper limits on the molecular gas masses (see
section \ref{sec:H2masses}). Certainly, as we have seen, all the CO-detected objects are resolved and cover more than one beam, in some cases several beams. Taking this into account, and also the factor
2 uncertainty in the dust masses, it is possible that the true gas-to-dust ratios in the apparently discrepant objects are closer to typical Galactic values (i.e. $\gta$100). However, deeper ALMA observations, that are more sensitive to low-surface-brightness CO emission that covers several beams, will be required to confirm that the
molecular gas reservoirs are truly extended in this way in the so-far undetected sources.

Finally, we should bear in mind that the values we quote are averaged values inside large structures, which do not take into account that the gas-to-dust ratio may change with the locations across the distribution of the molecular gas (e.g. with the distance from the AGN). This has been reported for Cen A by \citet{Parkin12}, who found average a gas-to-dust mass ratio of 103, but with an  increase in this ratio to approximately 275 towards the centre of Cen A.

\subsection{Molecular gas in radio AGN: the broader galaxy context}
\label{sec:discussionHost}

Over the last few decades there has been considerable progress in understanding the relationships
between properties such as stellar mass, star formation rate (SFR) and gas content for the general population of non-active galaxies. For well-resolved nearby galaxies, one of the clearest correlations is found between the surface density of the SFR and that of the gas mass, which encompasses normal disk galaxies
as well as  starbursts \citep{Schmidt59, Kennicutt98}; however, there are also strong trends between the global (spatially integrated) properties of galaxies. For example, there is a clear correlation between total stellar mass and SFR for star forming galaxies (the ``main sequence'': \citealt{Noeske07}), and the total SFR is also found to be correlated with total molecular gas mass for such galaxies \citep[e.g.][]{Saintonge22}. It is important to determine how radio AGN fit in with these global correlations, since this might provide information about the extent to which the jet and AGN activity affect the star formation
in the host galaxies -- the so-called AGN feedback effect.

In our previous {\it Herschel}-based work on the 2Jy sample, we found that the majority of radio 
AGN fall on or below the main sequence for star formation,
and from a comparison between the star formation rates and gas masses (estimated from dust masses), 
we deduced that their star formation efficiencies are comparable with, or higher than,
those of typical star-forming galaxies \citep{Bernhard22}. Our ALMA results provide an opportunity
to extend these results.

\begin{figure}
\vglue -1cm
\hspace*{-6mm}
\includegraphics[width=11.0cm]{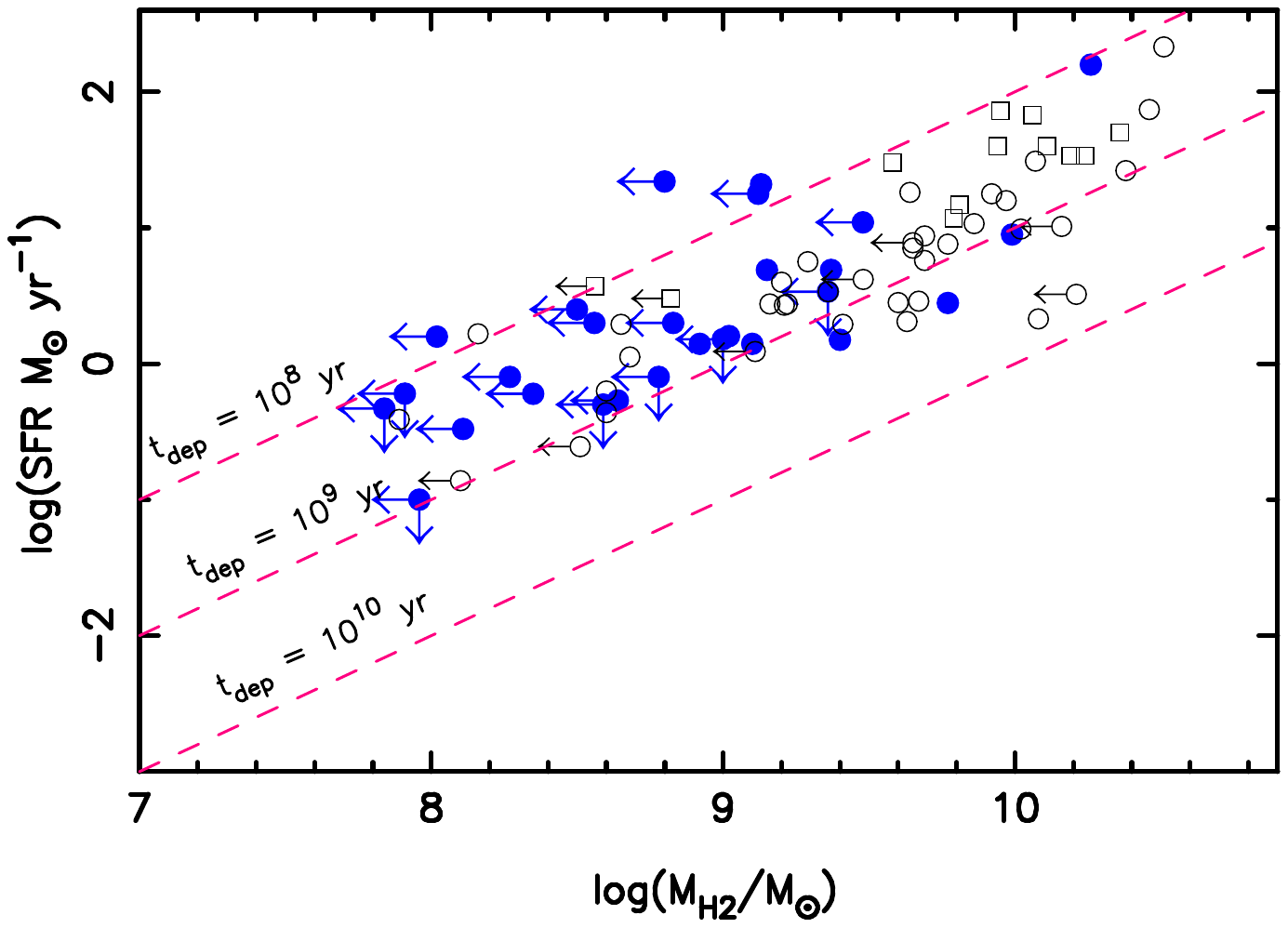}
\vglue -1cm
\hspace*{-6mm}
\includegraphics[width=11.0cm]{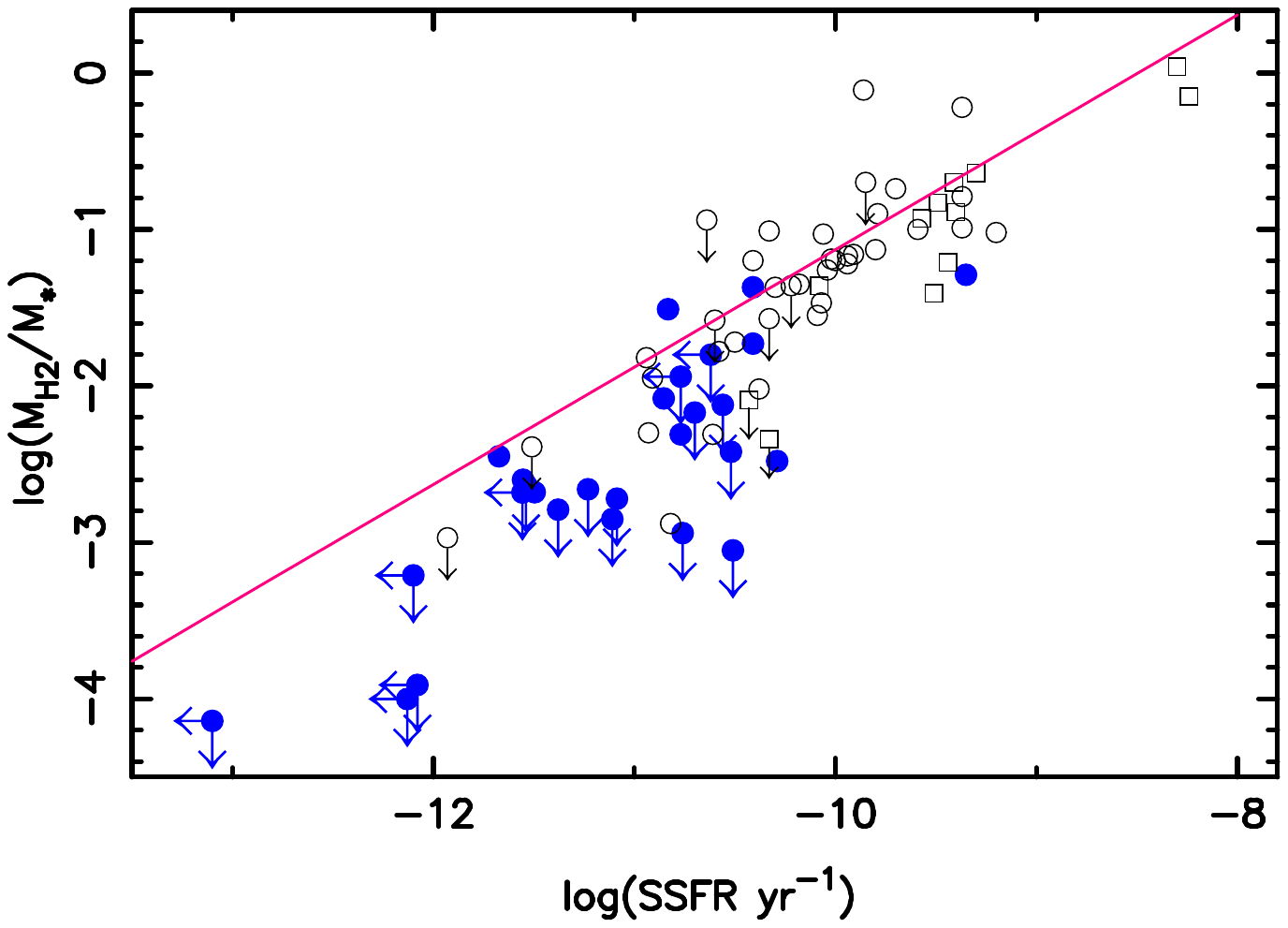} 
 \caption{The relationship between stellar mass, molecular gas mass and star formation rate for the 2Jy radio AGN sample (solid blue symbols). Top: star formation rate (SFR) vs molecular gas mass, with gas depletion times shown
 as diagonal dashed lines. Bottom: molecular-to-stellar mass ratio vs specific star 
 formation rate (SSFR), with the correlation derived by \citep{Saintonge22} for the
 general galaxy population in the local universe shown as the solid
 line. To make these plots we used the stellar masses and star formation
 rates for the 2Jy sources derived by \citet{Bernhard22} following correction for AGN 
 conatmination of the far-IR continuum using a template-fitting approach. For comparison, in 
 both plots we also show the results for type 1 quasars (open black
 circles) from \citet{Shangguan20}, and type 2 quasars (open black squares)
 with stellar mass estimates from \citet{Ramos22} and \citet{Jarvis20}.}
\label{fig:plotsHost}
\end{figure}

The top panel of Figure \ref{fig:plotsHost}  shows SFR plotted against molecular gas mass for the objects in our
ALMA study, with the dashed diagonal lines representing molecular gas depletion times -- a measure of
the star formation efficiency. For comparison, we also show results for local  type 1 and type 2
quasars \citep{Jarvis20,Shangguan20,Ramos22}. Consistent with previous results based on {\it Herschel}-derived dust masses, it clear that the radio AGN have relatively short gas depletion times $t_{dep} \lta 10^9$\,yr, suggesting high star formation efficiencies. For reference, galaxies on the star forming main sequence with similar
stellar masses to the radio AGN ($\gta 10^{11}$\,M$_{\odot}$), have molecular gas depletion times of
$t_{dep} \sim 10^9$\,yr \citep{Saintonge22}, whereas the disks of nearby normal spiral galaxies have depletion times that are  typically 
longer ($t_{dep} \sim 2\times10^9$\,yr; \citealt{Bigiel08}). Comparing with other AGN, on average the radio AGN gas depletion times are  similar to those of the
local quasar samples shown in Figure \ref{fig:plotsHost}, and Seyfert galaxies \citet{Salvestrini22}, but shorter than those
of low-luminosity AGN (LLAGN) \citep{Casasola15} and the molecular disk of Centaurus A \citep{Espada19}.

Perhaps the tightest correlation between gas and stellar properties of galaxies is that between
the ratio of molecular gas mass and stellar mass (i.e. $M_{H_2}/M_*$) and the specific star formation rate ($SSFR = SFR/M_*$) \citep[see][and references therein]{Saintonge22}. The bottom panel of Figure \ref{fig:plotsHost} shows where the radio AGN fall relative this correlation (the solid line represents
the general galaxy population). From this it is clear that both the radio AGN and radio-quiet quasars fall on or below the correlation. This implies that these luminous AGN populations have
similar or higher specific star formation rates for a given molecular-to-stellar mass ratio compared with the general galaxy population.

Overall, we find no evidence that the
star formation rates of the radio AGN host galaxies have been reduced compared 
with what one would expect for
non-active galaxies with similar properties; clearly, in these objects the effects of negative AGN feedback on star formation are not strongly manifested on the timescales of the AGN activity. This ties in with the idea that the most important impacts of AGN feedback on galaxy evolution are cumulative -- the combined effect of several cycles of AGN activity over long timescales --  rather than clearly detectable in a single activity cycle (see \citealt{Harrison24} and references therein).

\section{Summary and Conclusions}

\vspace{0.5cm}
\noindent
We have reported the results of an ALMA \coOne\ survey of the complete sample of 29 2Jy radio AGN
with redshifts $0.05 < z < 0.3$. The key results and conclusions are as follows.

\begin{itemize}
    \item The detection rate of CO(1-0) molecular emission  for the 2Jy sample (34$\pm$9\%) is comparable to, or lower than, that of pre-ALMA CO surveys of samples of radio AGN (27 -- 62\%), despite 
    the considerably higher sensitivity of the ALMA observations. This apparently surprising result
    may be explained by the fact that (a) the pre-ALMA surveys covered samples that were at notably lower redshifts on average than the 2Jy sample; (b) with their relatively small beam size ($FWHM\sim2$ arcsec) the ALMA observations may have missed diffuse emission on large scales around the host galaxies; (c) as well as any large-scale diffuse emission, the larger beams of the pre-ALMA survey observations ($FWHM \sim$5 -- 20\,arcsec)  may have admitted emission from nearby companion galaxies in some cases.

    \item All the early-type host galaxies of the CO-detected 2Jy sources have relatively large masses of molecular hydrogen ($8\times10^8 < M_{H_2} < 2\times10^{10}$\,M$_{\odot}$), and the proportion (34$\pm$9\%) of the sample showing such large molecular masses is notably higher than for the nearby ATLAS$^{3D}$ and MASSIVE samples of early-type non-active galaxies (1.1 and 9\% respectively: \citealt{Young11,Davis19}), and the sample of lower luminosity radio AGN observed with ALMA by Ruffa et al.  (2019a: 11\%).

    \item Although the detection rate of relatively large molecular gas masses is higher for the 2Jy radio AGN than for samples of nearby ETG, it is lower than those derived for samples of local, radio-quiet quasars observed by ALMA \citep{Shangguan20,Ramos22,Molyneux24}. This suggests that luminous, radio-loud AGN are less gas-rich on average than their radio-quiet counterparts.

    \item The molecular gas in many of the 2Jy objects is distributed in large (4.5 -- 25\,kpc diameter), rotating disk structures that show signs of distortion and irregularity in their outer reaches. In contrast, the molecular disks in nearby early-type galaxies, and radio AGN with lower luminosities, tend to be more compact (typical diameters $\sim$2 -- 5\,kpc) and show less evidence for significant structural distortion \citep{Alatalo13,Ruffa19a}. Moreover, the molecular disks of the 2Jy objects are not always perpendicular to the radio axes, but show a
    wide range of projected alignments.

    \item Given the 1 -- 2 orders of magnitude spread in molecular gas masses for a given AGN luminosity, it is unlikely that the mass of the molecular gas reservoir is the sole determinant of the level of AGN activity; other factors (e.g. the degree to which the cold gas is dynamically settled) must also be important.

    \item At the sensitivity and resolution of our observations we see no evidence for molecular outflows in \coOne\ emission in any of the detected 2Jy sources. However, one source -- PKS1814-63 -- shows a relatively broad \coOne\ absorption
    component against a compact synchrotron-emitting lobe close to its nucleus. This may
    be a sign of turbulence induced by the interaction between the radio jets and the circum-nuclear ISM.

    \item The 2Jy radio AGN have relatively small gas depletion times ($t_{dep} \lta 10^9$ yr) for their stellar masses ($M_* \gta 10^{11}$\,M$_{\odot}$) and fall on, or close to, the correlation  between molecular-to-stellar mass ratio and specific star formation rate found for the general non-active galaxy population. Therefore, there is no evidence that negative AGN feedback has affected
    the star formation efficiencies in their host galaxies on the timescale of the AGN lifecycle.
\end{itemize}

Together, the high masses, rotational kinematics, large extents, and distorted outer structures of the molecular gas in many of the CO-detected 2Jy objects suggest an external origin for the gas. This is consistent with the results of deep imaging observations, and supports the idea that the powerful AGN and jet activity has been triggered by galaxy interactions, as the accreted gas settles into a stable dynamical configuration. However, a minority of the detected sources situated in relatively rich cluster environments show molecular gas properties that are more akin to those of cool core clusters. In such cases, the accretion of gas that has cooled from the hot X-ray halos is a plausible alternative to galaxy interactions as a triggering mechanism.

\section*{Acknowledgements}
 This paper makes use of the following ALMA data: ADS/JAO.ALMA\#2019.1.01022.S (PI Tadhunter), ADS/JAO.ALMA\#2018.1.00739.S (PI Balmaverde), ADS/JAO.ALMA\#2016.1.01214.S (PI McNamara) and ADS/JAO.ALMA\#2017.1.00629.S (PI McNamara).
 ALMA is a partnership of ESO (representing its member states), NSF (USA) and NINS (Japan), together with NRC (Canada), MOST and ASIAA (Taiwan), and KASI (Republic of Korea), in cooperation with the Republic of Chile. The Joint ALMA Observatory is operated by ESO, AUI/NRAO and NAOJ. CT acknowledges support from STFC grants ST/R000964/1 and ST/V000853/1. CRA acknowledges funding from the State Research Agency (AEI-MCINN) of the Spanish Ministry of Science and Innovation under the grant ``Tracking active galactic nuclei feedback from parsec to kiloparsec scales'', with reference PID2022-141105NB-I00. MVM's research has been funded by grant Nr. PID2021-124665NB-I00 of the 
Spanish Ministry of Science and Innovation/State Agency of Research 
MCIN/AEI/ 10.13039/501100011033 and by "ERDF A way of making Europe".

\section*{Data Availability}

The reduced data underlying this article will be shared on reasonable request to the corresponding author, and the raw data are publicly available via the ALMA archive.

{}

\begin{appendix}
\section{Additional figures for detected sources}
\label{sec:FiguresAppendix}



\vglue -1.0cm\noindent
\begin{figure*}
   \centering  
\includegraphics[height=5.7cm]{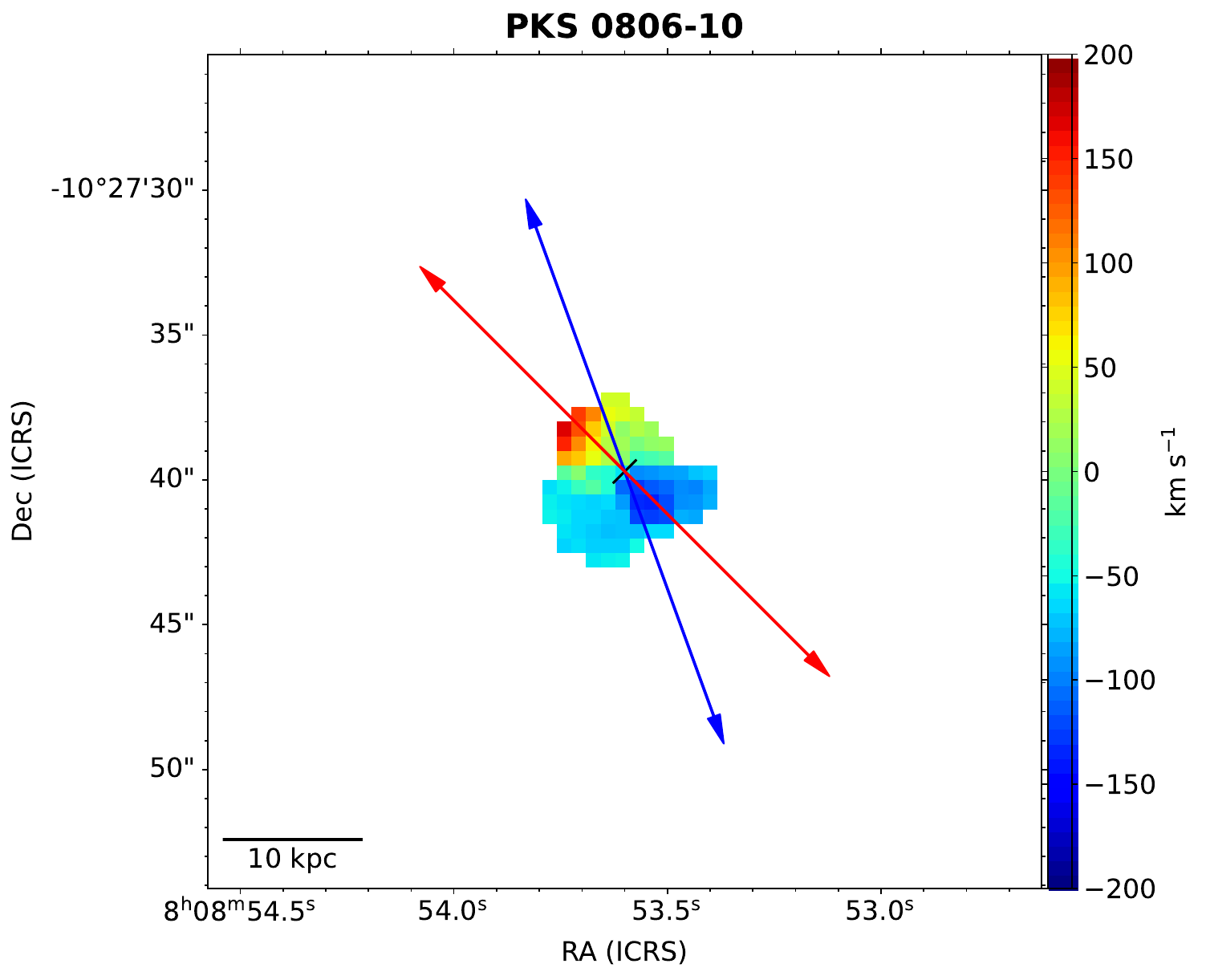}
\includegraphics[height=5.7cm]{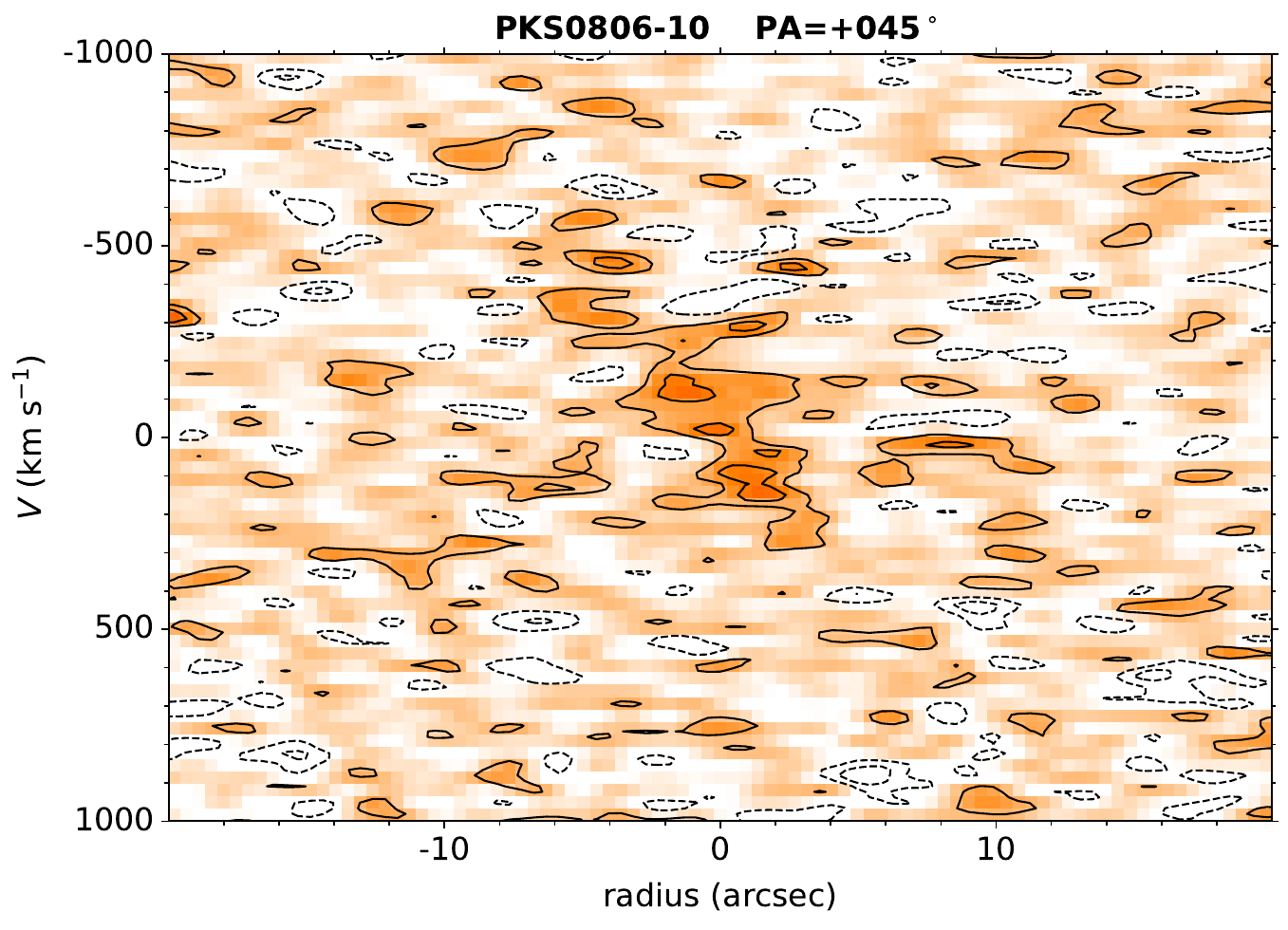} \\

\includegraphics[height=5.7cm]{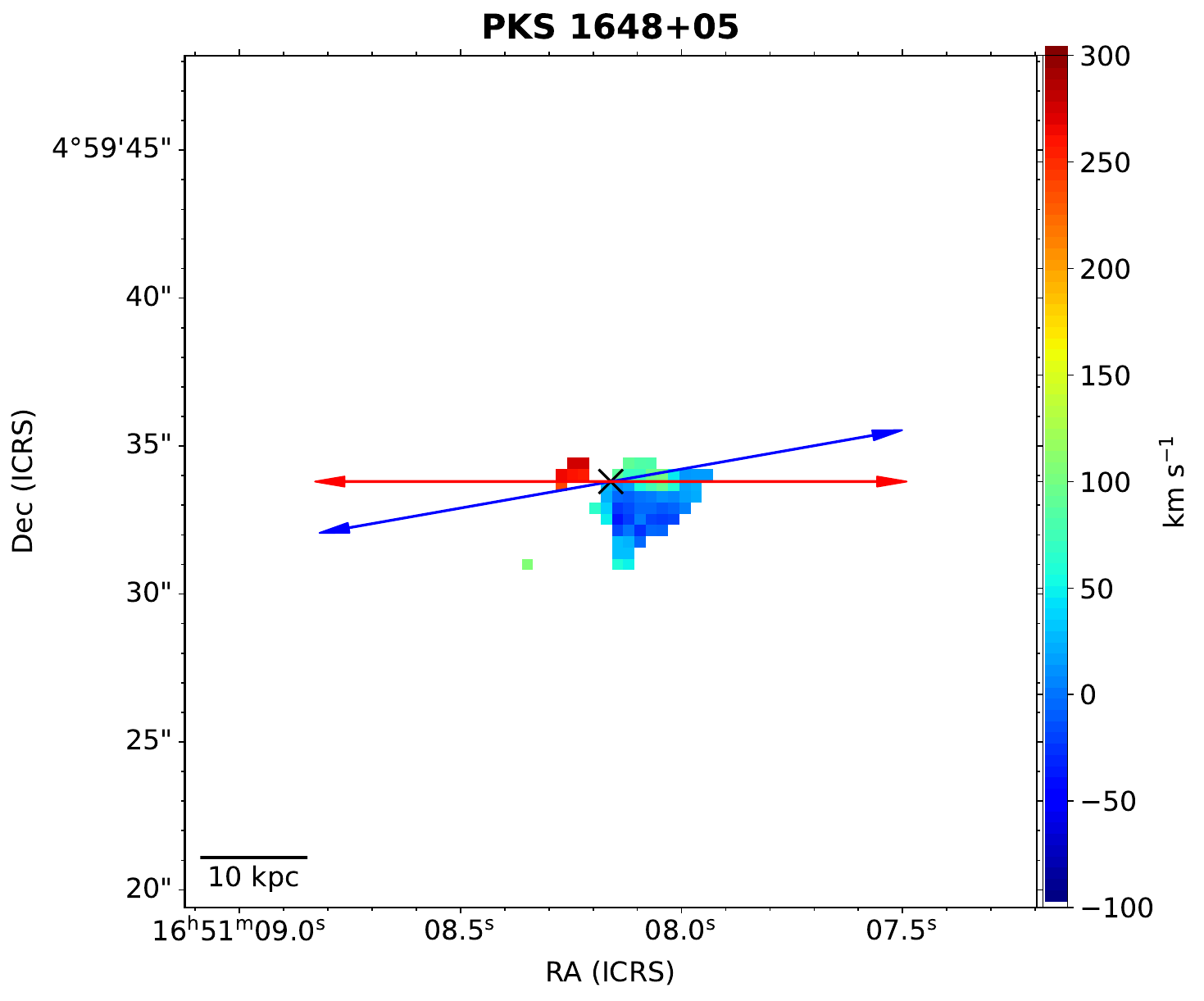} 
\includegraphics[height=5.7cm]{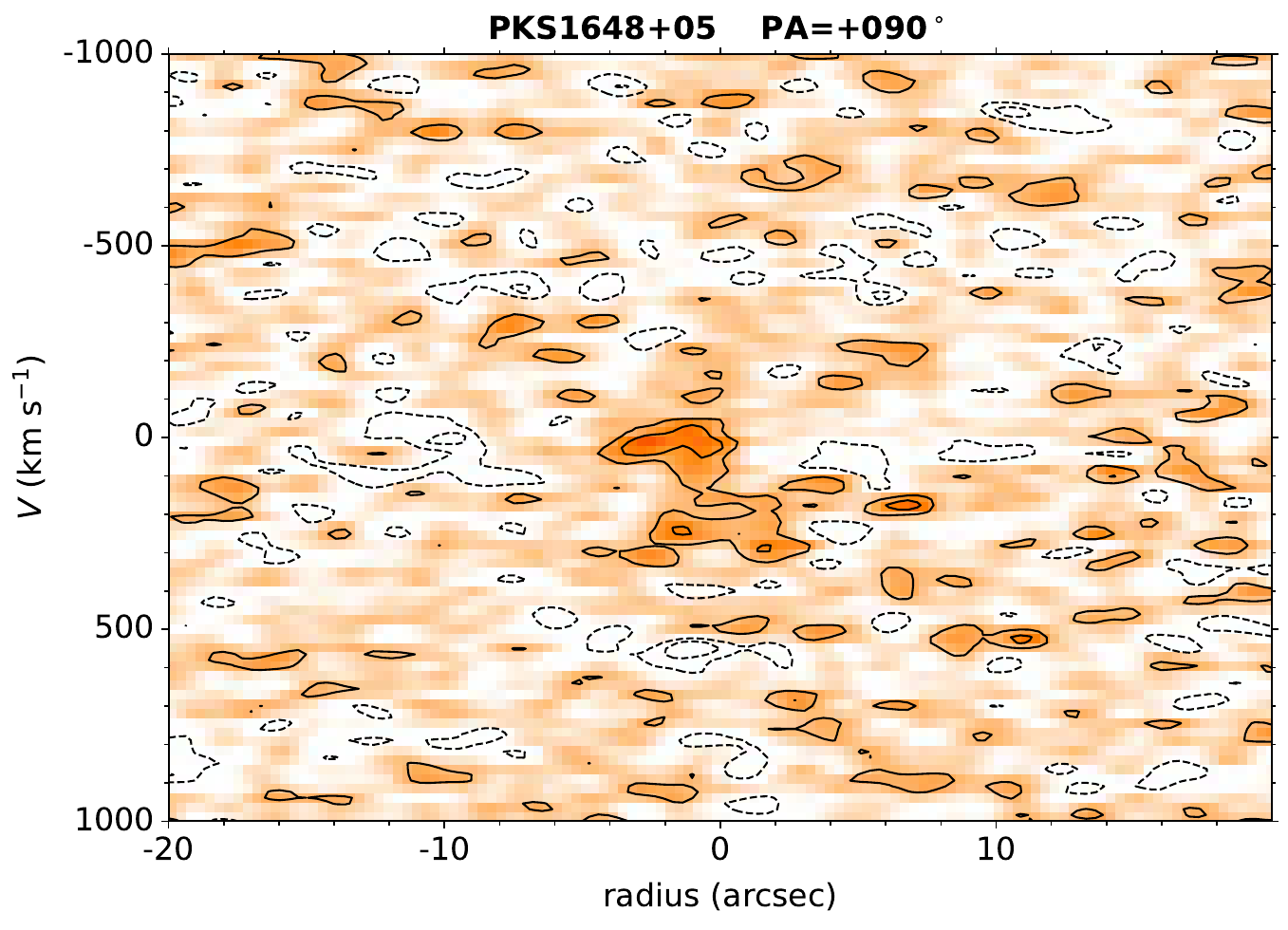} \\

\includegraphics[height=5.7cm]{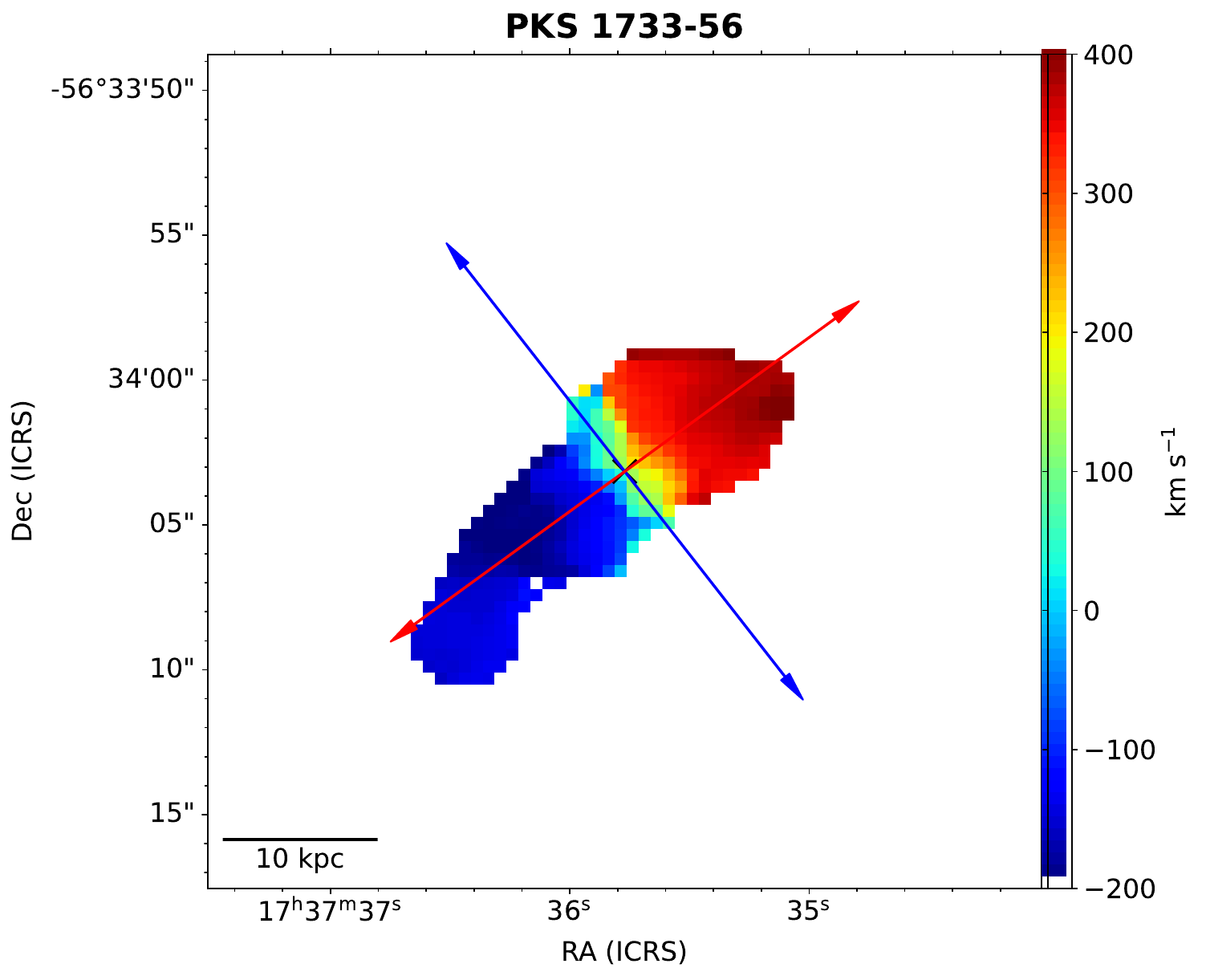}
\includegraphics[height=5.7cm]{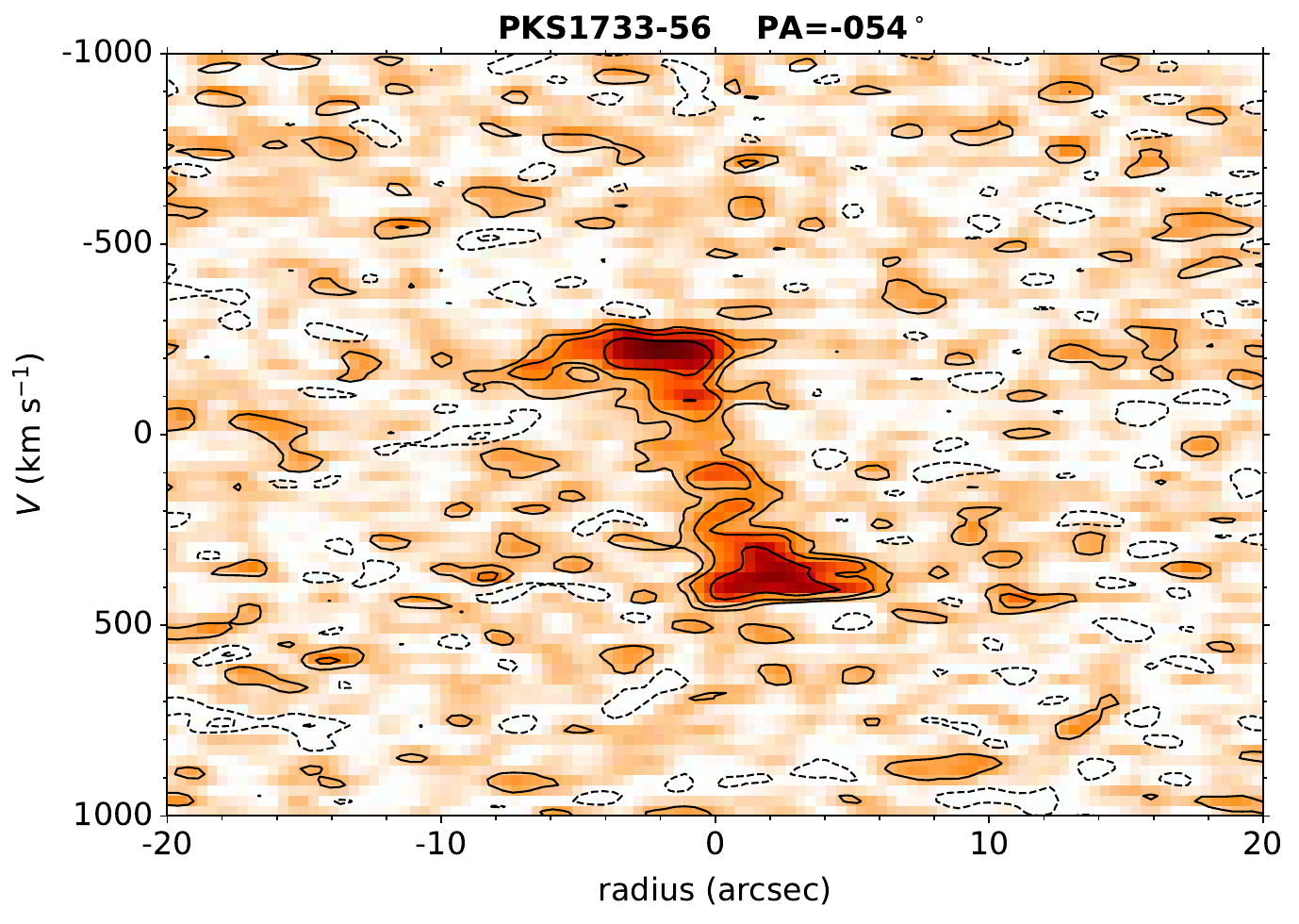} \\

\includegraphics[height=5.7cm]{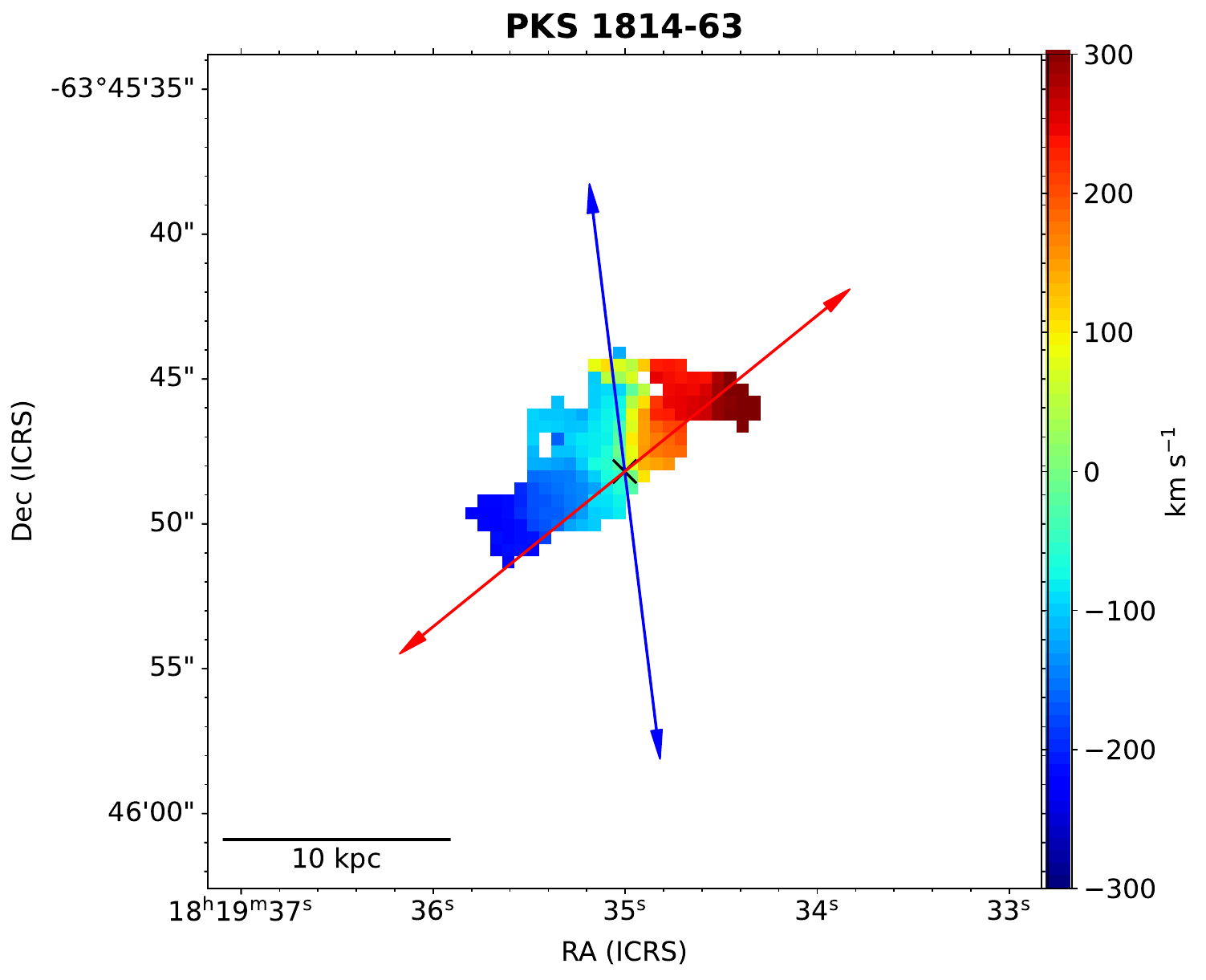}
\includegraphics[height=5.7cm]{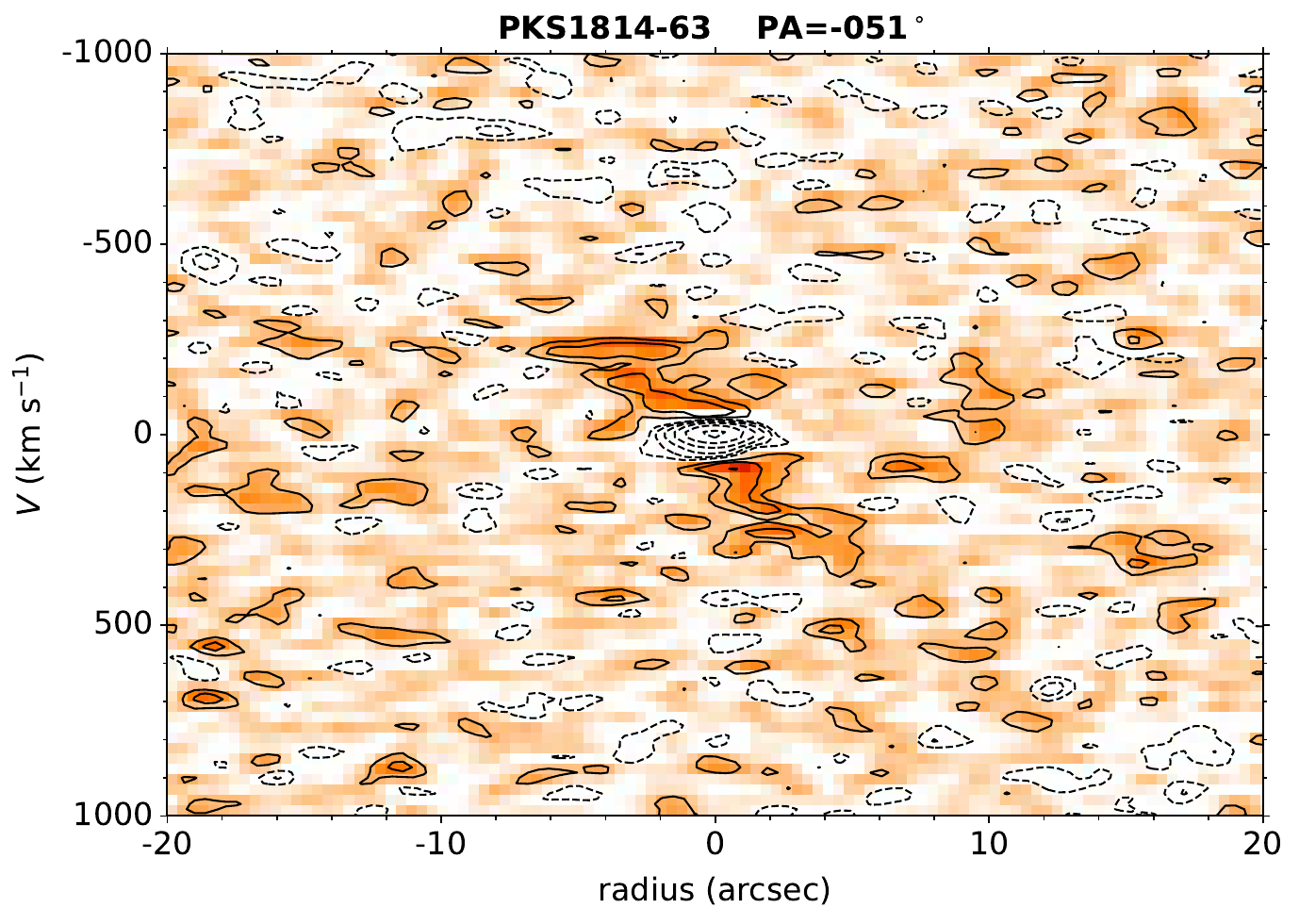} \\

\caption{Velocity fields of the CO detected targets (left) and position velocity plots along the major axis of the emission (right). The cross indicates the position of the AGN. The blue arrow indicates the direction of the radio jets.  The red arrow indicate the position angle along which the position-velocity plot was made. Contour levels in the PV plots are --2, 2, 4, 6,.... $\sigma$.}
\label{fig:VeloField}
\end{figure*}

\begin{figure*}
   \centering
\includegraphics[height=5.7cm]{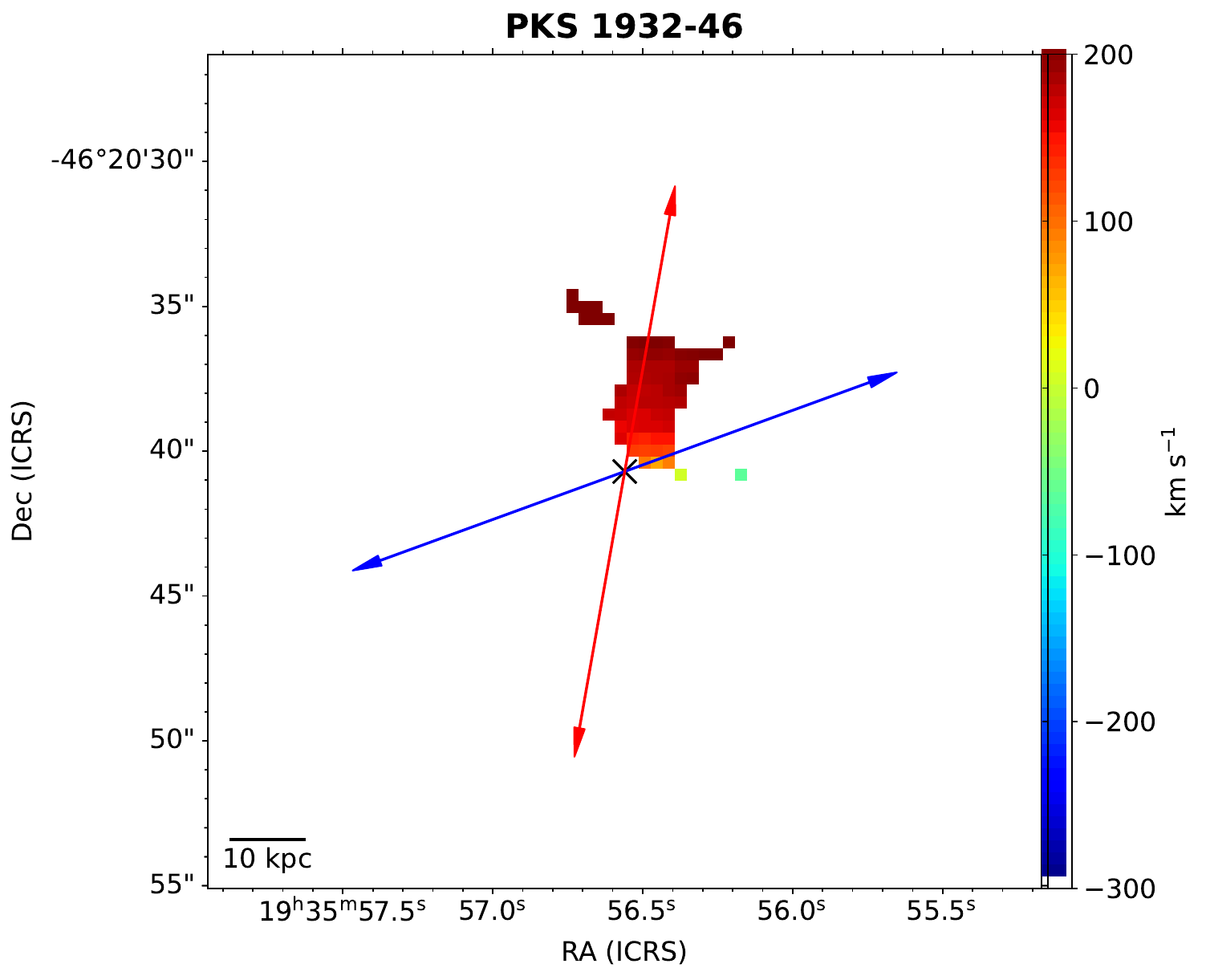}
\includegraphics[height=5.7cm]{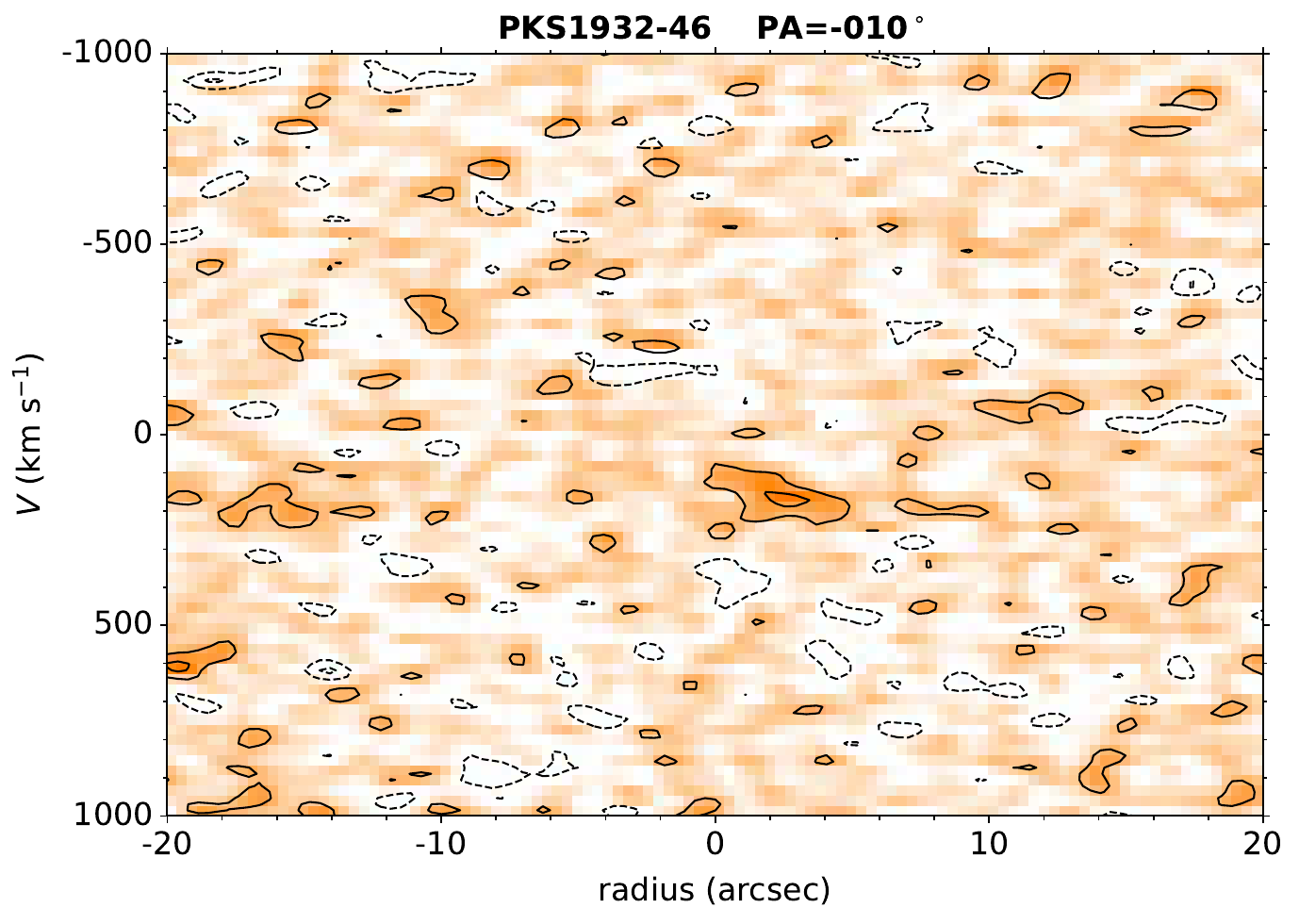} \\

\includegraphics[height=5.7cm]{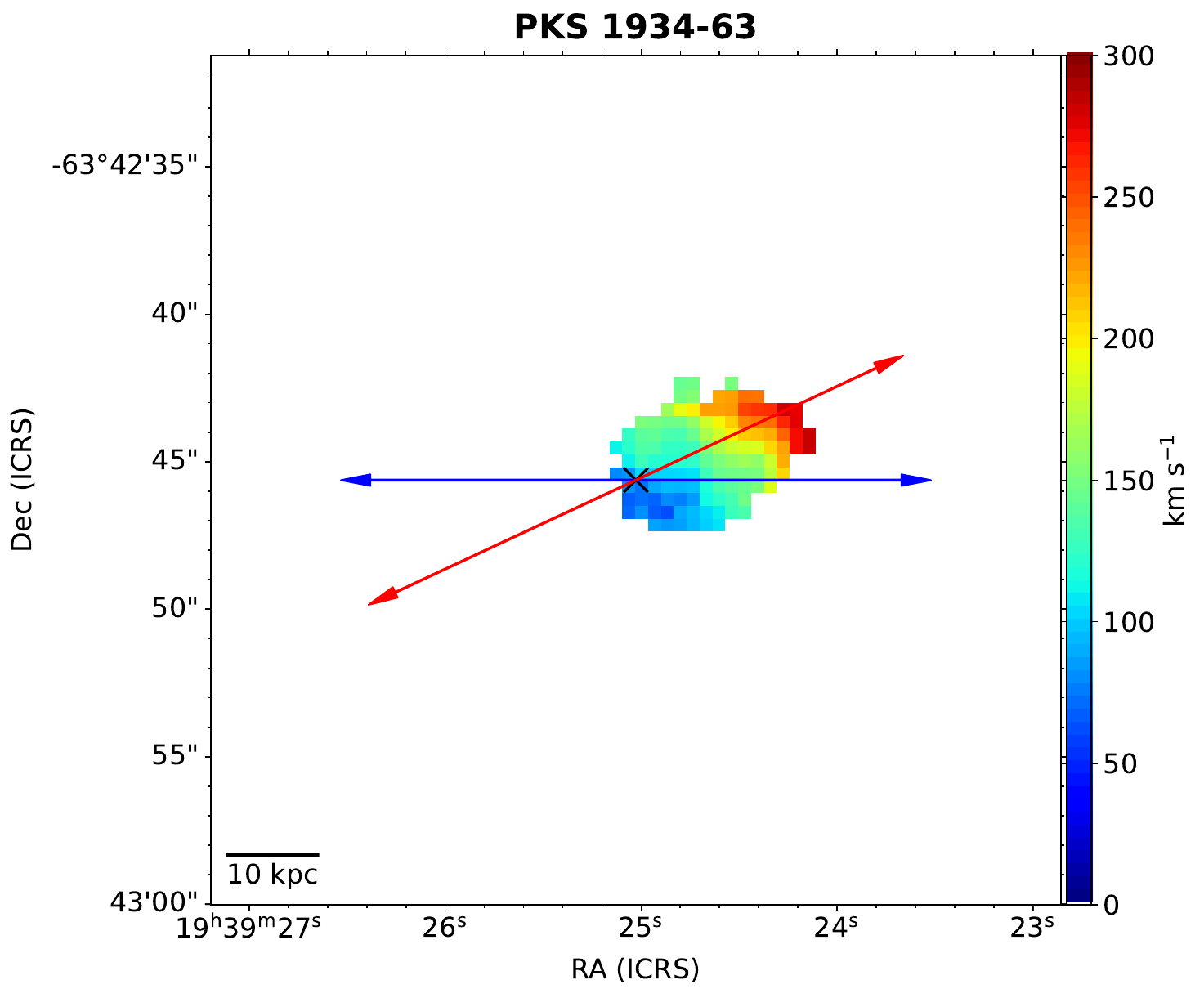}
\includegraphics[height=5.7cm]{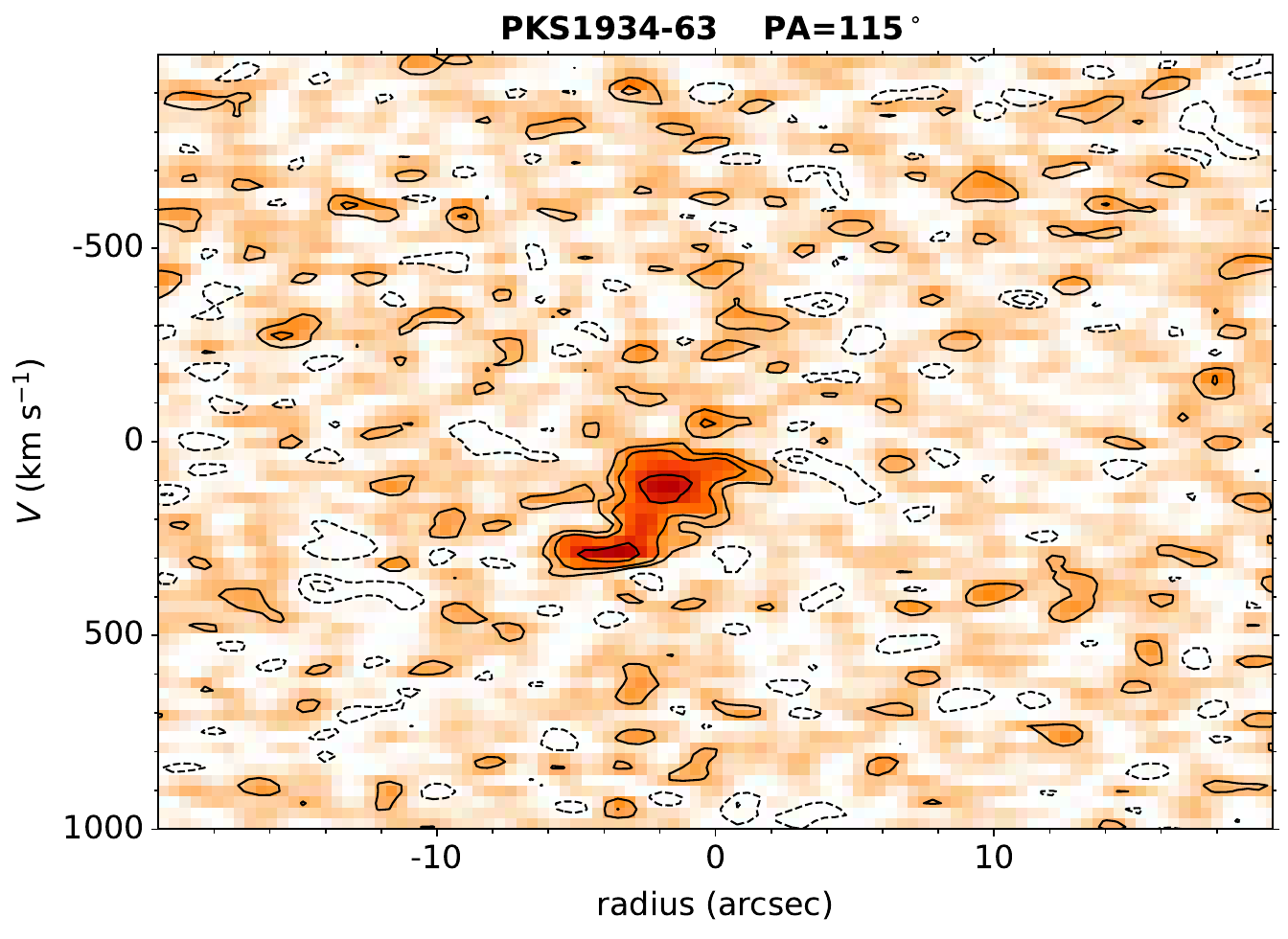} \\

\includegraphics[height=5.7cm]{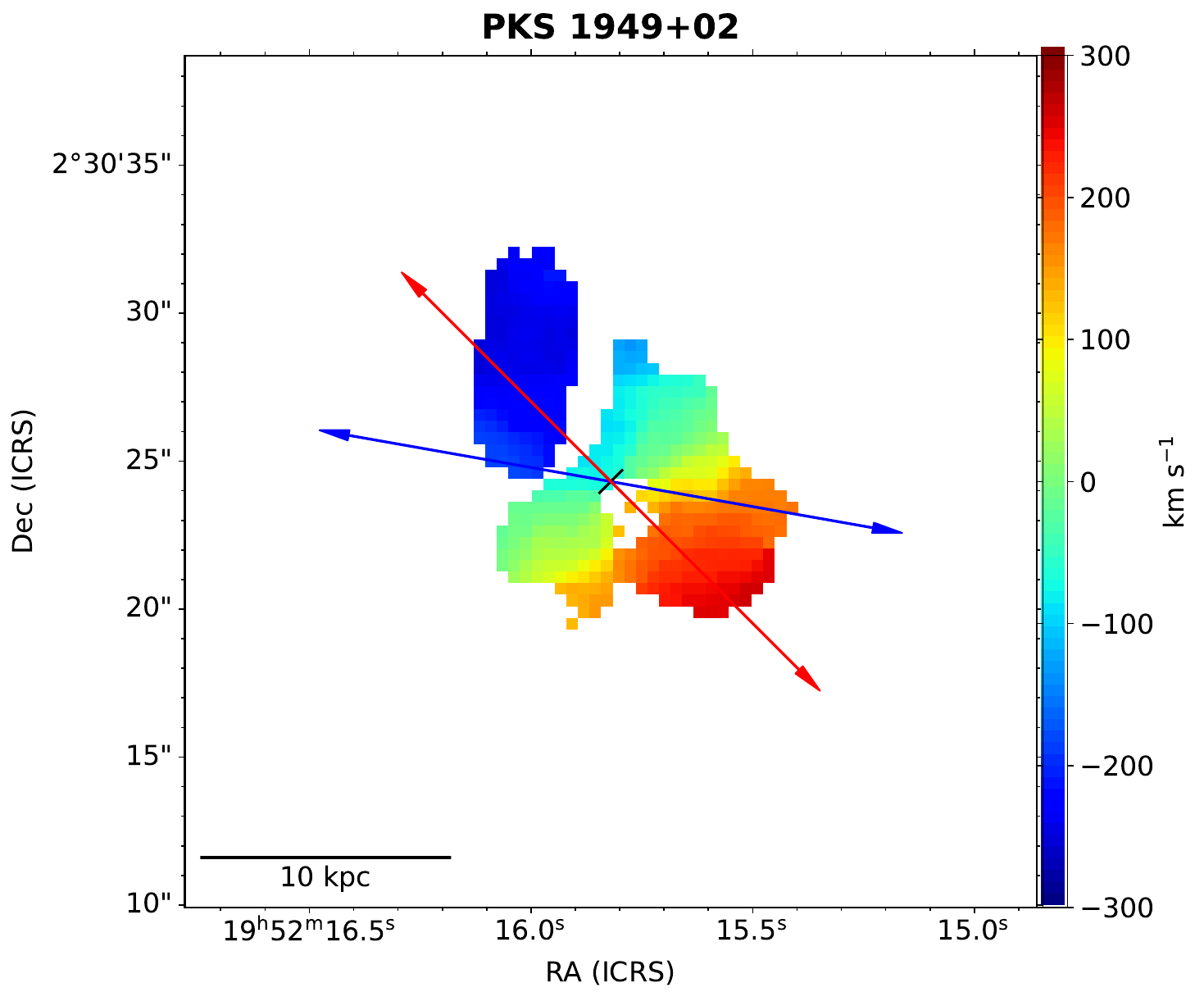}
\includegraphics[height=5.7cm]{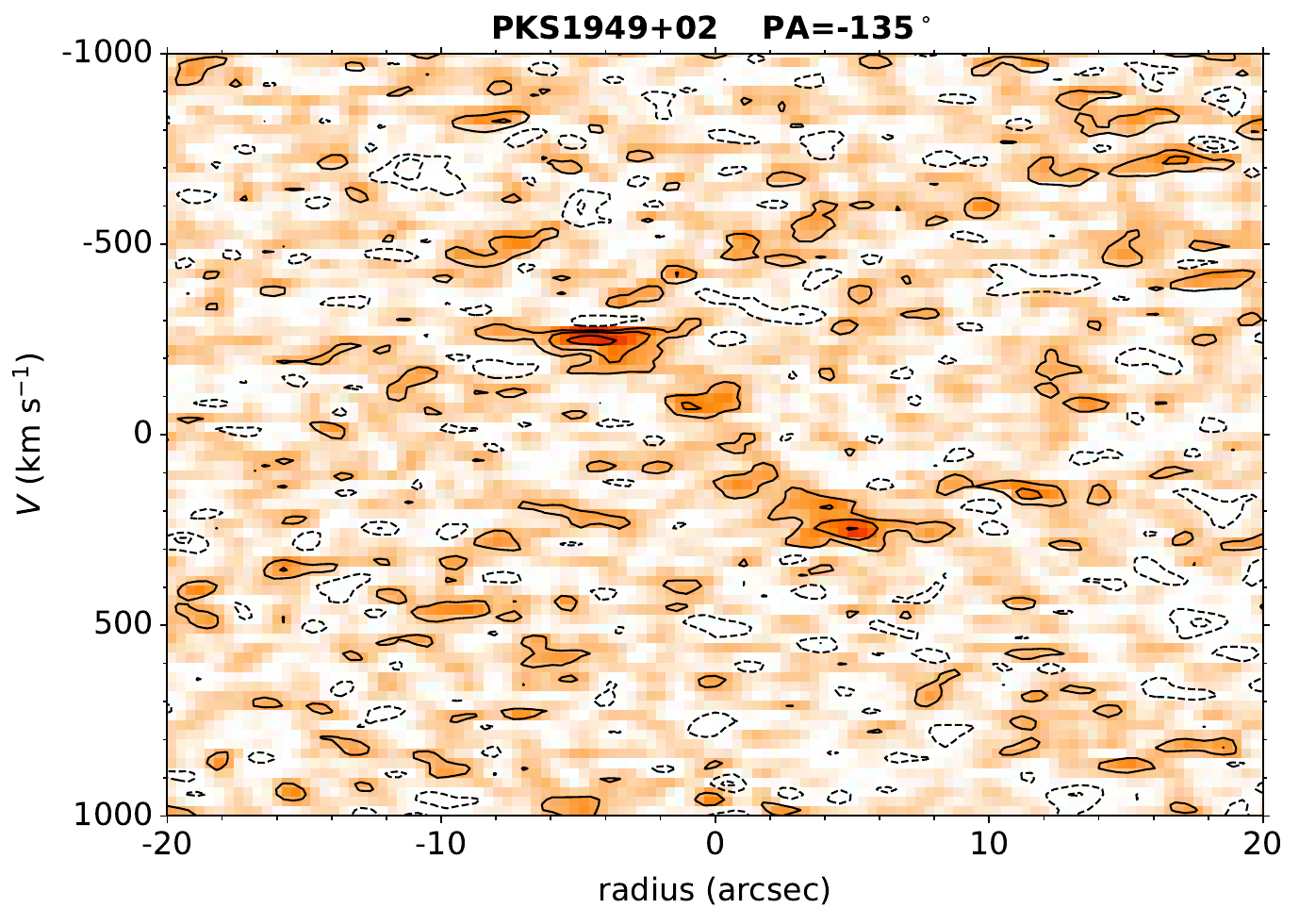} \\

\includegraphics[height=5.7cm]{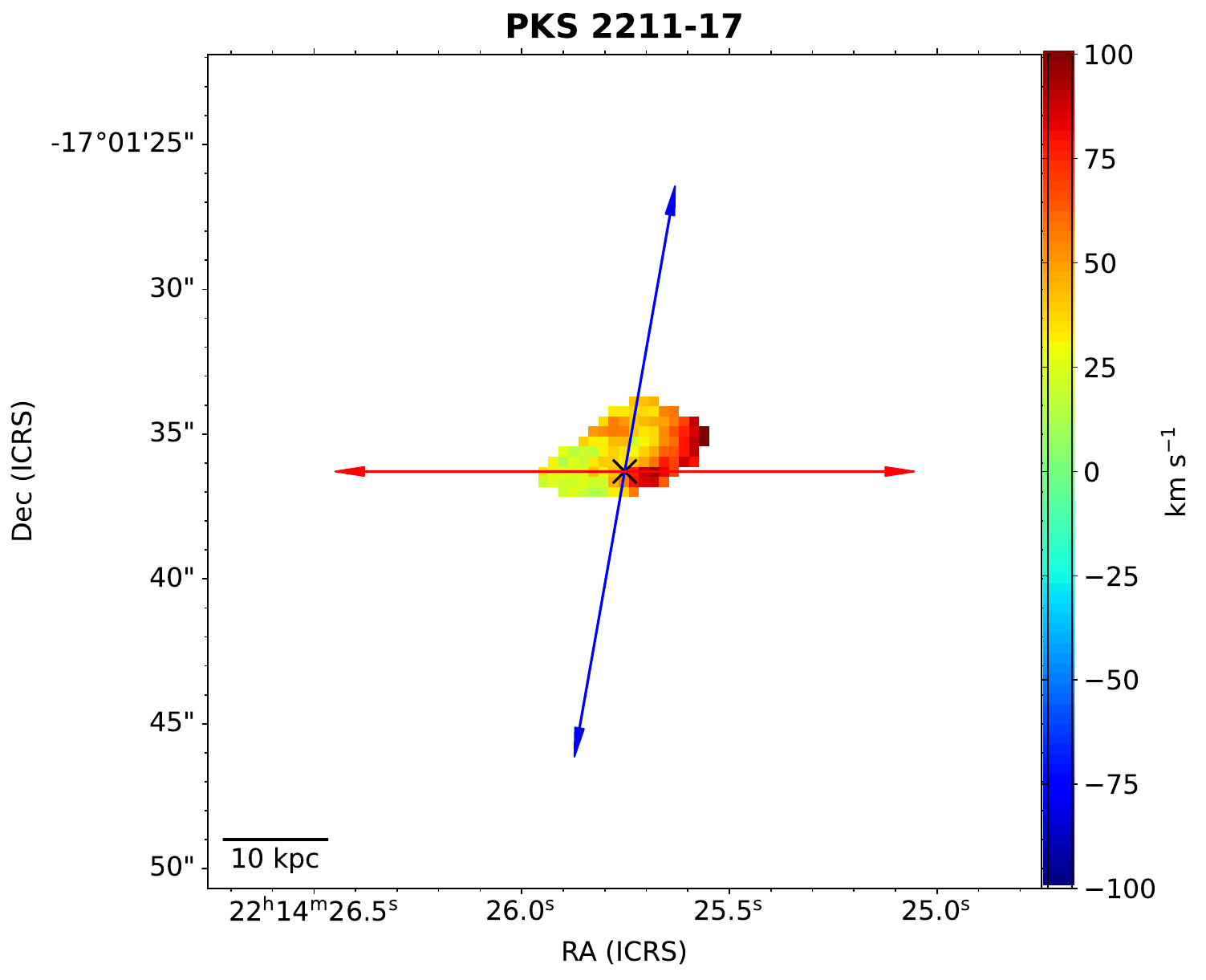}
\includegraphics[height=5.7cm]{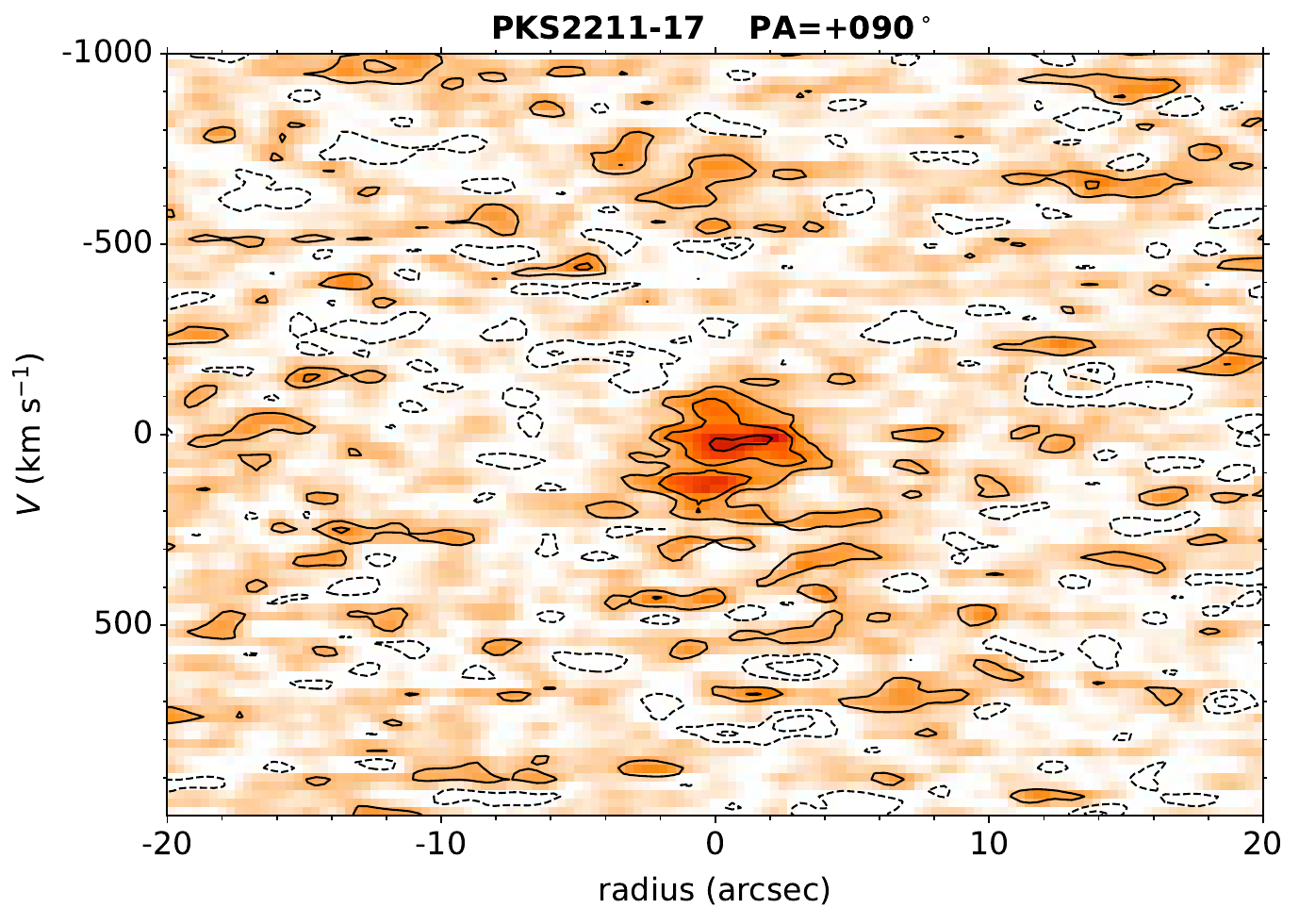} \\

\contcaption{ }
\end{figure*}

\begin{figure*}
   \centering
\includegraphics[height=5.9cm]{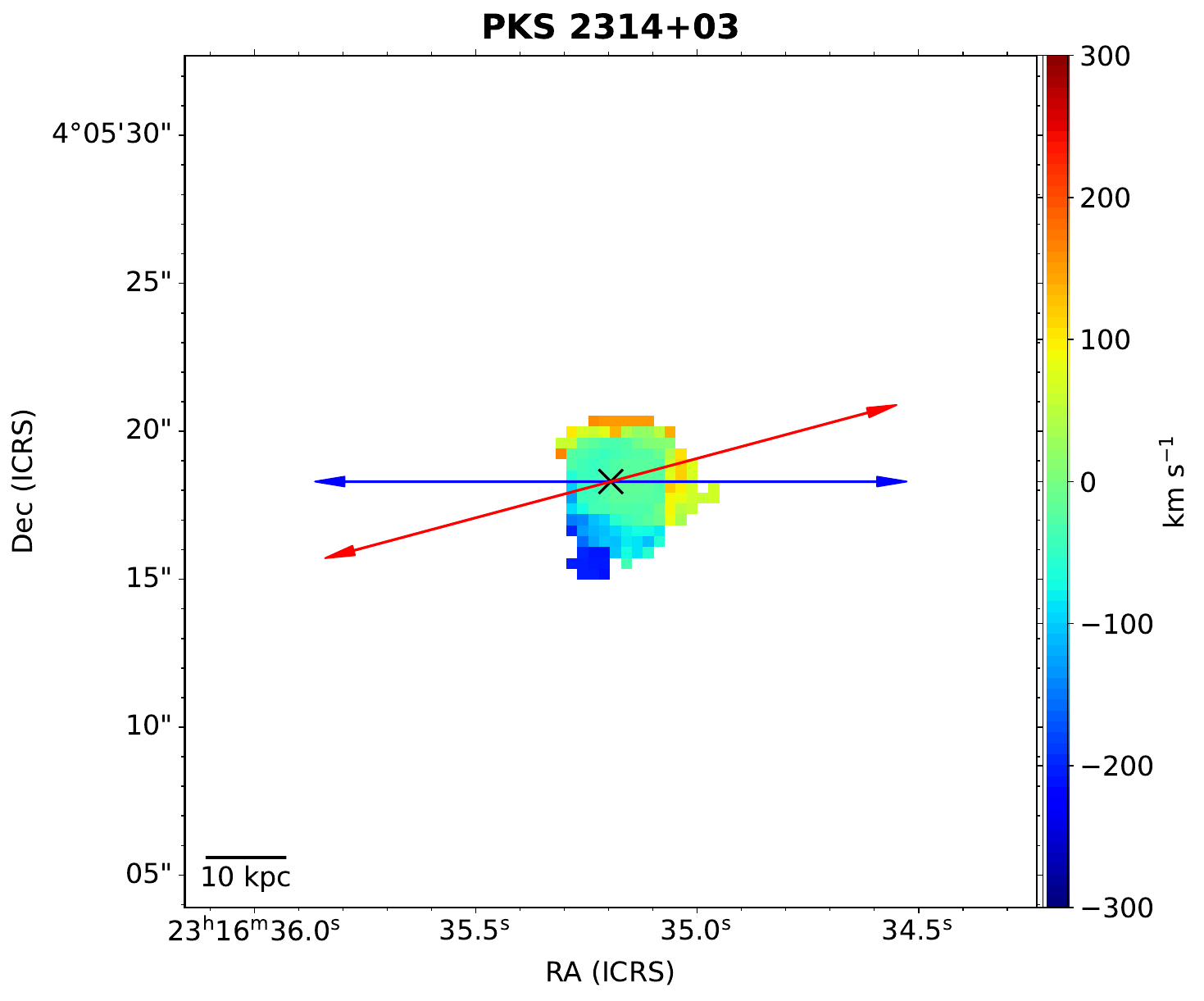}  
\includegraphics[height=5.9cm]{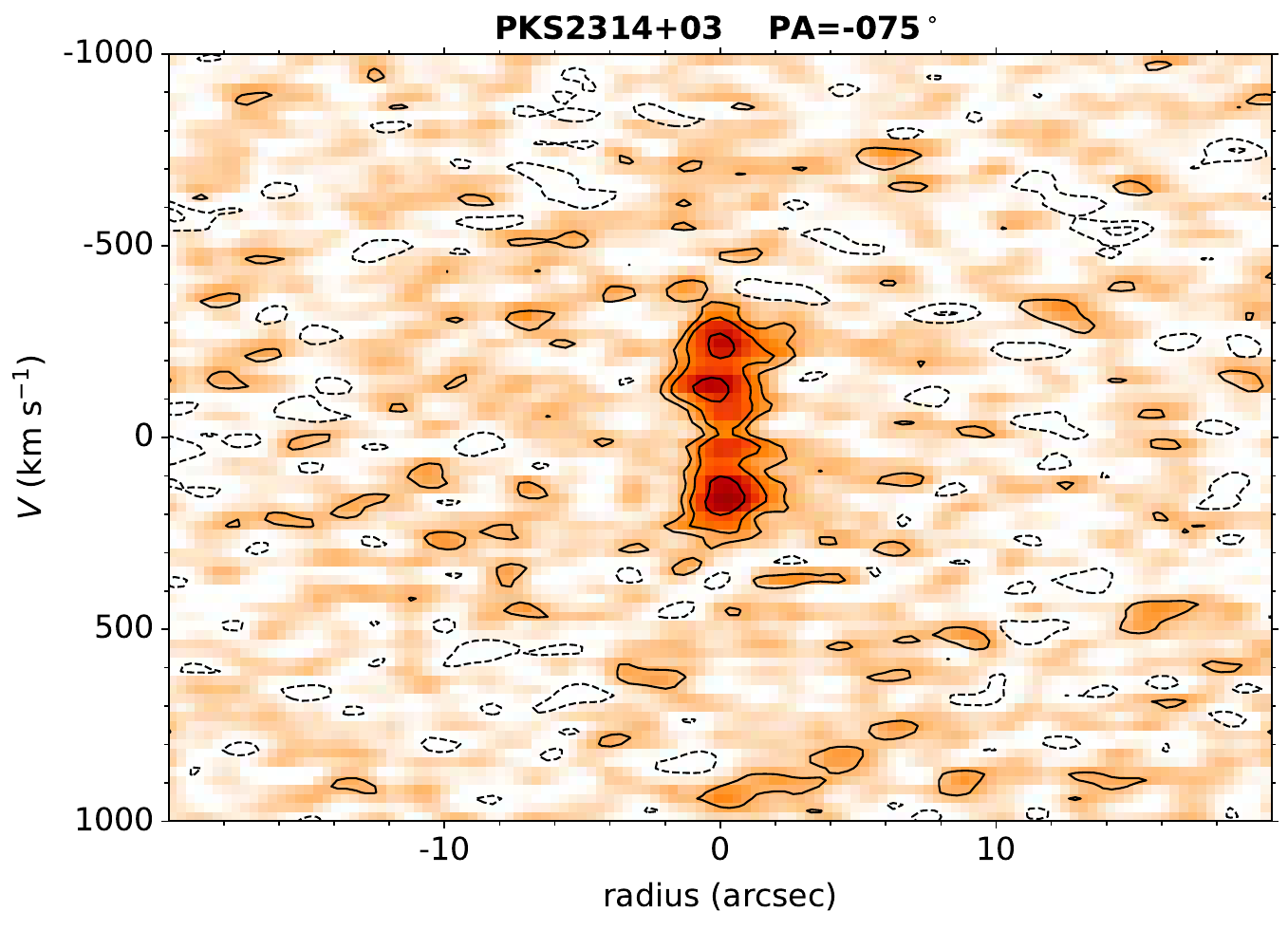} \\
\contcaption{}
\end{figure*}


\begin{figure*}
   \centering
\includegraphics[width=4.2cm]{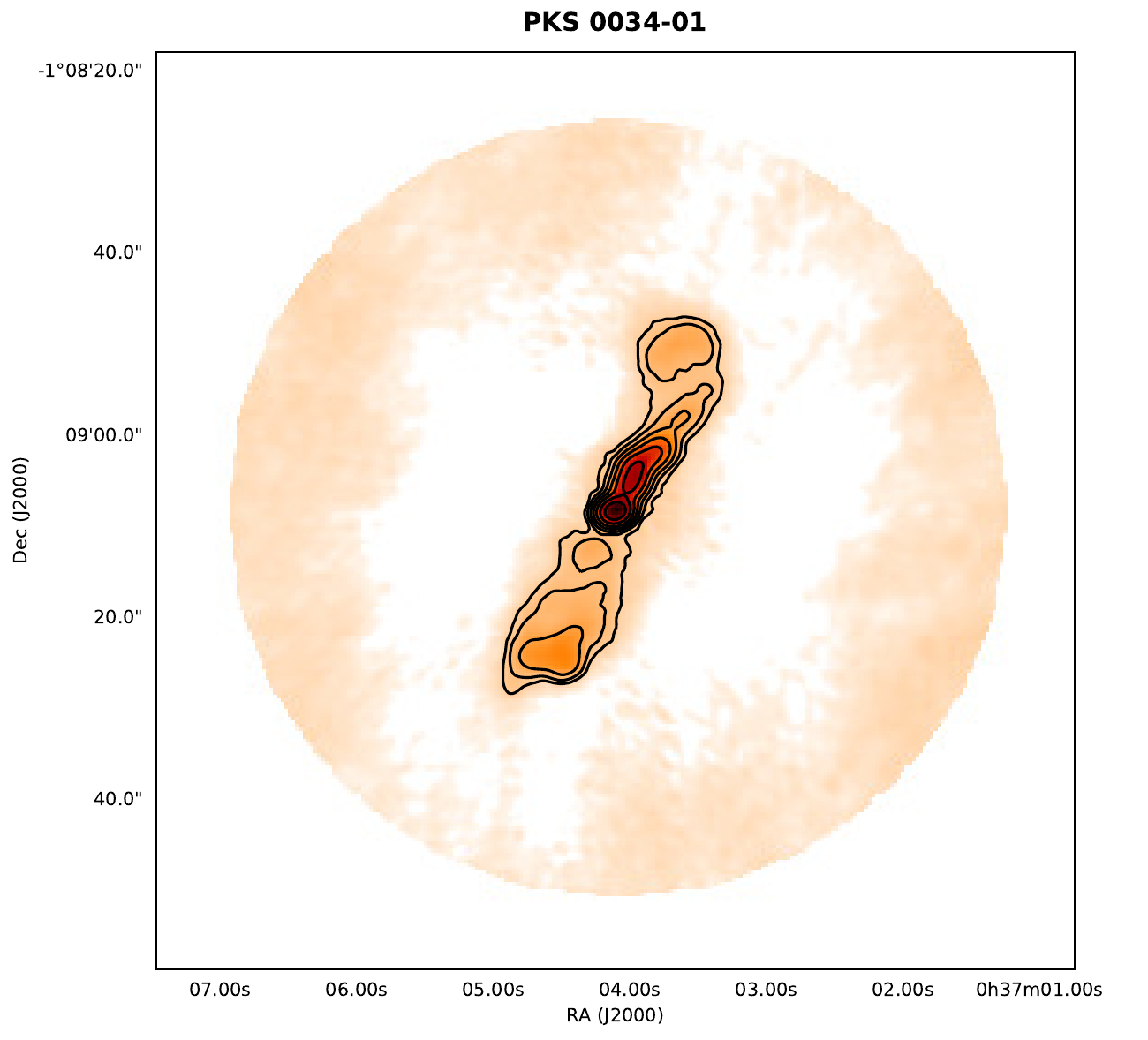} 
\includegraphics[width=4.2cm]{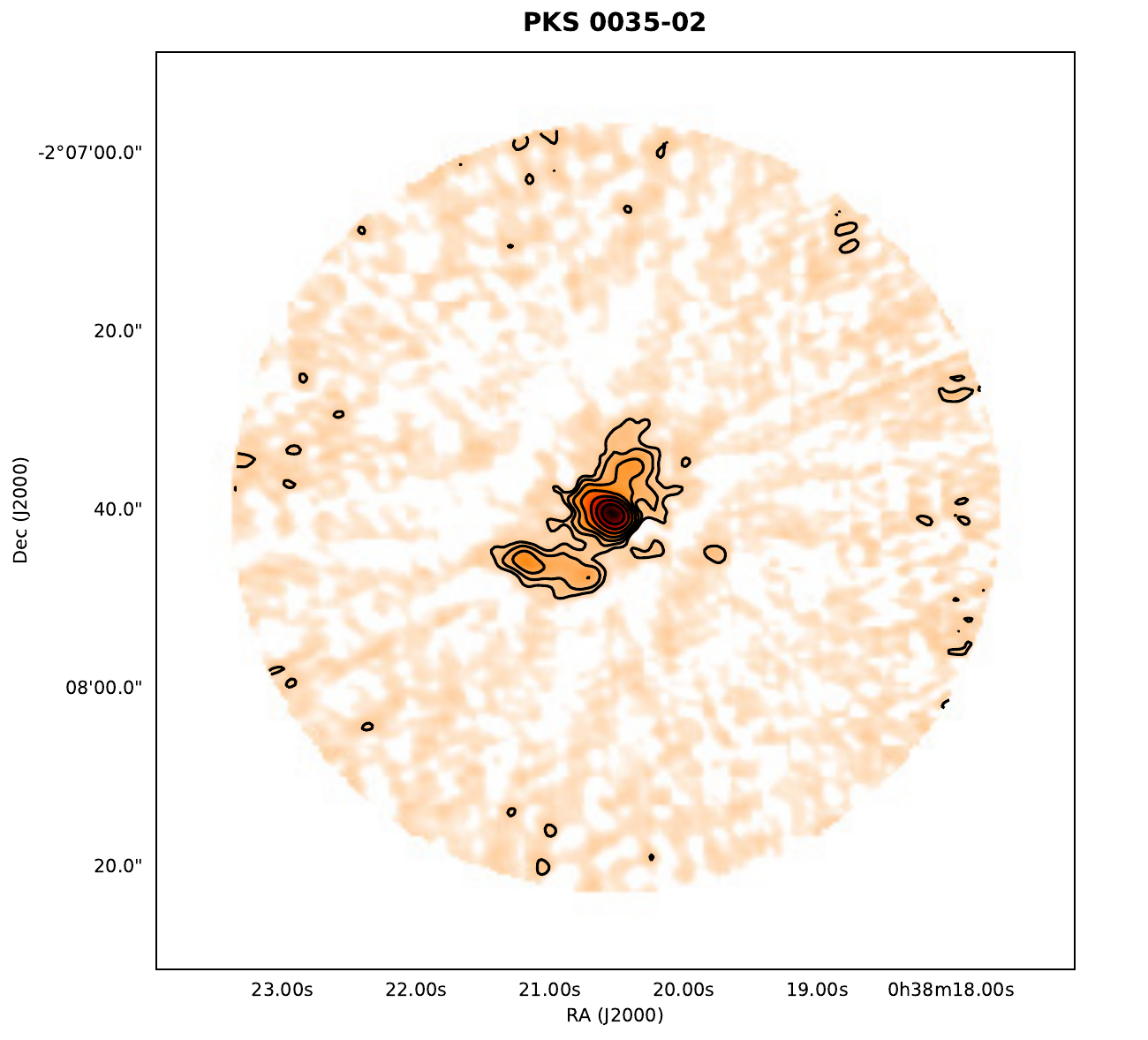} 
\includegraphics[width=4.2cm]{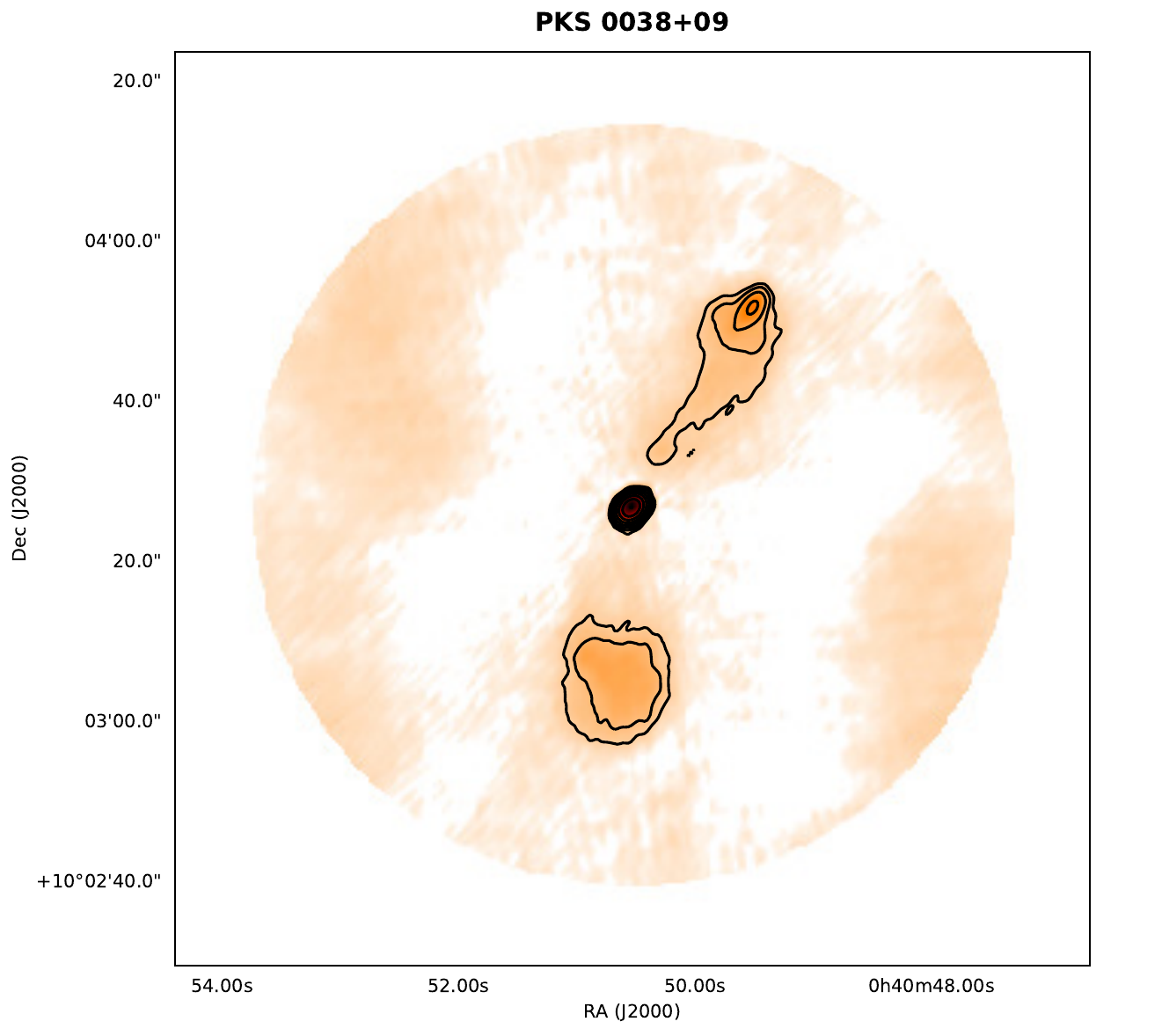} 
\includegraphics[width=4.2cm]{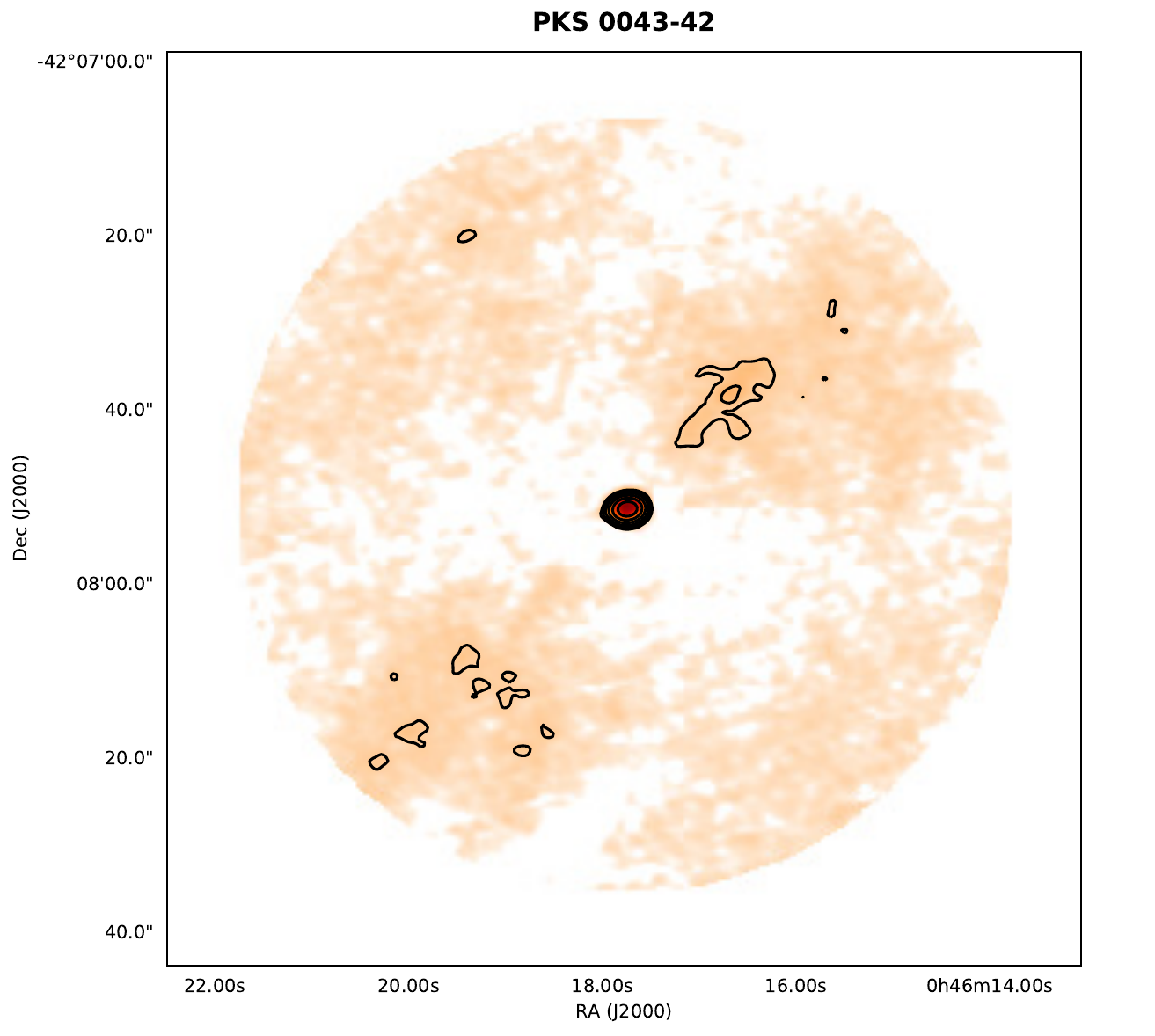} \\
\includegraphics[width=4.2cm]{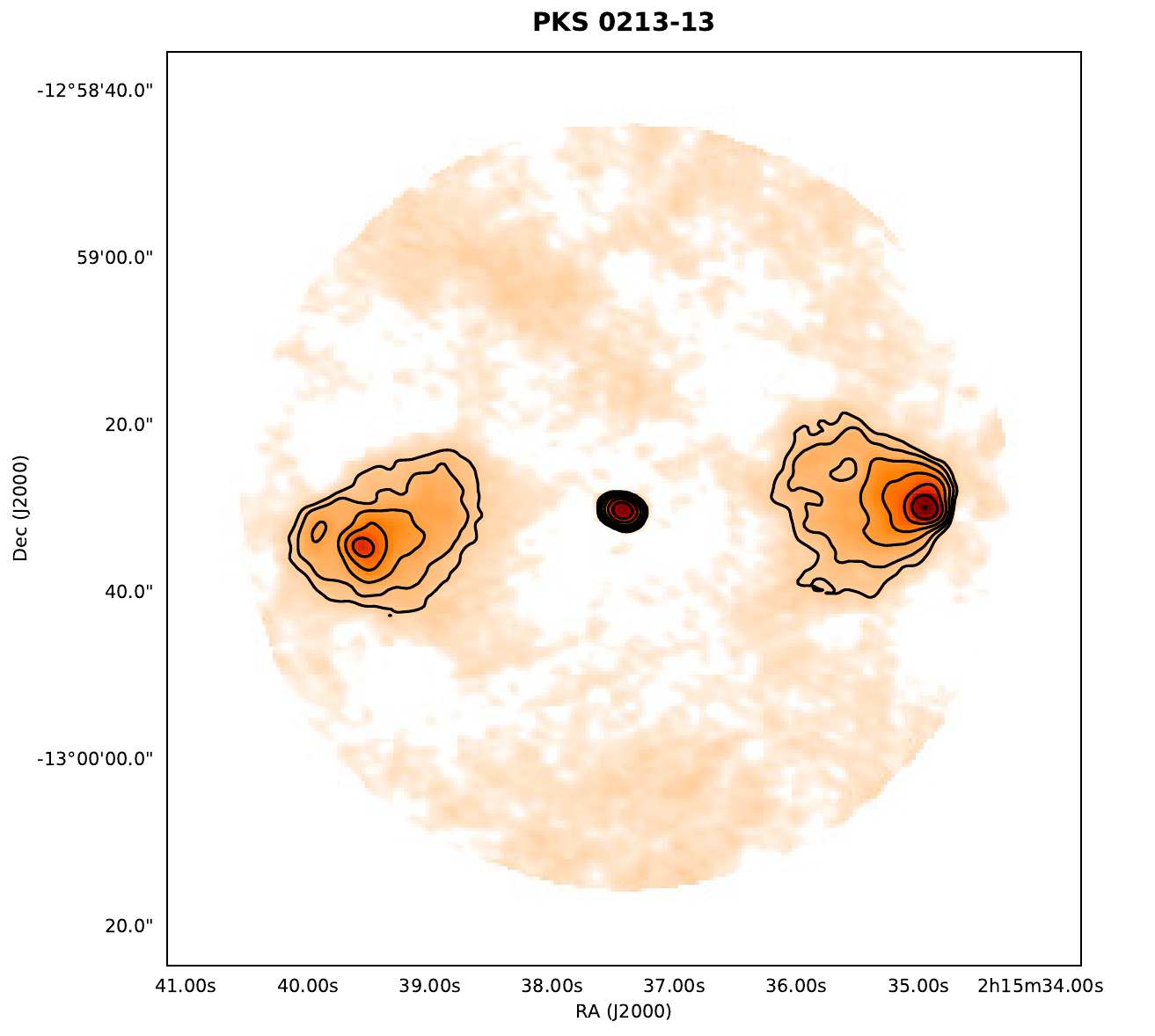} 
\includegraphics[width=4.2cm]{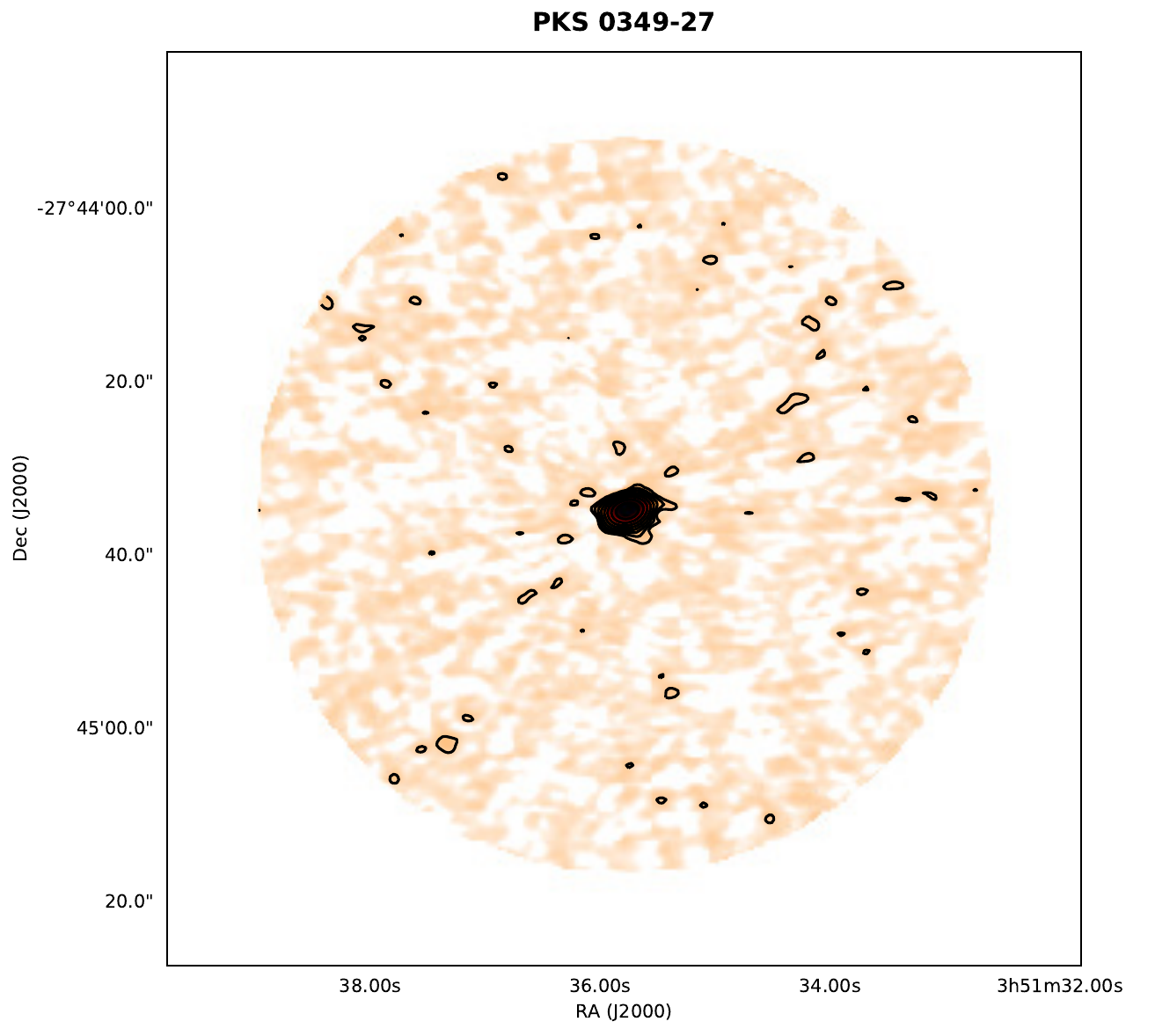} 
\includegraphics[width=4.2cm]{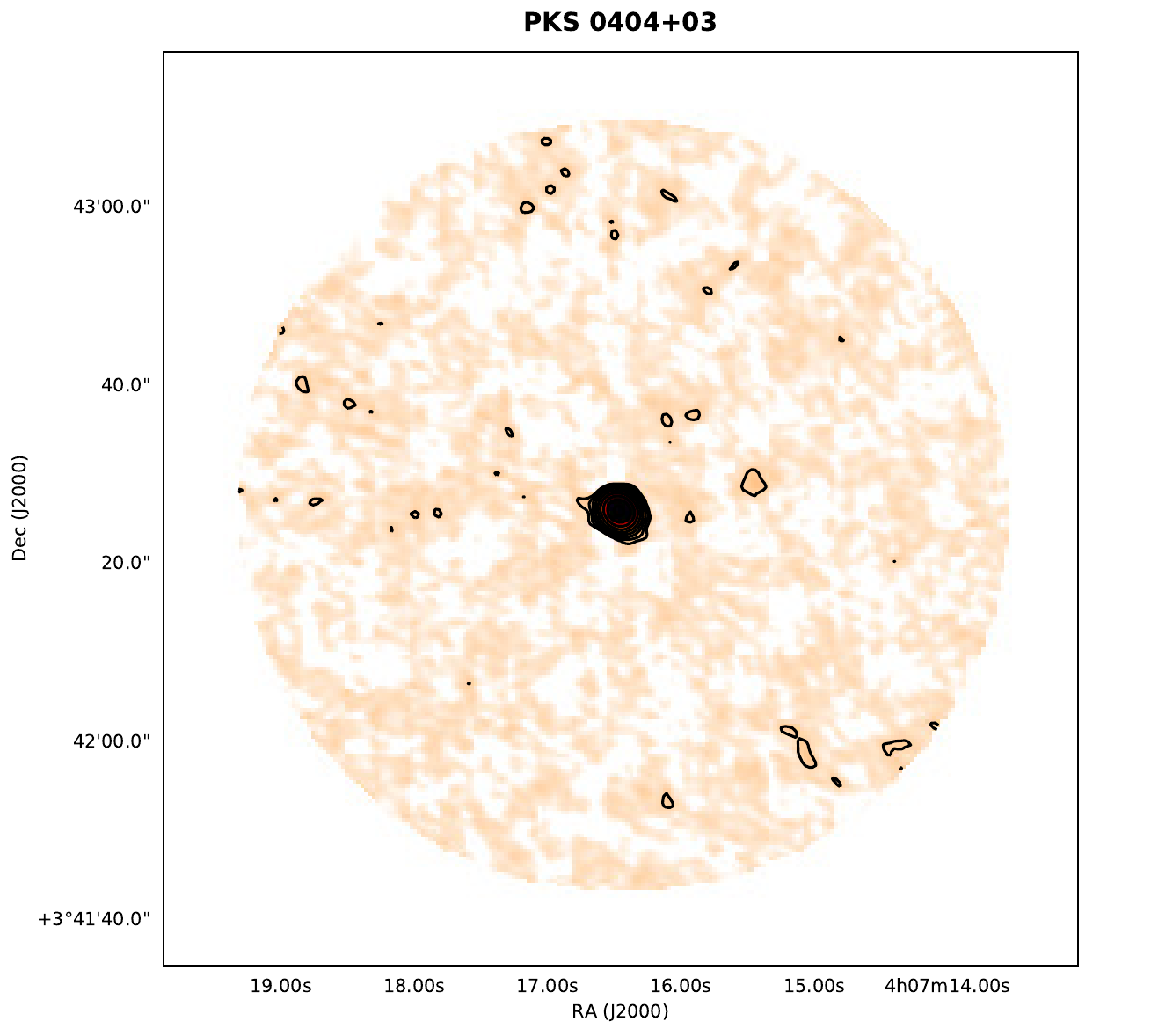} 
\includegraphics[width=4.2cm]{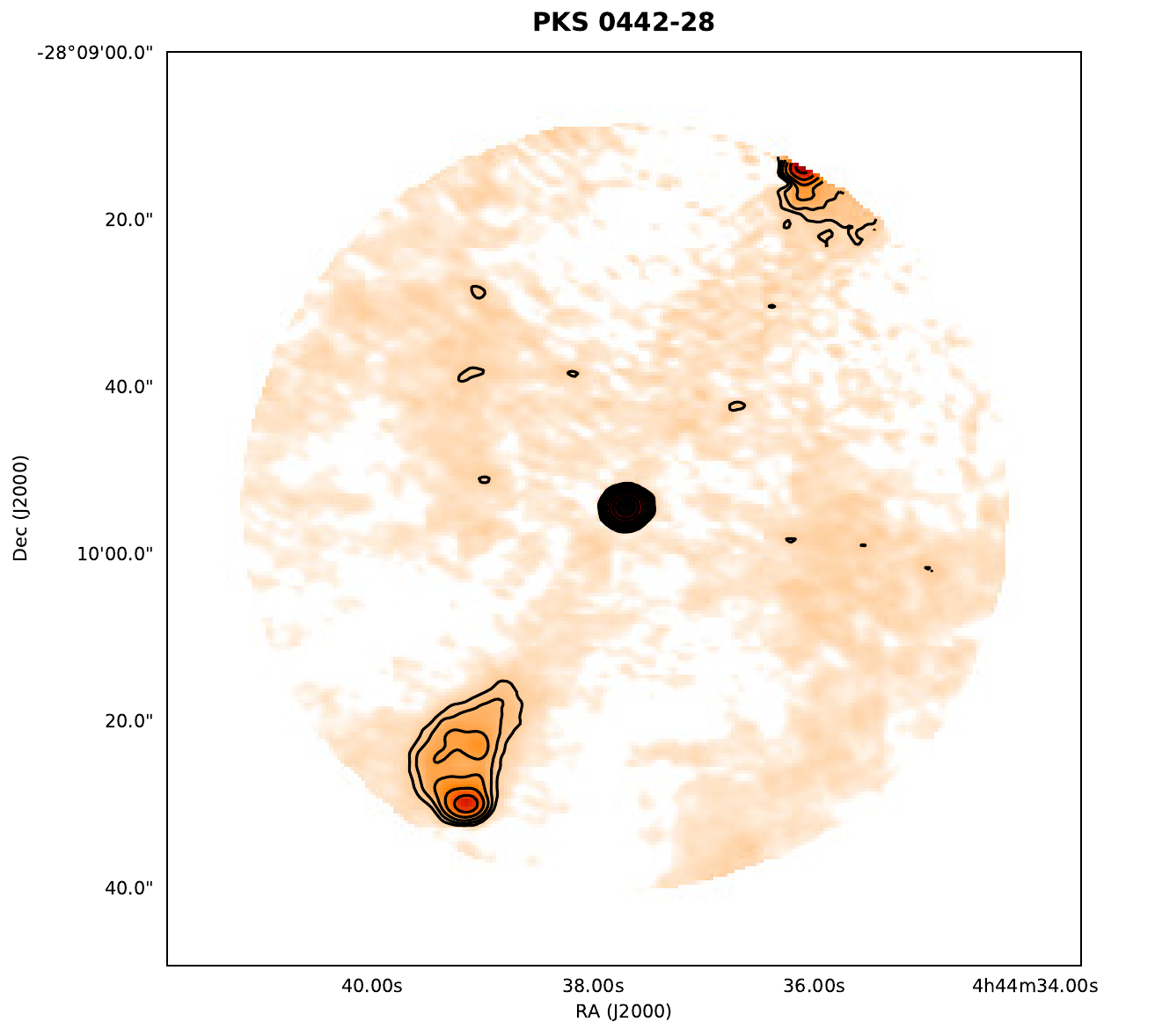} \\
\includegraphics[width=4.2cm]{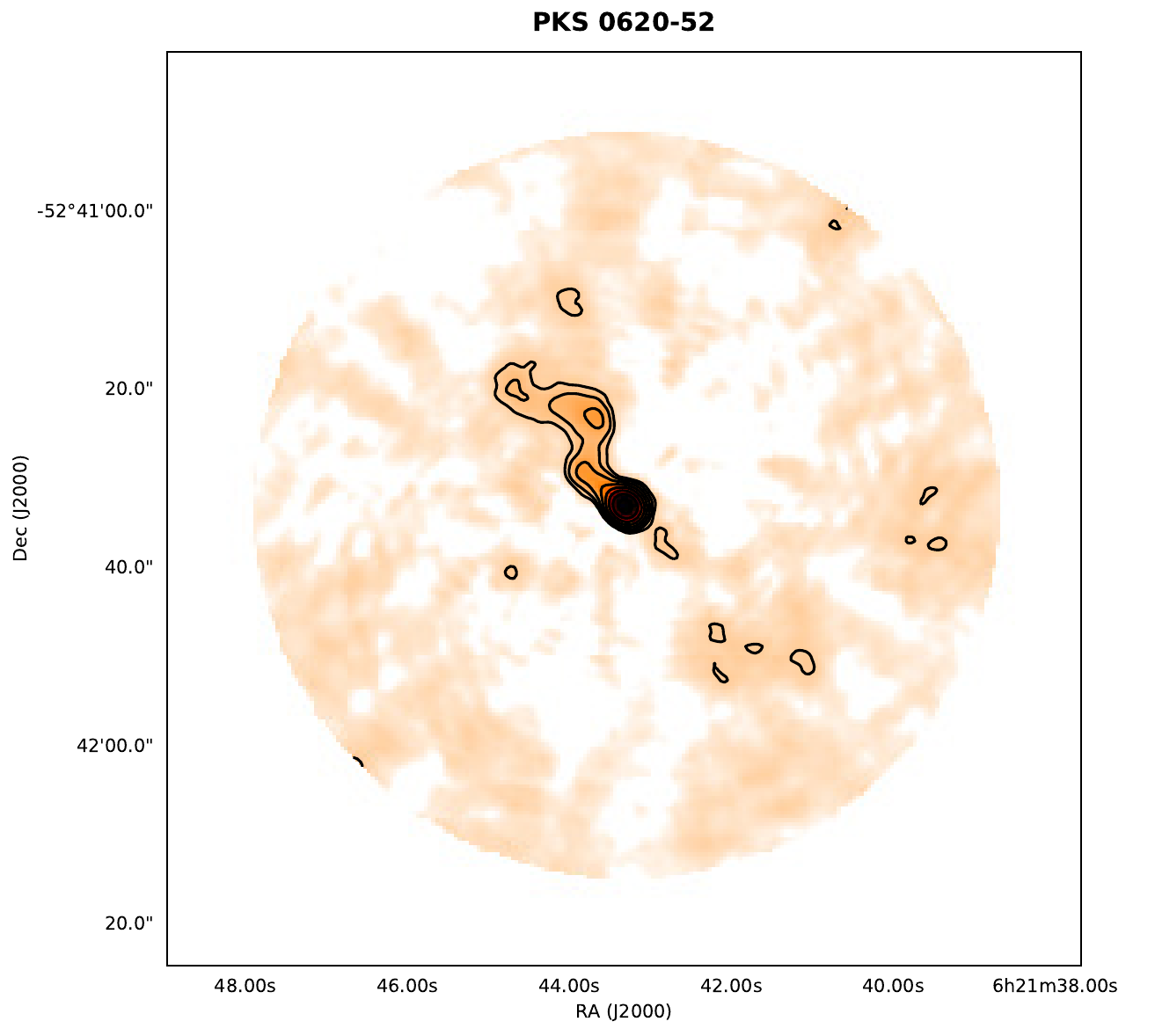} 
\includegraphics[width=4.2cm]{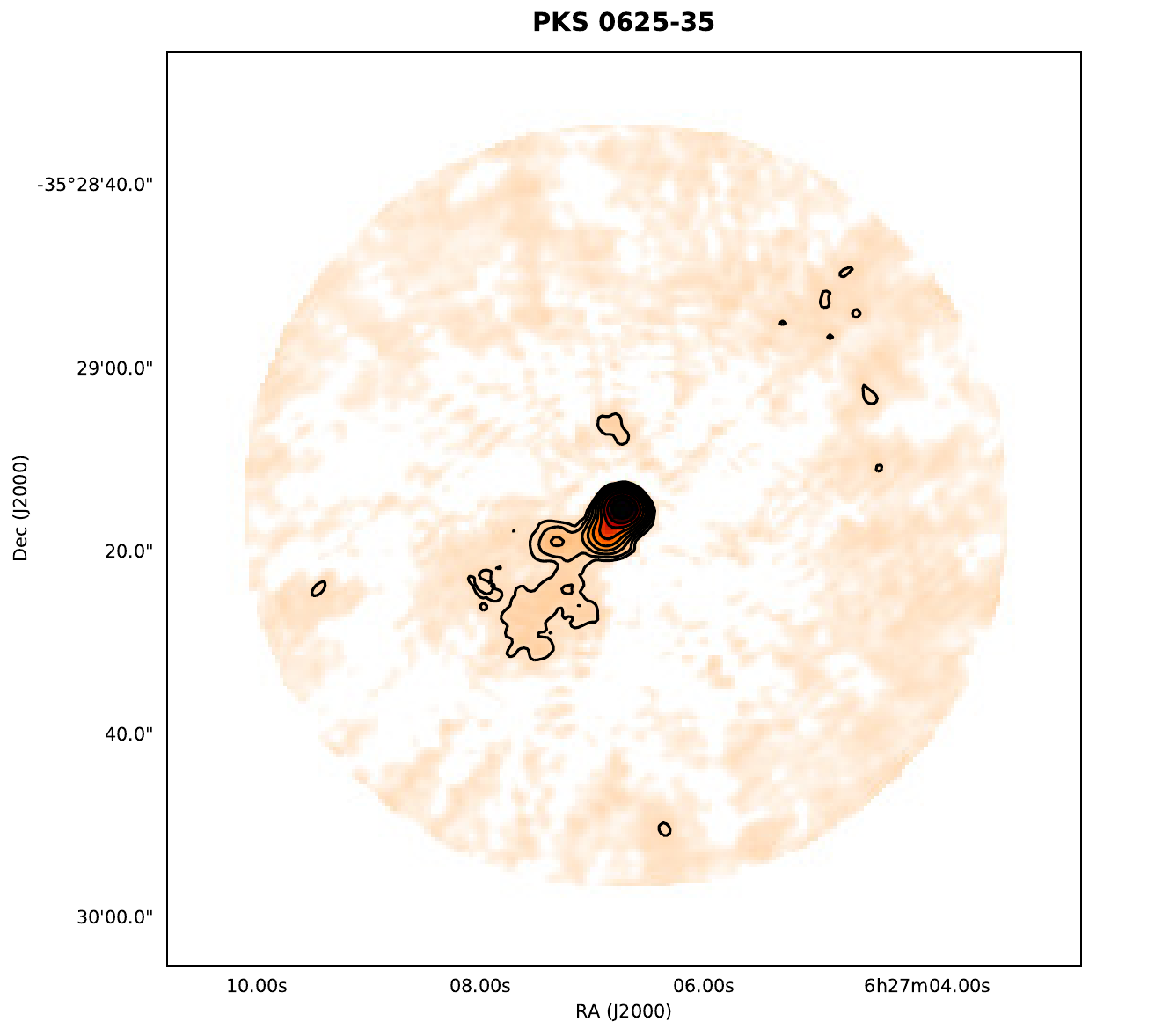} 
\includegraphics[width=4.2cm]{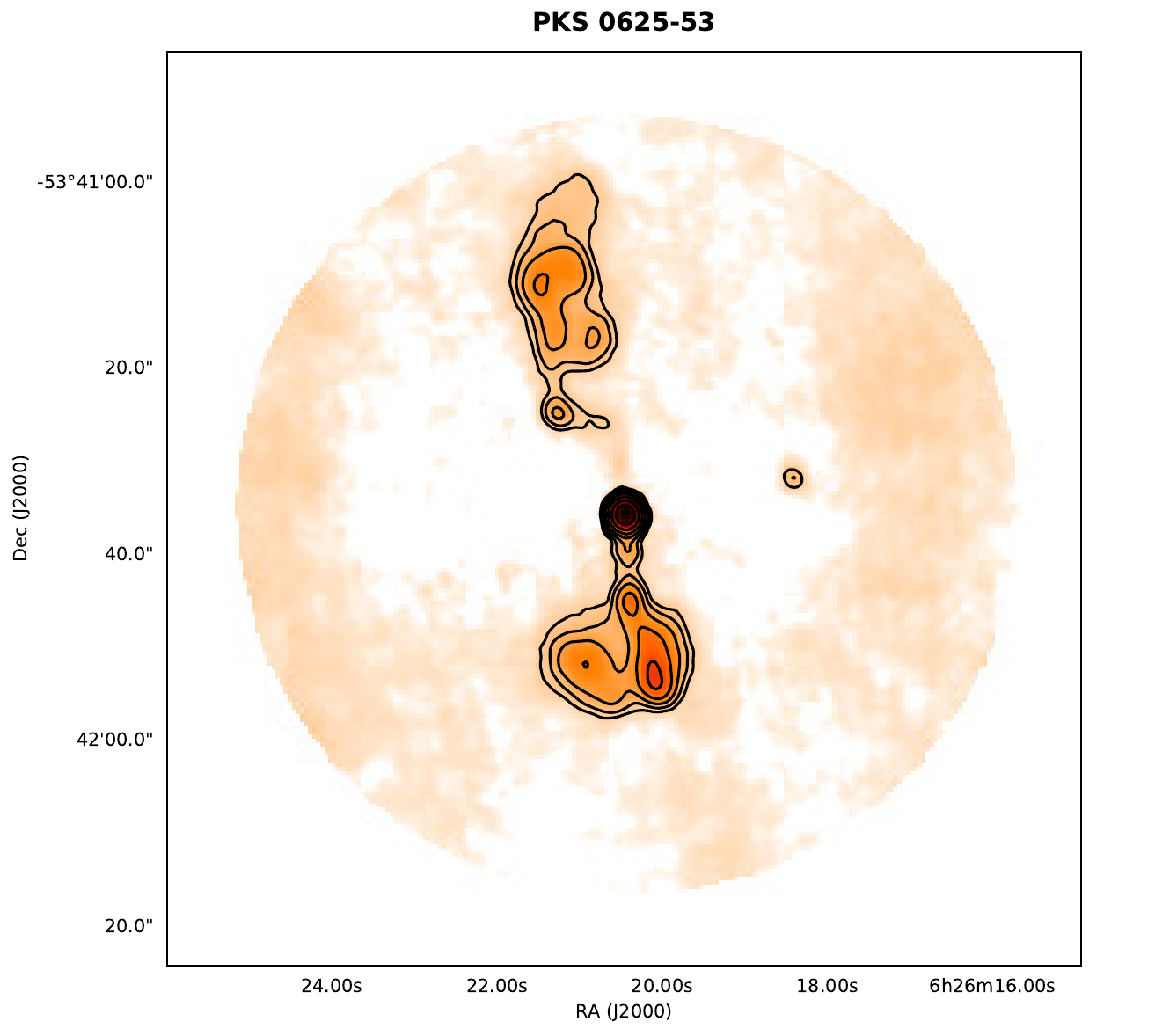} 
\includegraphics[width=4.2cm]{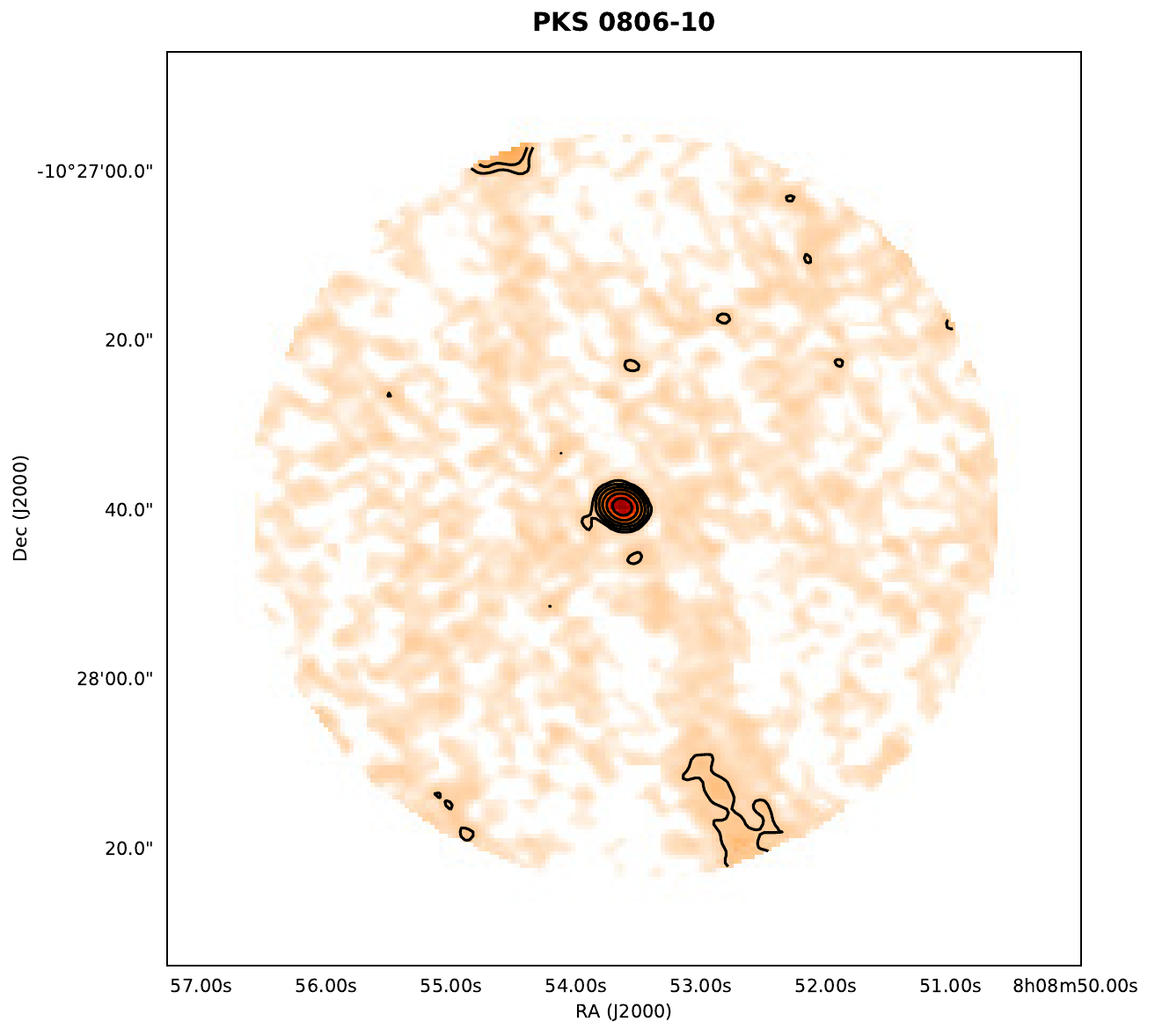} \\
\includegraphics[width=4.2cm]{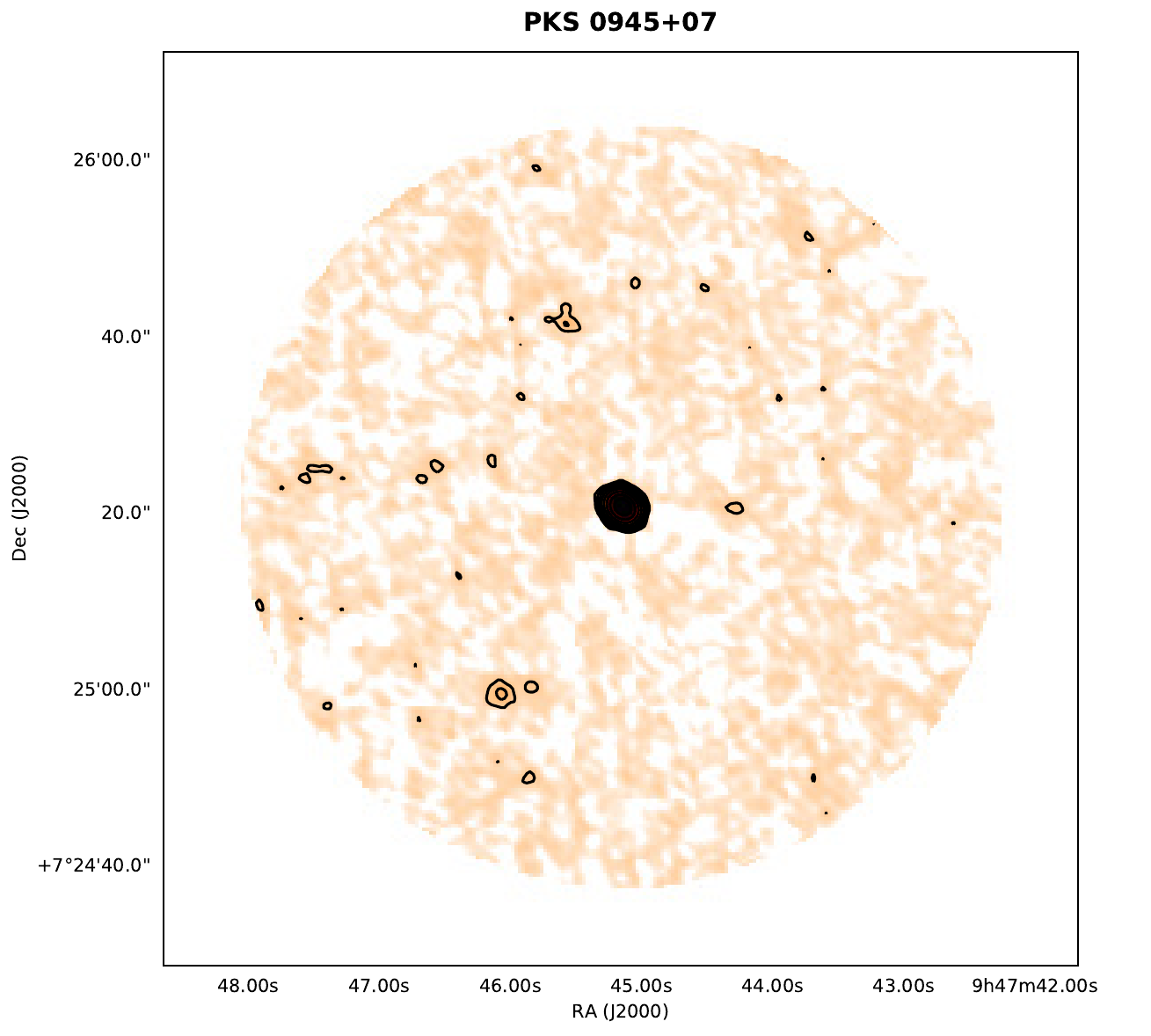} 
\includegraphics[width=4.2cm]{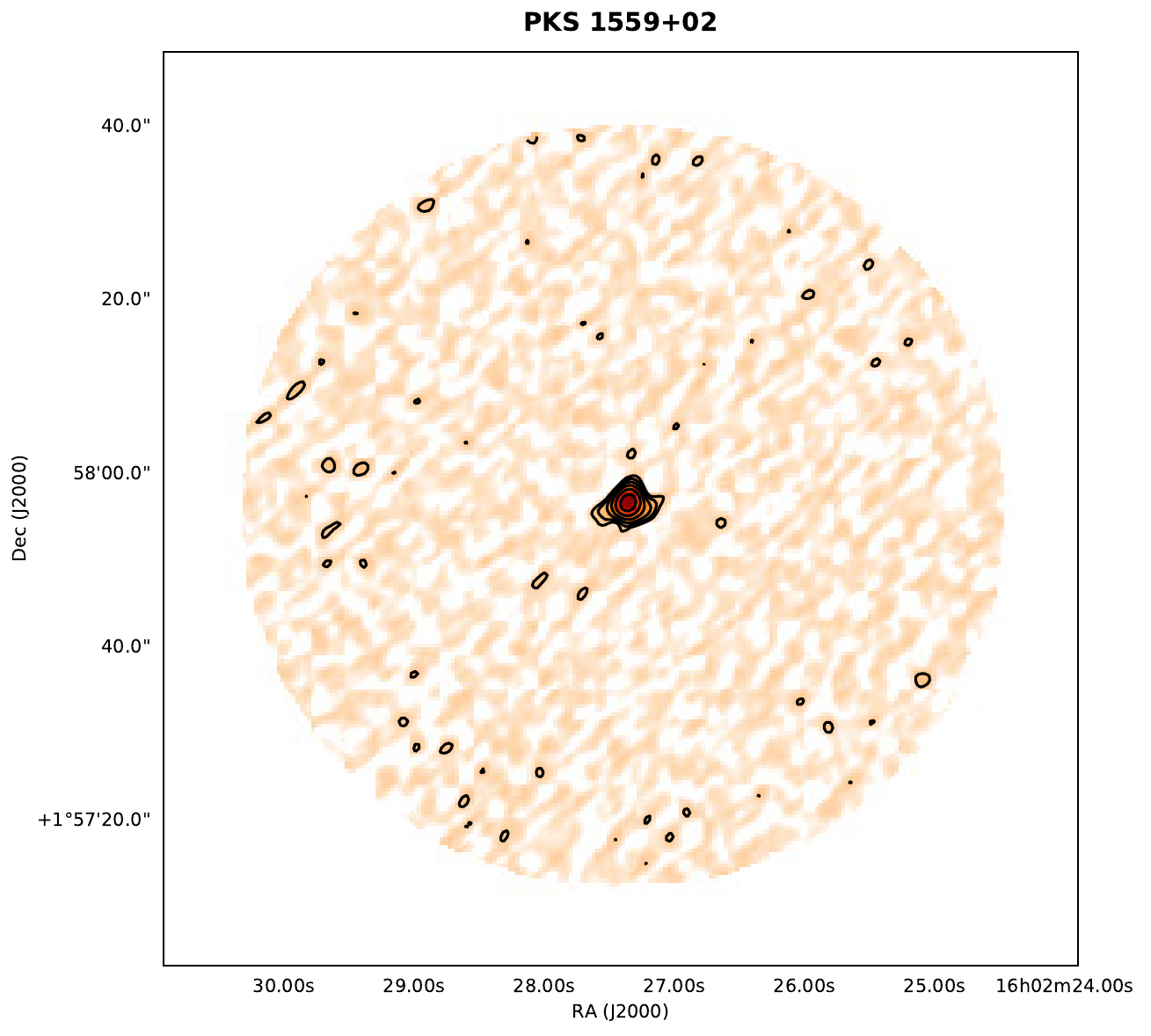} 
\includegraphics[width=4.2cm]{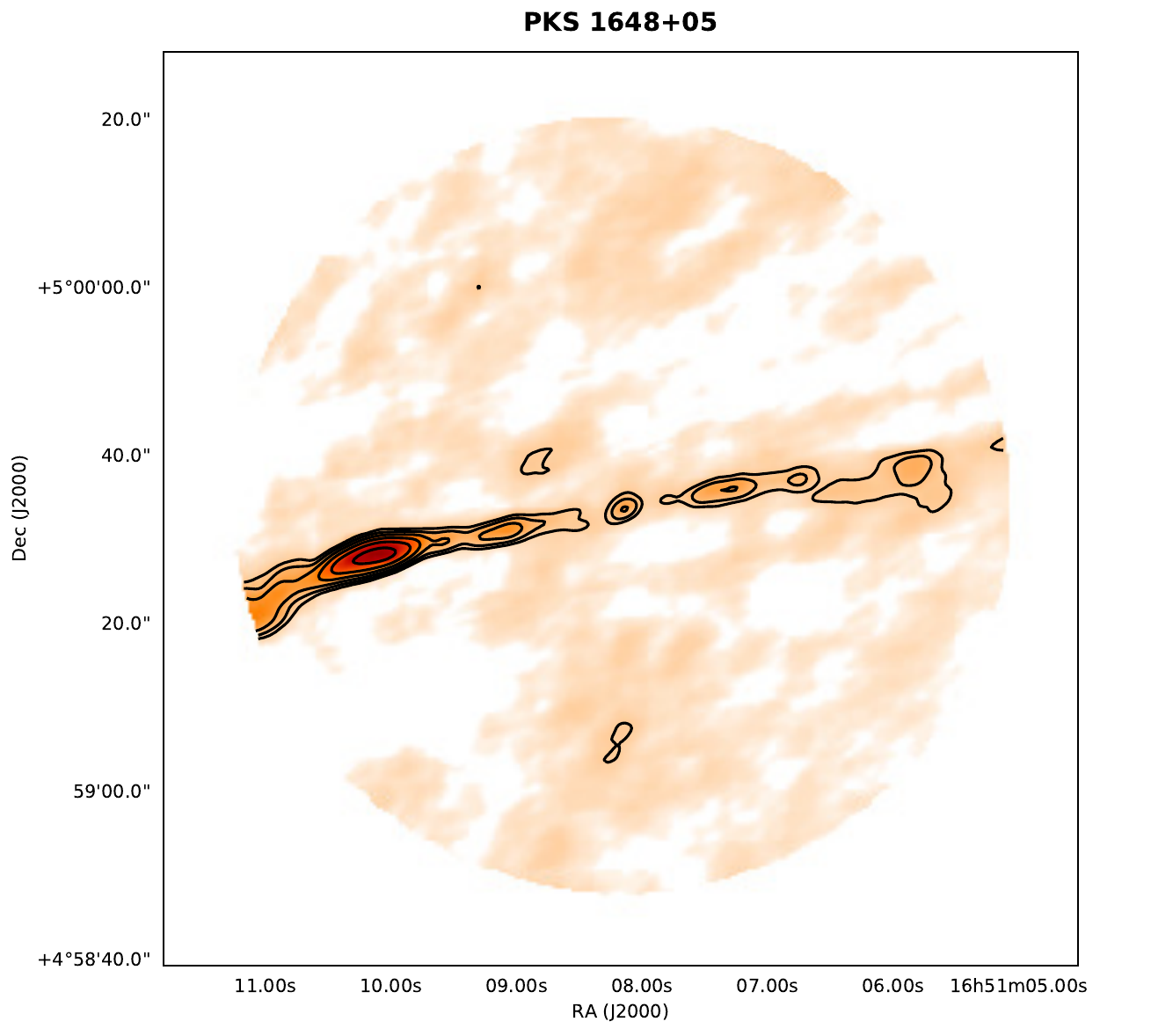} 
\includegraphics[width=4.2cm]{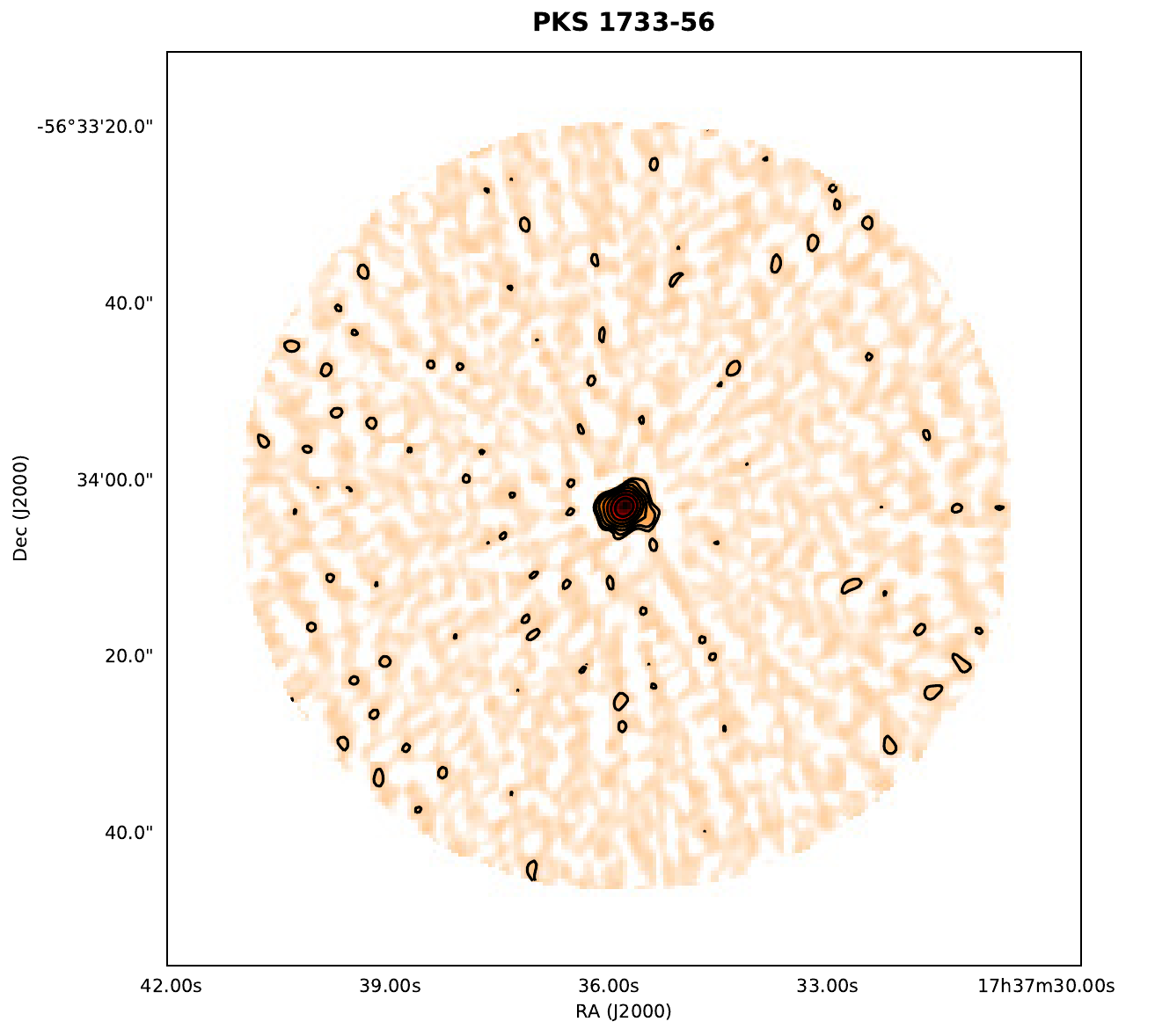} \\
\includegraphics[width=4.2cm]{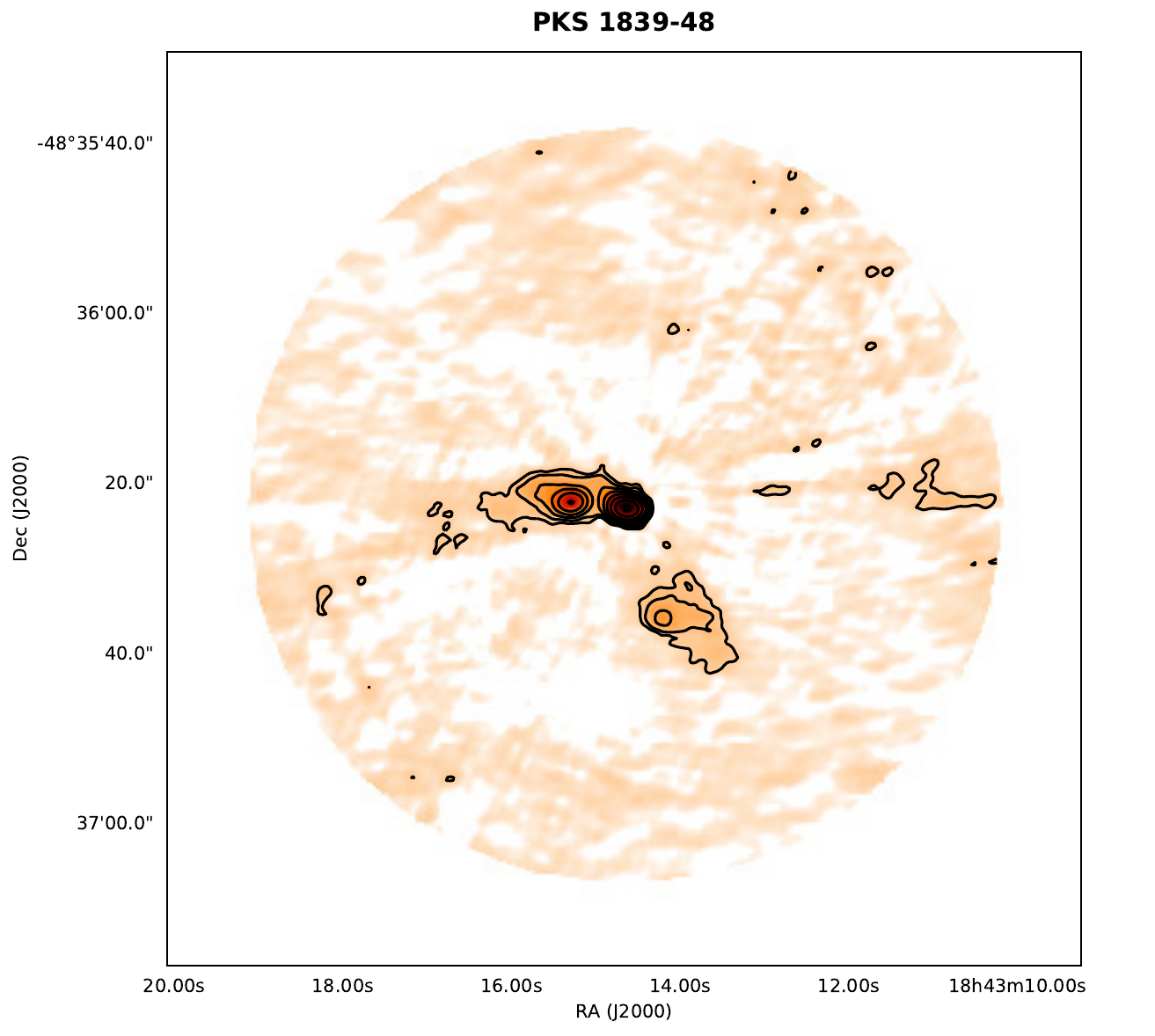} 
\includegraphics[width=4.2cm]{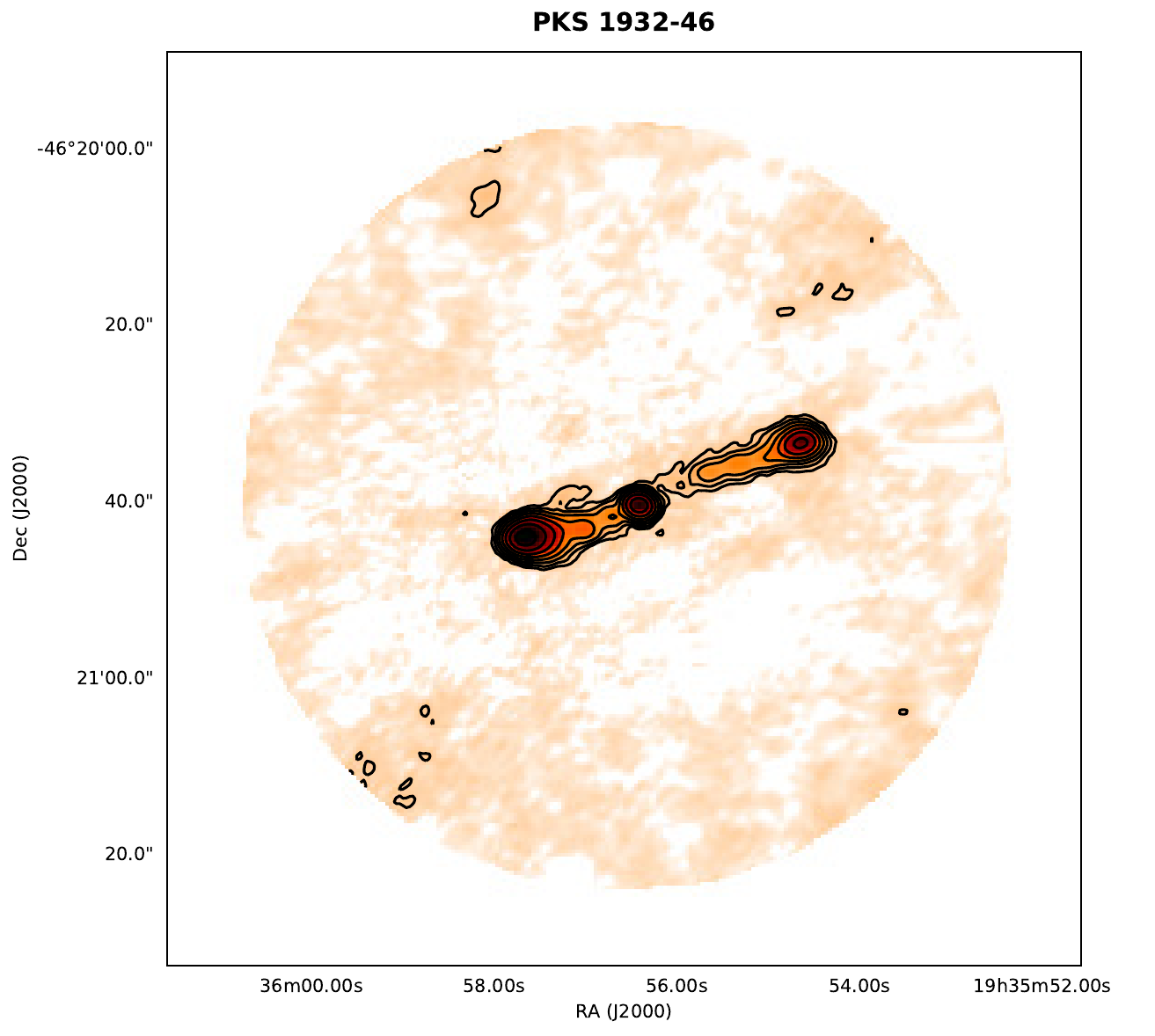} 
\includegraphics[width=4.2cm]{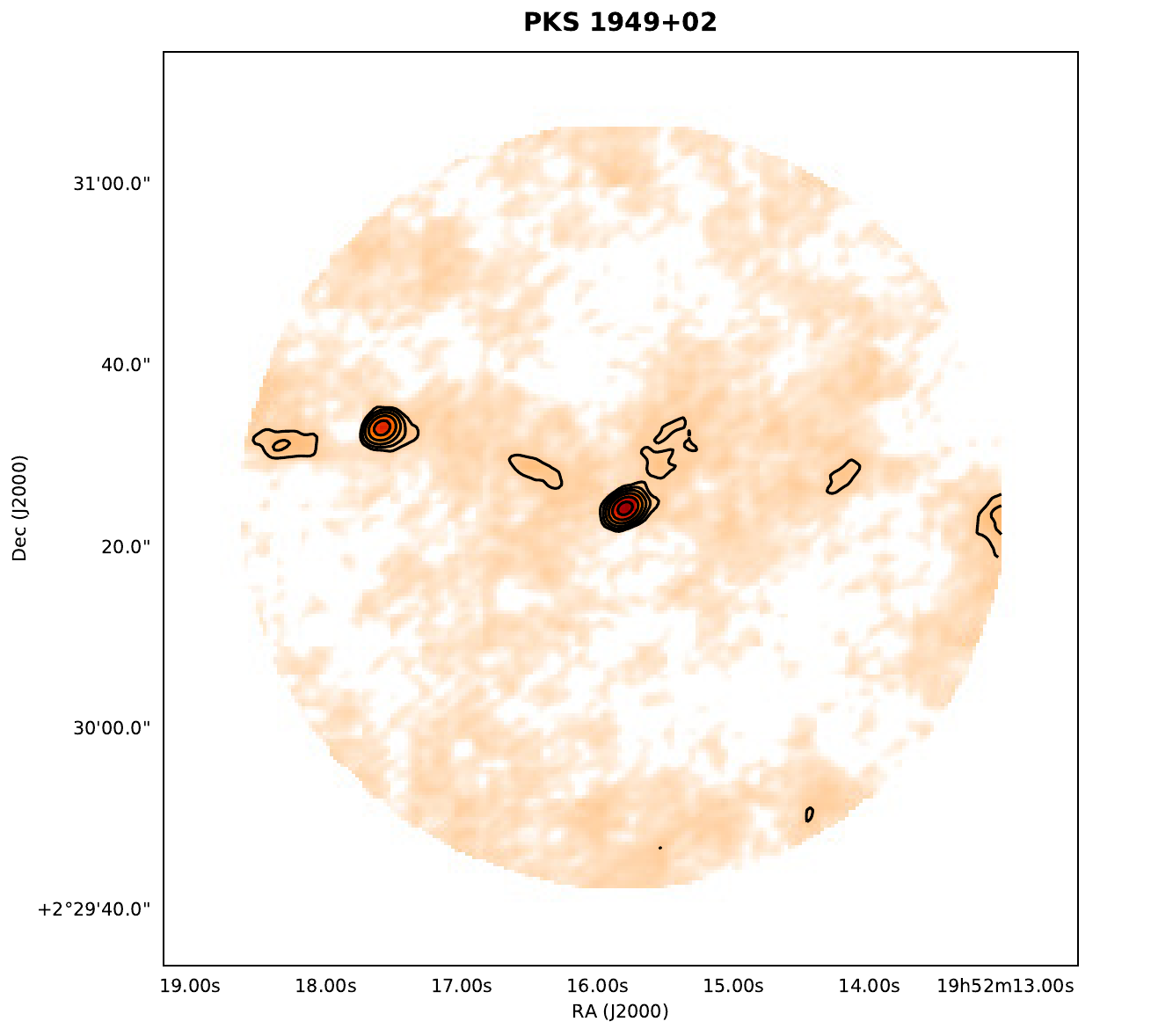} 
\includegraphics[width=4.2cm]{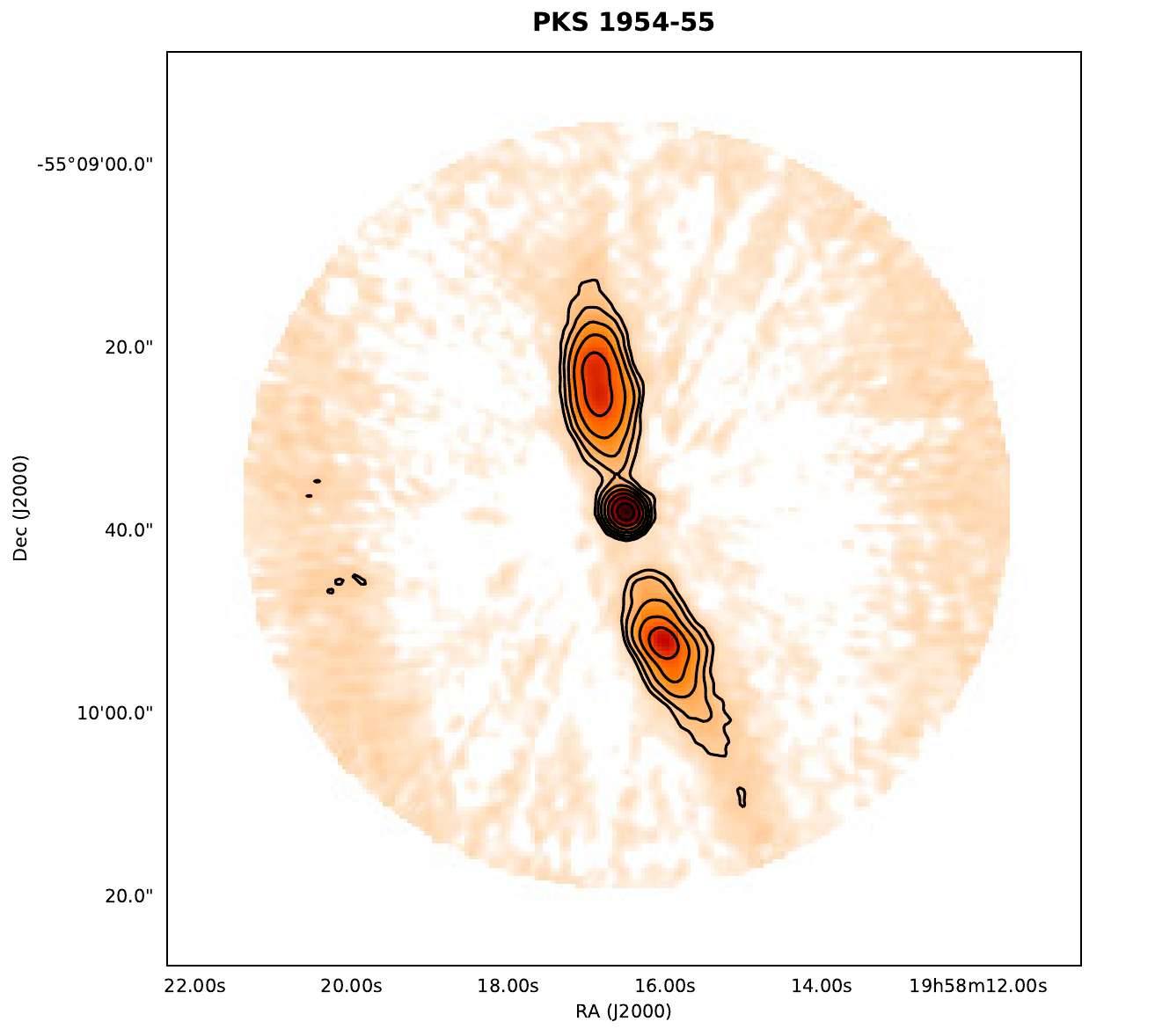} \\
\includegraphics[width=4.2cm]{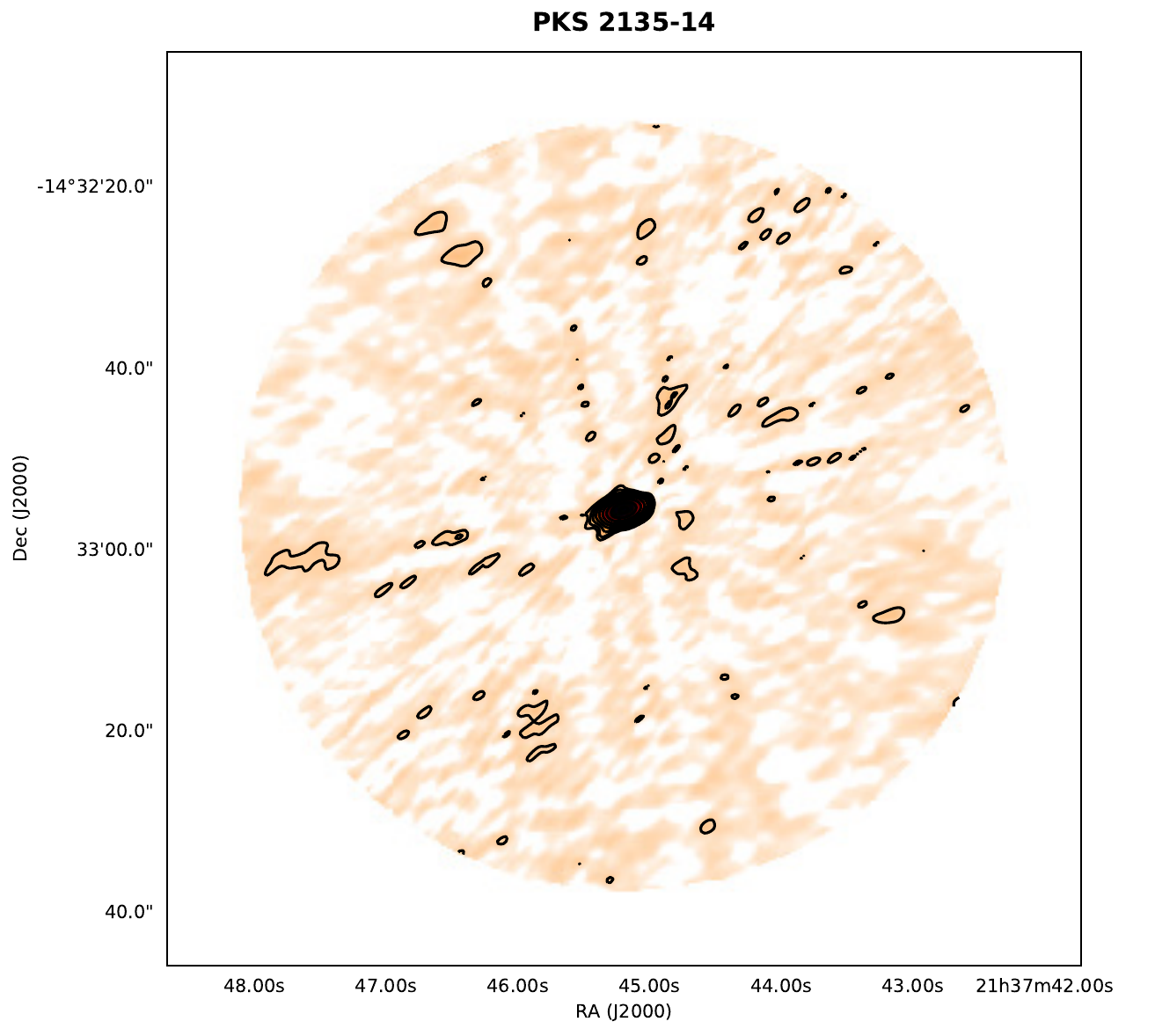} 
\includegraphics[width=4.2cm]{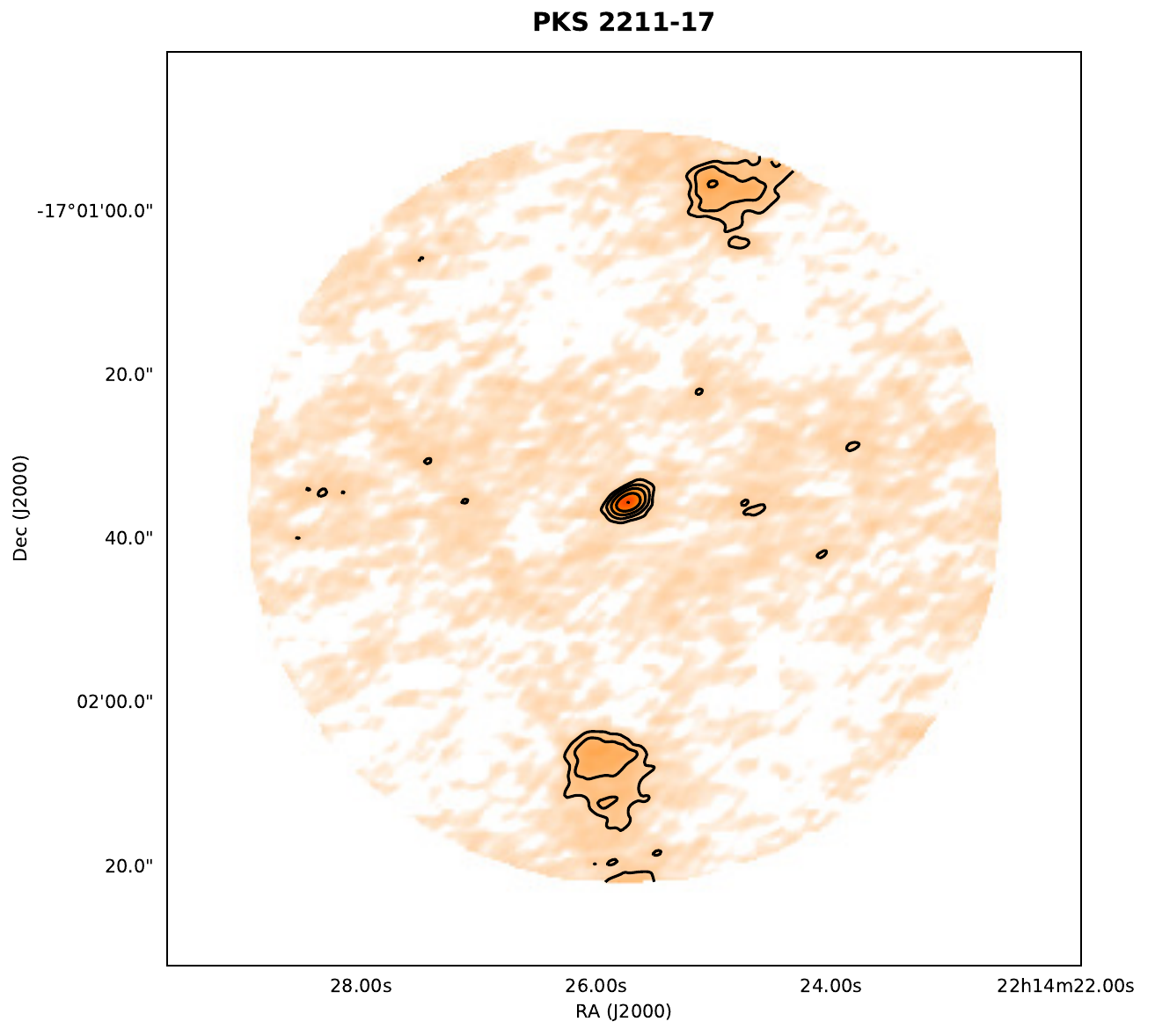} 
\includegraphics[width=4.2cm]{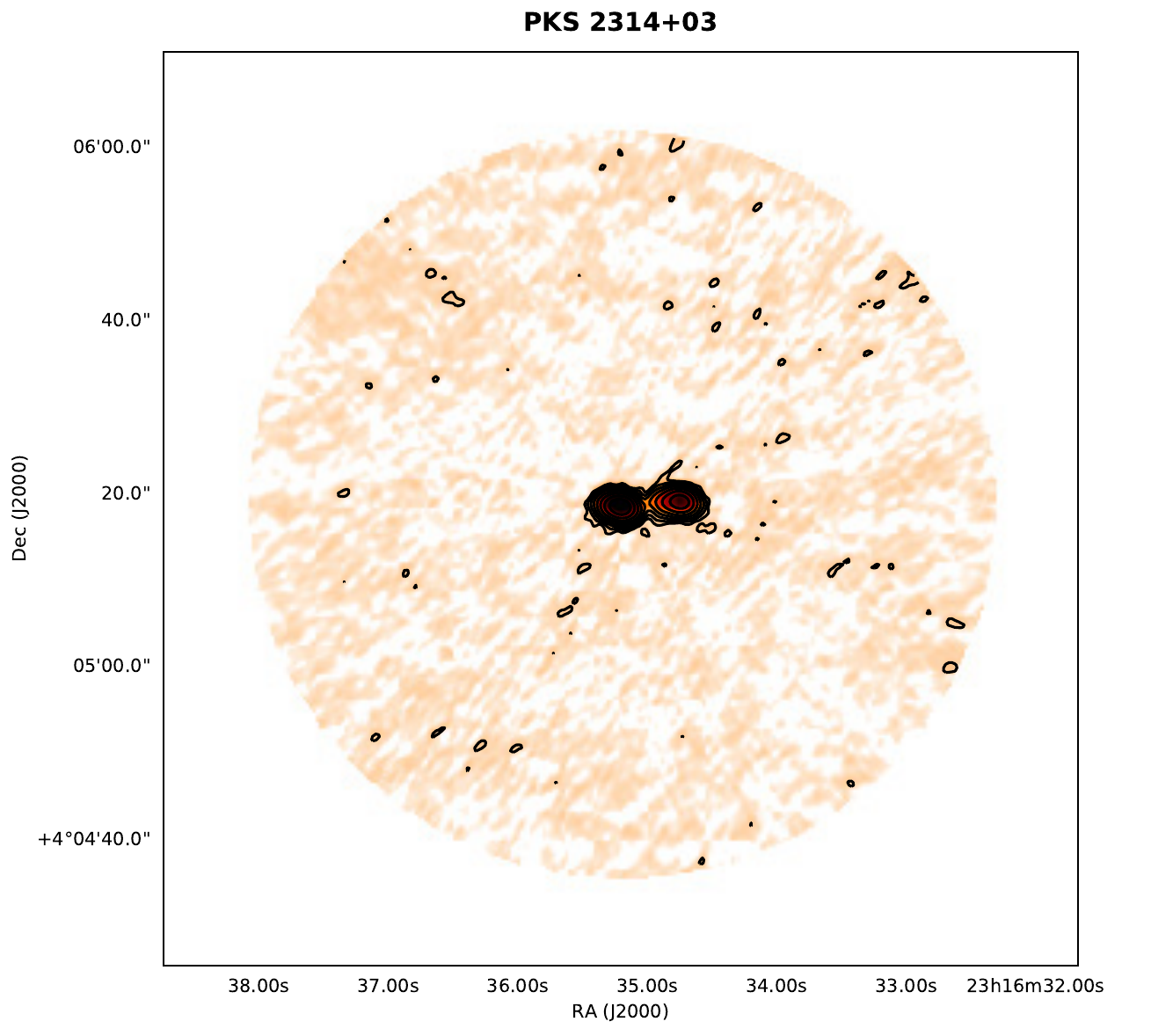} 
\includegraphics[width=4.2cm]{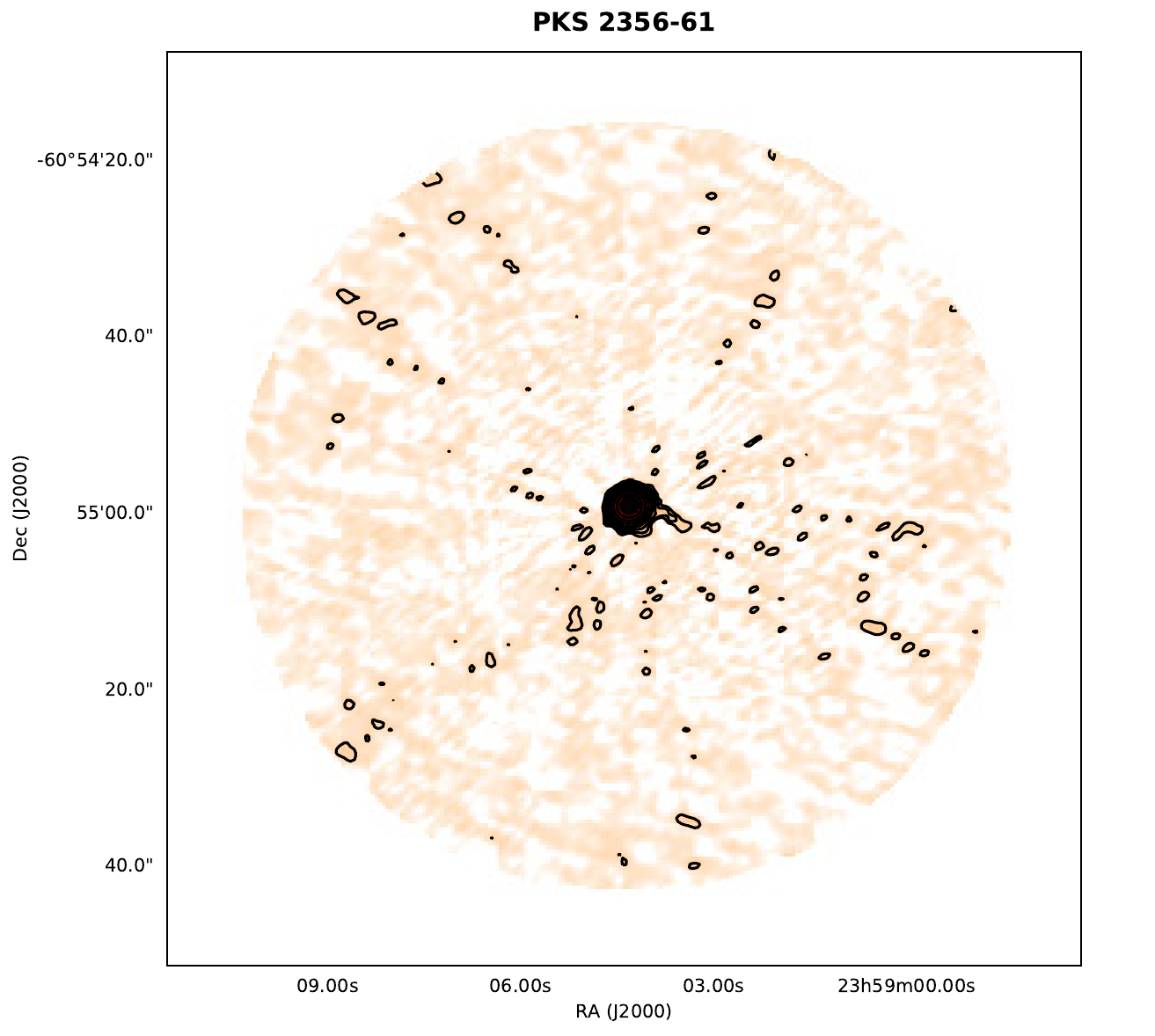} 

\caption{Images of the mm continuum emission of the 25 radio galaxies in the sample (the CSS/GPS objects - PKS1151-34, PHS1814-63, PKS1934-63 - are not shown, and the plot of PKS2221-02 is missing but the continuum is unresolved). Note that these images have not been corrected for the primary beam.
}
\label{fig:imagesCont}
\end{figure*}

\newpage

\begin{figure*}
   \centering
\includegraphics[height=7cm]{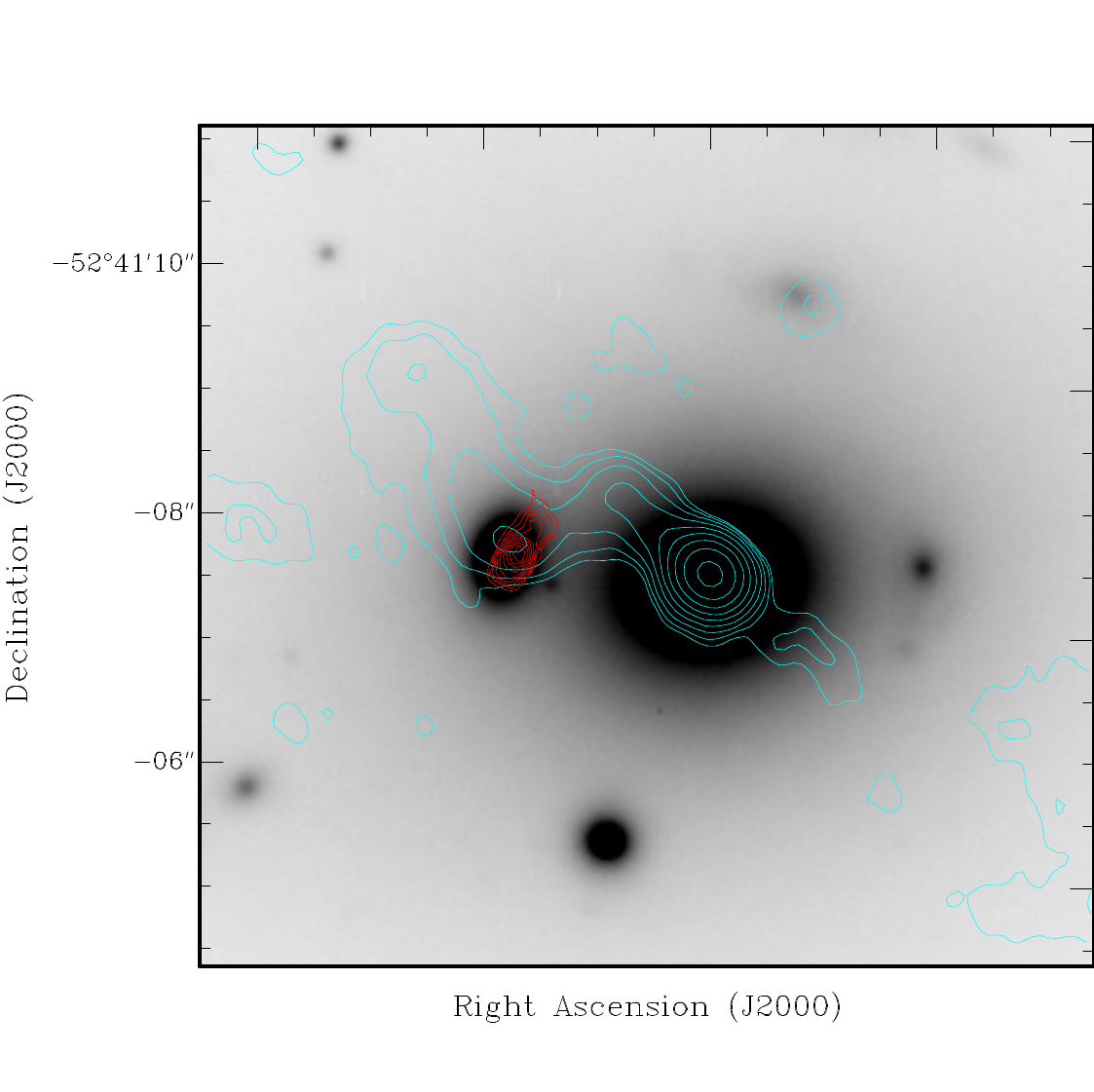}
\includegraphics[height=5cm]{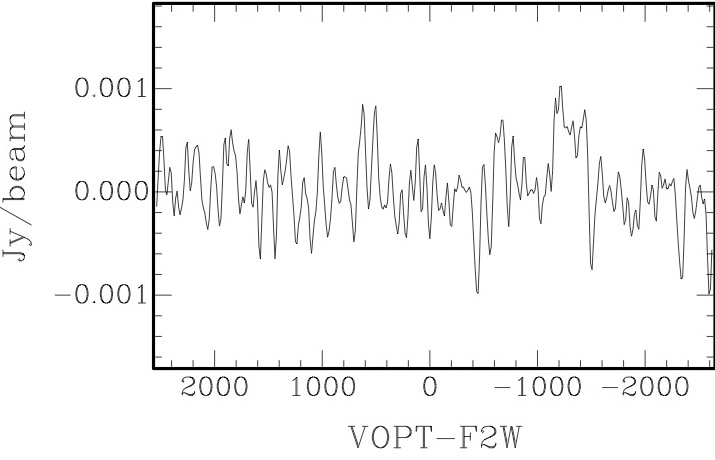} \\
\caption{A possible jet-cloud interaction in PKS0620--52. Left: of H$_2$ surface brightness derived from the CO observations (red) overlayed on the r' optical image of \citet{Ramos11} (grey scle), and the 3-mm continuum
(cyan) for PKS0620--52. The molecular gas appears to be associated with a companion located, in projection, where the radio jet bends. Right:  the \coOne\  profile at the location of the emission.  
}
\label{fig:CompanionsTentative}
\end{figure*}


\begin{figure*}
   \centering
\includegraphics[height=8.0cm]{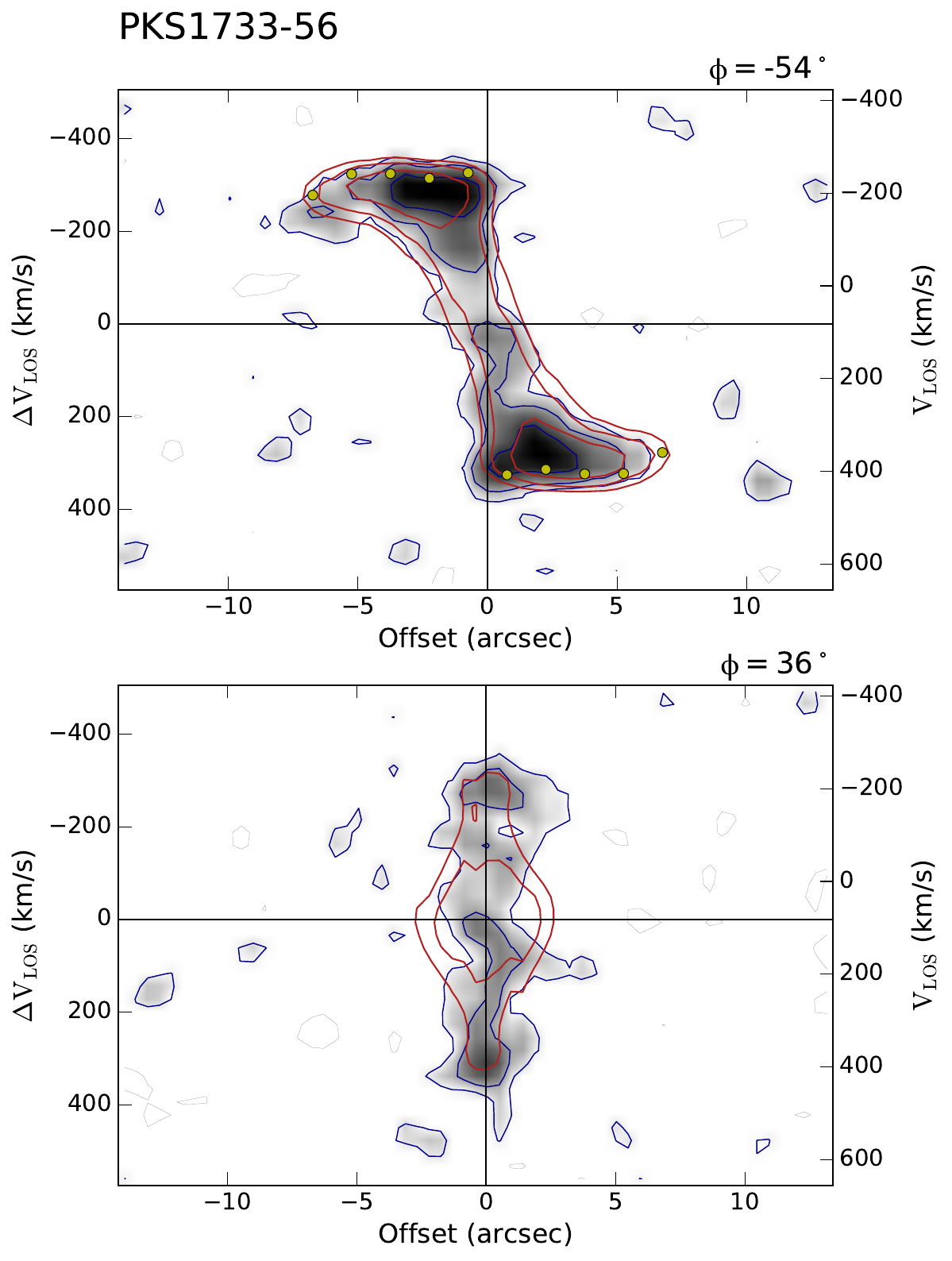} 
\includegraphics[height=8.0cm]{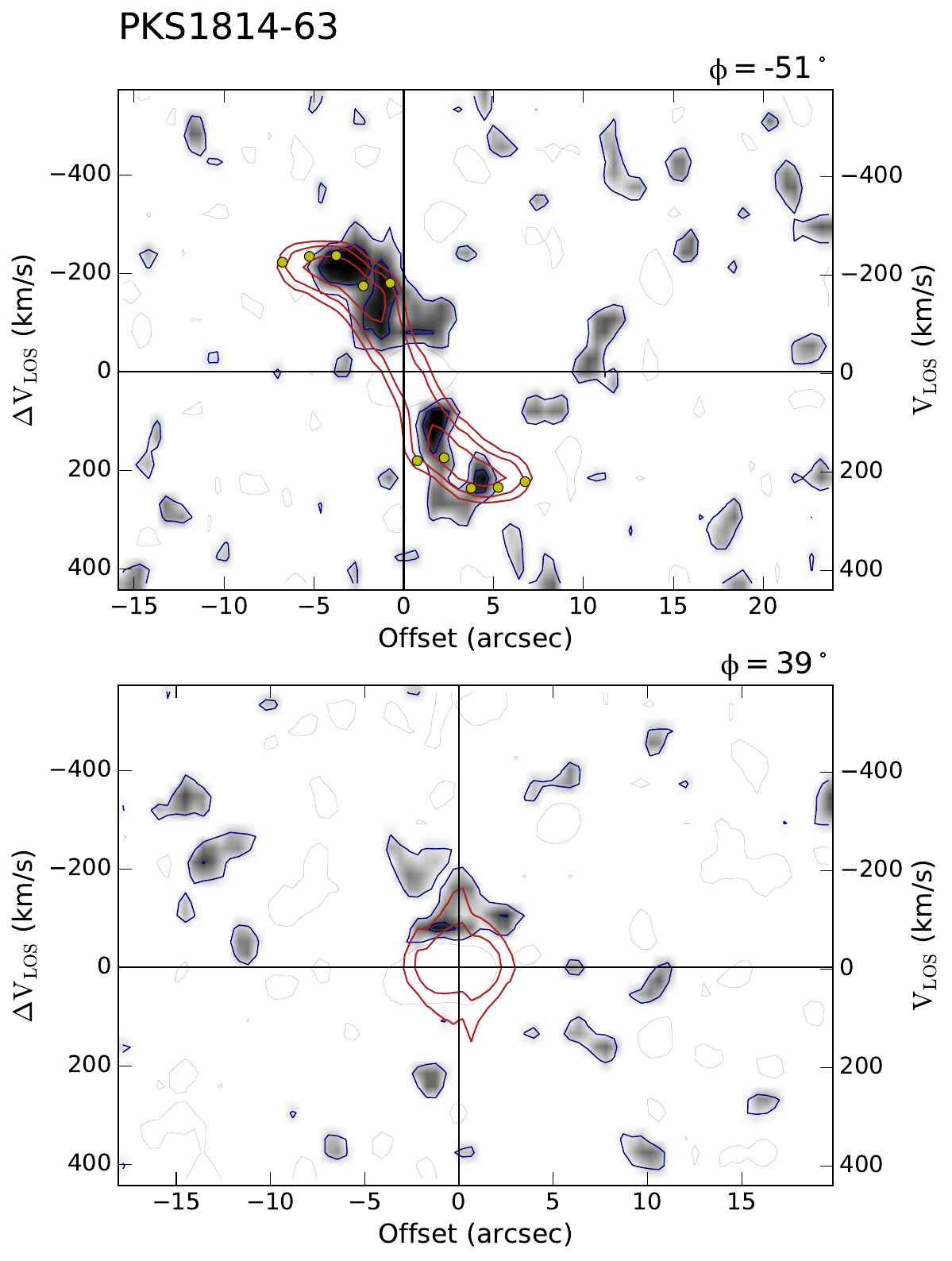} \\
\includegraphics[height=8.0cm]{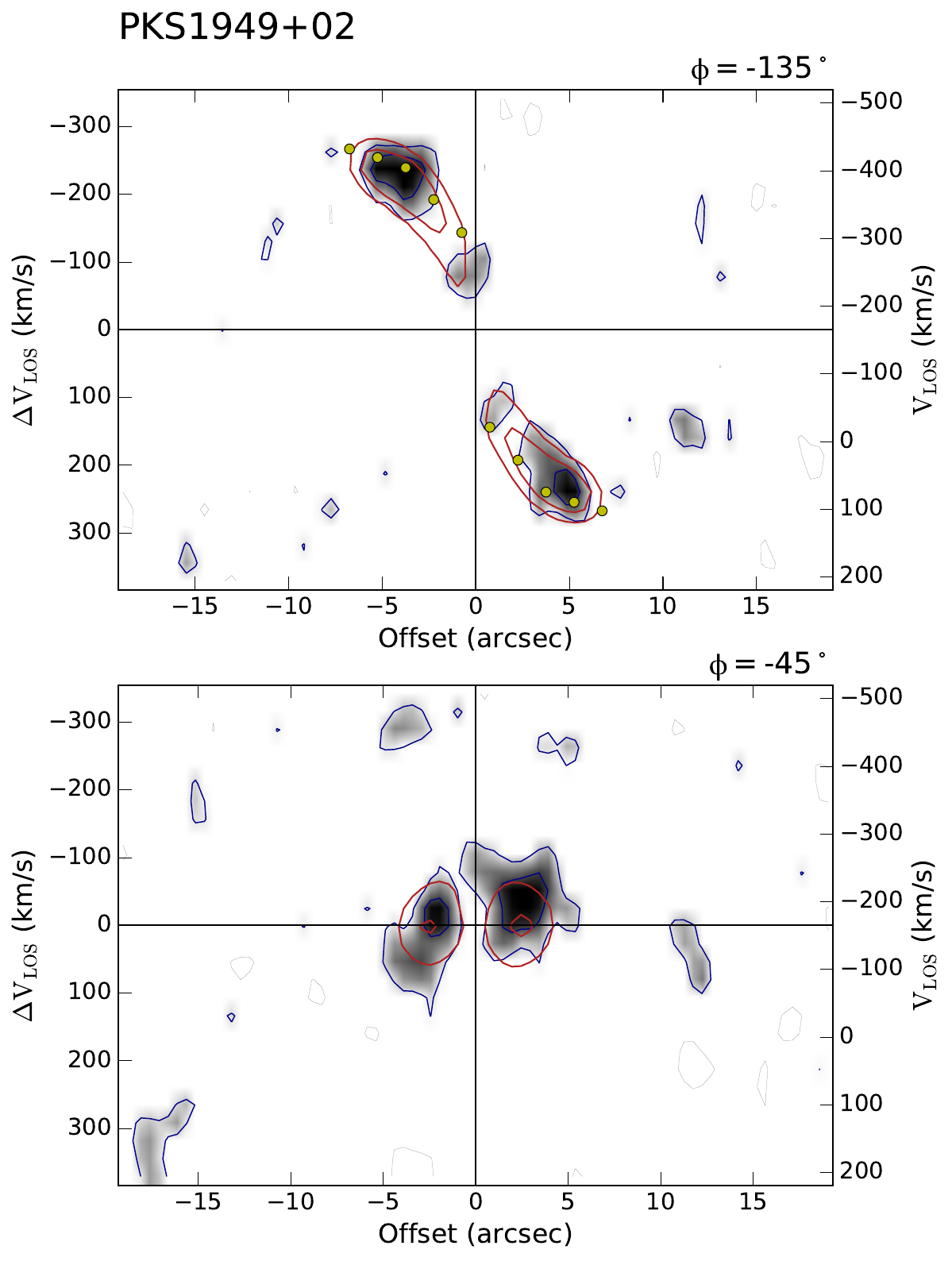}
\includegraphics[height=8.0cm]{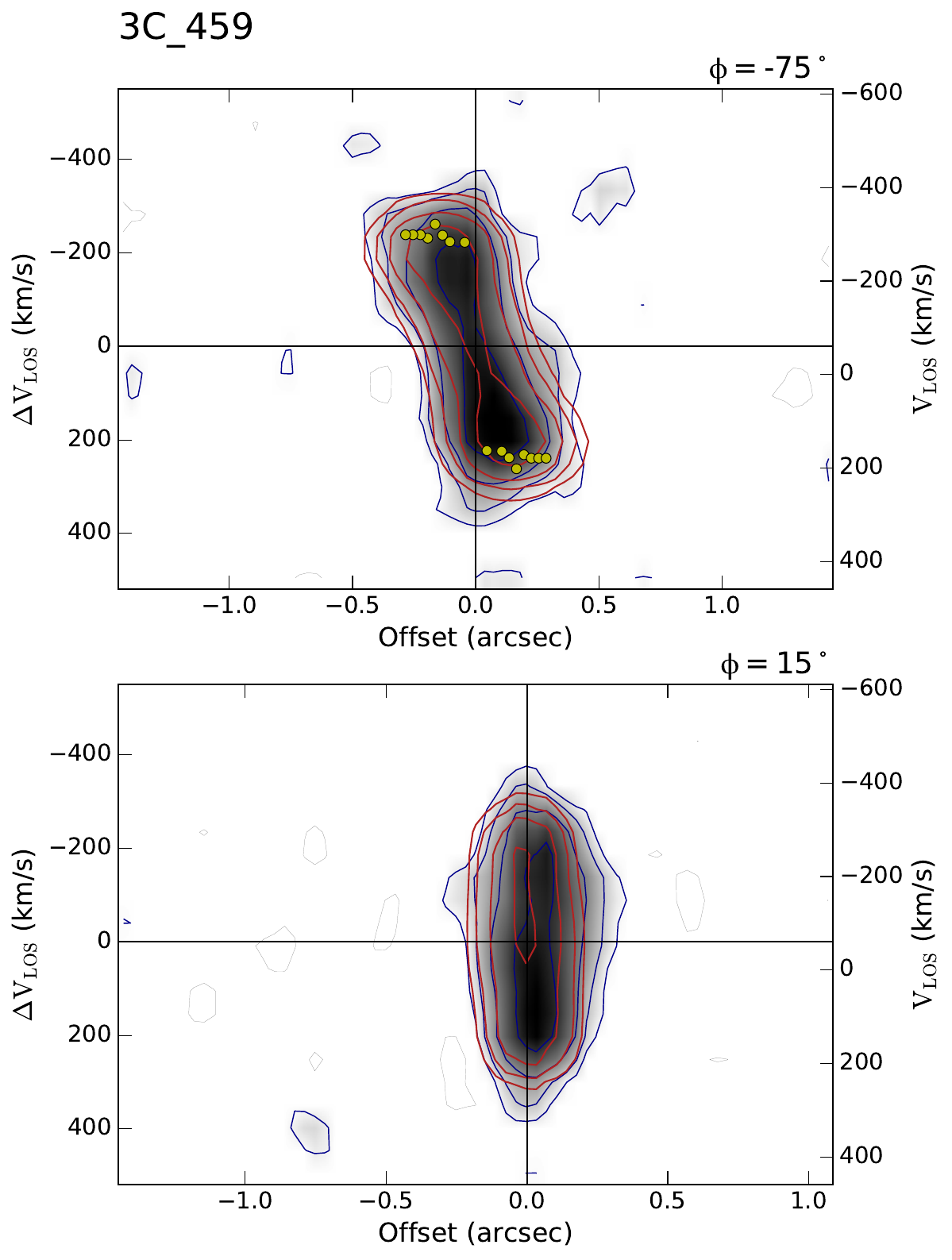}
\caption{Position-velocity plots, taken along major and minor axis, from the modelling of the CO distribution using the 3D-Barolo software \citet{DiTeodoro15}.  The grey scale represent the data and the contours the model. The cube of 3C~459 is the one at higher spatial resolution from the ALMA archive: ADS/JAO.ALMA\#2018.1.00739.S, PI Balmaverde.}

\label{fig:Barolo}

\end{figure*}
\clearpage
\end{appendix}

\bsp	
\label{lastpage}
\end{document}